\documentclass[conference]{IEEEtran}
\usepackage{filecontents}
\usepackage{multirow}
\usepackage{multicol}
\usepackage{tikz}
\usepackage{amsmath}
\usepackage{amsfonts}
\usepackage{nicefrac}
\usepackage{microtype}
\usepackage{lipsum}
\usepackage{graphicx}
\usepackage{float}
\usepackage{algpseudocode}
\usepackage{algorithm}
\usepackage{subcaption}
\usepackage{booktabs}
\usepackage{makecell}
\usepackage{colortbl}
\usepackage{mathabx}
\usepackage{contour}

\usepackage{cite}               
\usepackage{breakurl}           
\usepackage{url}                
\usepackage{xcolor}             
\usepackage[]{hyperref}         
\hypersetup{                    
  colorlinks,
  linkcolor={green!80!black},
  citecolor={red!70!black},
  urlcolor={blue!70!black}
}

\usepackage{mathptmx}  

\usepackage[T1]{fontenc}
\usepackage[utf8]{inputenc}
\usepackage{pslatex}

\newcommand{\Tech}{{{\sc Beagle}{}}}

\usepackage{amsmath,amsfonts,bm}

\def\eqref#1{equation~\ref{#1}}

\def\1{\bm{1}}

\DeclareMathAlphabet{\mathsfit}{\encodingdefault}{\sfdefault}{m}{sl}
\SetMathAlphabet{\mathsfit}{bold}{\encodingdefault}{\sfdefault}{bx}{n}

\newcommand{\normlone}{L^1}
\newcommand{\normltwo}{L^2}

\DeclareMathOperator*{\argmin}{arg\,min}

\usepackage[normalem]{ulem}
\usepackage{soul}

\newcommand*\circled[1]{\tikz[baseline=(char.base)]{
            \node[shape=circle,draw,inner sep=1pt] (char) {#1};}}

\definecolor{Gray}{gray}{0.935}
\newcolumntype{g}{>{\columncolor{Gray}}c}
\newcolumntype{G}{>{\columncolor{Gray}}r}

\begin{document}

\title{\Tech: Forensics of Deep Learning Backdoor Attack for Better Defense\vspace{-20pt}}


\author{
\IEEEauthorblockN{Siyuan Cheng, Guanhong Tao, Yingqi Liu, Shengwei An, Xiangzhe Xu,
\\ Shiwei Feng, Guangyu Shen, Kaiyuan Zhang, Qiuling Xu, Shiqing Ma$^{\dagger}$, Xiangyu Zhang}
\IEEEauthorblockN{Purdue University, $^{\dagger}$Rutgers University}
\IEEEauthorblockA{\{cheng535, taog, liu1751, an93, xu1415, feng292, shen447, zhan4057, xu1230, xyzhang\}@cs.purdue.edu\\
$^{\dagger}$sm2283@cs.rutgers.edu}
\vspace{1pt}
}

\IEEEoverridecommandlockouts
\makeatletter\def\@IEEEpubidpullup{6.5\baselineskip}\makeatother
\IEEEpubid{\parbox{\columnwidth}{
    Network and Distributed System Security (NDSS) Symposium 2023\\
    28 February - 4 March 2023, San Diego, CA, USA\\
    ISBN 1-891562-83-5\\
    https://dx.doi.org/10.14722/ndss.2023.24944\\
    www.ndss-symposium.org
}
\hspace{\columnsep}\makebox[\columnwidth]{}}

\maketitle

\begin{abstract}
Deep Learning backdoor attacks have a threat model similar to traditional cyber attacks. Attack forensics, a critical counter-measure for traditional cyber attacks, is hence of importance for defending model backdoor attacks.
In this paper, we propose a novel model backdoor forensics technique. 
Given a few attack samples such as inputs with backdoor triggers, which may represent different types of backdoors, our technique automatically decomposes them to clean inputs and the corresponding triggers. It then clusters the triggers based on their properties to allow automatic attack categorization and summarization. Backdoor scanners can then be automatically synthesized to find other instances of the same type of backdoor in other models.
Our evaluation on 2,532 pre-trained models, 10 popular attacks, and comparison with 9 baselines show that our technique is highly effective. The decomposed clean inputs and triggers closely resemble the ground truth. The synthesized scanners substantially outperform the vanilla versions of existing scanners that can hardly generalize to different kinds of attacks.
\end{abstract}

\section{Introduction} \label{sec:intro}

Deep Learning (DL) backdoor attacks~\cite{badnet,trojnn} leverage vulnerabilities in pre-trained models such that inputs stamped with a specific (small) input pattern (e.g., a polygon patch) or undergone some fixed transformation (e.g., applying a filter) induce intended model misbehaviors, such as misclassification to a {\em target label}. The misbehavior-inducing input patterns/transformations are called {\em backdoor triggers}. The vulnerabilities are usually injected through various data poisoning methods~\cite{dynamic,inputaware,composite,blend,invisible,wanet,saha2020hidden}. Some even naturally exist in normally trained models~\cite{moth,deck}.

\begin{figure}[t]
    \centering
    \begin{tabular}{lcr}
    \begin{minipage}[b]{.133\textwidth}
        \includegraphics[height=45mm]{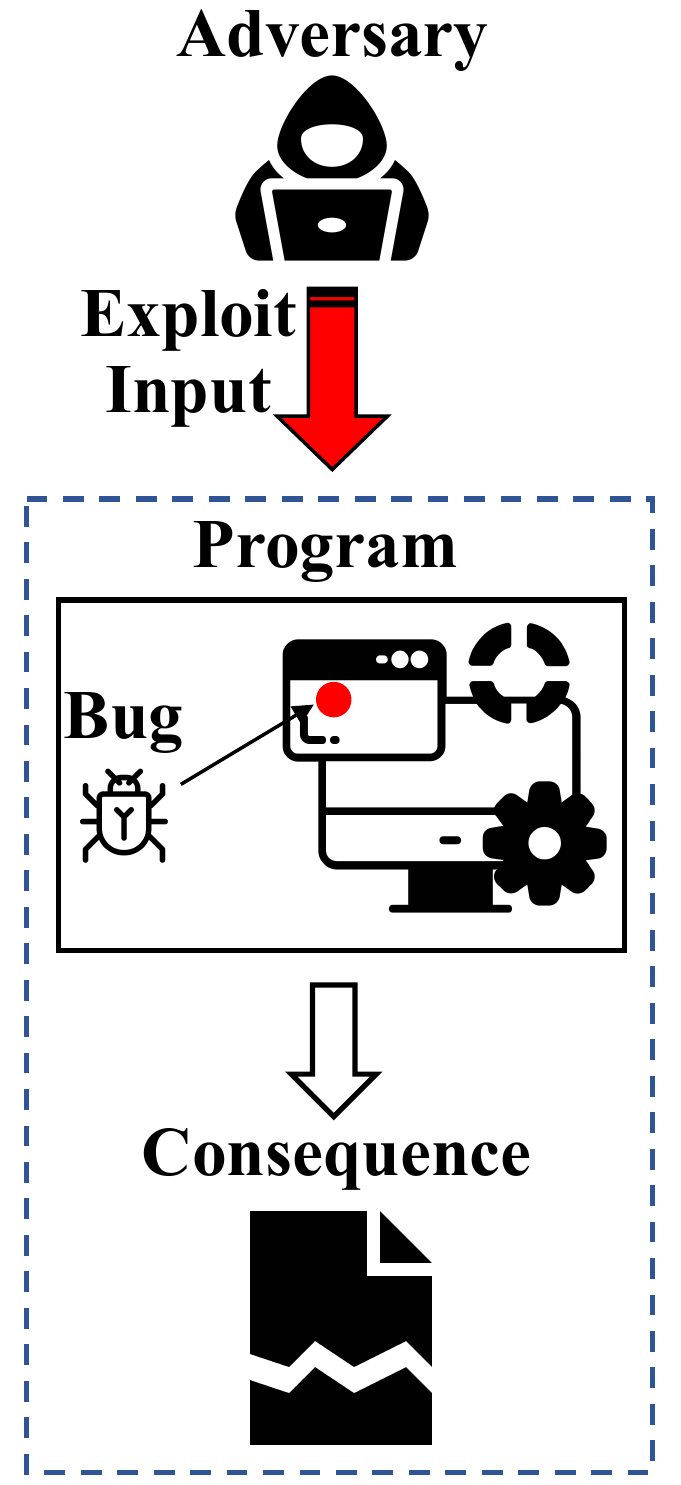}
        \captionsetup{width=1.2\textwidth}
        \caption{Cyber attack}
        \label{fig:trad_attack}
    \end{minipage}
    &&
    \begin{minipage}[b]{.337\textwidth}
        \includegraphics[height=45mm]{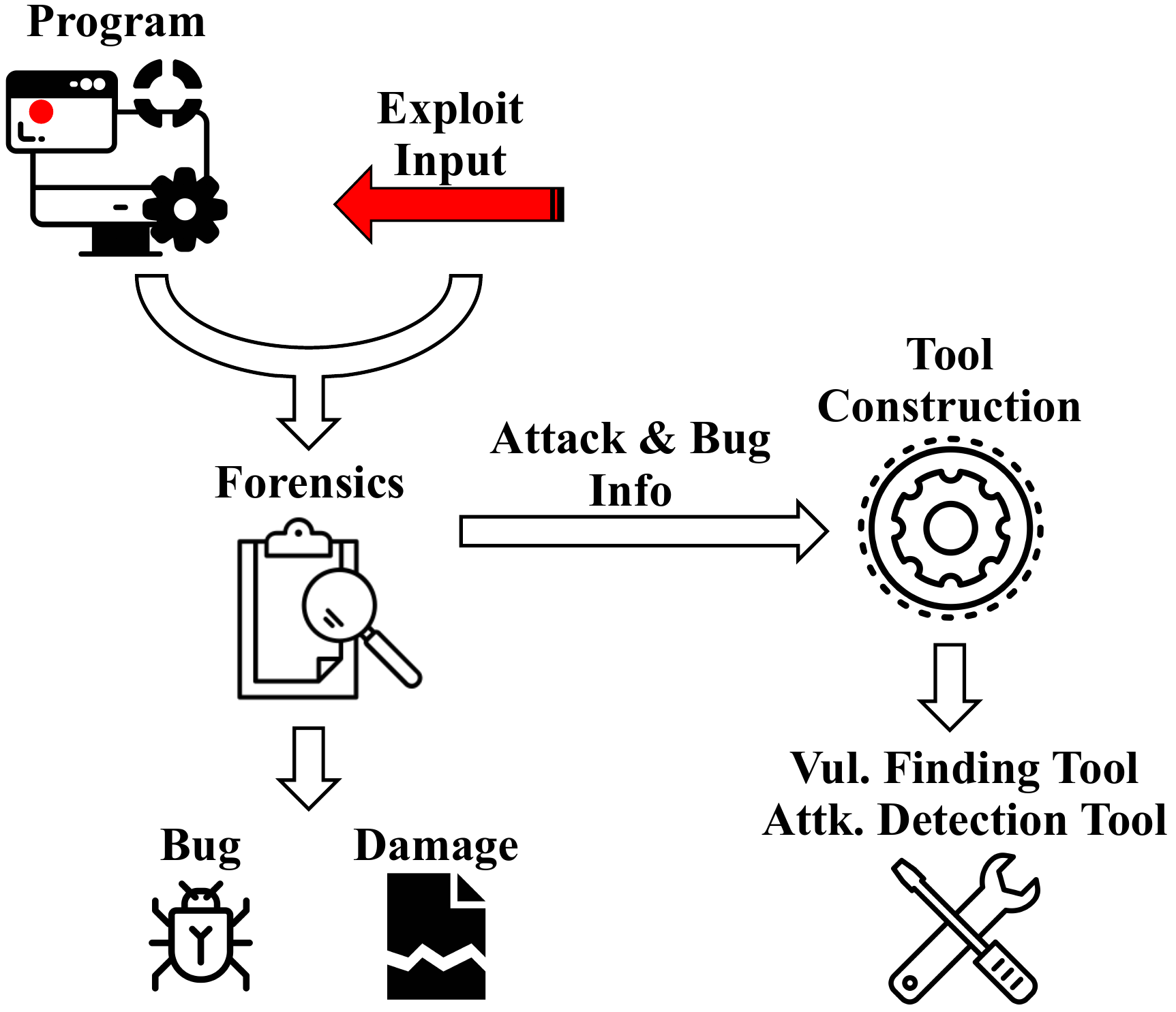}
        \caption{Forensics}
        \label{fig:trad_forensic}
    \end{minipage}
    \end{tabular}
\end{figure}

The attack model of DL backdoors becomes increasingly similar to that of traditional cyber attacks (on software), and in the meantime DL models have more and more applications in critical tasks such as autonomous driving and ID recognition (for access control). 
Defending model backdoors hence becomes a pressing need. Figure~\ref{fig:trad_attack} shows the traditional cyber attack model. Vulnerabilities exist in applications (e.g., due to implementation bugs). 
The adversary exploits a vulnerability by crafting a special input, e.g., an extremely long input to exploit a buffer overflow vulnerability. 
The exploit could lead to a wide range of damage (e.g., hijacking a system, leaking information, and corrupting services).
The adversary has no control of the execution environment of application on the user side (the dotted box in Figure~\ref{fig:trad_attack}).
He can only manipulate the input to achieve his goal. Many inputs can be easily crafted to exploit the same vulnerability. And vulnerabilities can be patched by fixing bugs.

Analogously in DL backdoor attacks, vulnerabilities are model properties such that (any) inputs can be transformed in a specific way to exploit them. The process of crafting inputs does not require access to model execution (on the user side). The input crafting efforts are minimal as triggers are known by the adversary beforehand (because he injected them). In contrast, traditional adversarial attack~\cite{carlini2017towards,madry2018towards,brendel2018decision} usually requires much more computing efforts to generate exploit perturbations. Some even do that on-the-fly.  
Moreover, backdoors can be effectively removed by model hardening with negligible model accuracy degradation~\cite{li2021neural,wu2021adversarial,liu2018fine,zhao2020bridging}. Model misbehaviors can have a lot of downstream ramifications. For example, misclassifying a stop sign to something else could have catastrophic consequences in an auto-driving system. 

Forensics~\cite{mpi,nodoze,clarion,omega,alchemist} is an important countermeasure for traditional cyber attacks. As shown in Figure~\ref{fig:trad_forensic}, given attack instances (including the application and a small set of exploit inputs), forensics techniques aim to identify their root causes (e.g., the bug), assess damage, and provide critical information to build vulnerability/malware scanners to identify similar attack instances and similar bugs.
They also greatly facilitate attack prevention and program repair~\cite{liu2019tbar, nguyen2013semfix, le2012systematic, wen2018context}. 
Due to the similarity of DL backdoor attacks and traditional
cyber attacks, we argue that  forensics is an important step in 
fighting against DL backdoor attacks as well.
There are existing efforts in detecting inputs that contain backdoor triggers~\cite{februus,huang2020one} and recognizing, cleansing poisonous inputs from training data using evidence collected from a few attack instances~\cite{strip,chen2018detecting}. The former aims to decide if a given input contains any backdoor trigger. Existing techniques usually leverage the observation that such inputs manifest themselves by having out-of-distribution values in the input or feature space~\cite{tran2018spectral,februus}.
The latter searches for a subset of training samples such that training on the subset reduces the {\em attack success rate} (ASR) to almost 0 without causing model accuracy degradation.
\textit{Februus}~\cite{februus} aims to cleanse individual trojaned inputs by removing stamped triggers. It first identifies the trigger in a given input using GradCAM~\cite{gradcam} based on the assumption that the classification output is dominated by the trigger area. It then removes the entire trigger area and uses GAN to fill in the space.
These existing works focus on specific sub-problems in forensics, inspiring a more comprehensive forensics workflow.
For example, backdoor input detection techniques can be used to capture attack instances for downstream forensics analysis.

In this work, we propose a novel DL backdoor forensics method \Tech{} ({\it Forensics of
\underline{\sc B}ackdoor attack on deep l\underline{\sc EA}rnin\underline{\sc G} mode\underline{\sc L}s for better defens\underline{\sc E}}). Given a few attack instances, each including the model and a few inputs likely containing backdoor triggers, \Tech{} automatically decomposes each trojaned input to a clean input and a trigger. The trigger could be a patch-like input pattern or an input space transformation function. The decomposed clean input should closely resemble the original clean input (which is unknown to \Tech{}), and the decomposed trigger should be very similar to the injected trigger (which is also unknown to \Tech{}). The decomposed trigger will be able to flip a large set of clean inputs to the same target label, if applied.
This is analogous to the root cause analysis stage in traditional cyber attack fornensics.
More importantly, \Tech{} will automatically cluster these attack instances leveraging the  decomposition results such that each cluster denotes a specific type of backdoor.  It further summarizes each cluster to a set of distributions, and automatically synthesizes a corresponding scanner to find the same type of backdoor {\em in other models}.   
Note that the instantiations of a type of backdoor on different models are largely different. For example, different patch attack instances (on different models) may have different patch shapes, sizes, pixel patterns, and different positions to stamp the patches. It is unlikely 
that we can detect other instances of the same type of attack by simply stamping the raw decomposed triggers produced by the forensics analysis. Instead, \Tech{} abstracts the given instances such that other instantiations can be detected. This is analogous to building vulnerability and malware detection tools based on forensics results in traditional cyber security.

Our method formulates the attack decomposition step as a cyclic optimization problem. At the beginning, the decomposed clean input and the decomposed trigger are of very low quality, for instance, some random disintegration of the trojaned input. The cyclic optimization ensures that any improvement on the decomposed clean input leads to improvement of the decomposed trigger, and vice versa. High quality decomposition can be achieved when the process converges. 
We formulate backdoor attacks in two mathematical forms: {\em patching attacks} that induce localized input perturbations and {\em transforming attacks} that induce pervasive perturbations. As such, the existing wide range of different attacks can be modeled by different coefficient distributions for the mathematical forms, allowing us to achieve automatic categorization.
The formulas and their coefficient distributions are then used to synthesize loss functions for scanners.
A scanner determines if a model contains any backdoor, without requiring any trojaned inputs, analogous to a traditional malware/vulnerability scanner, which scans without (exploit) inputs.
Given a model to scan, the synthesized loss functions are used to invert a backdoor trigger for the model. If such inversion succeeds, the model is considered trojaned.
The inversion process essentially generates small input perturbation patterns or transformation functions by gradient descent based on the synthesized loss functions such that the generated perturbation/transformation (i.e., trigger) can induce model misclassification.
Our contributions are summarized in the following.

\begin{itemize}
    \item We propose a novel model backdoor attack forensics technique that contains automatic attack root cause analysis, attack summarization, and scanner synthesis.
    \item Our root cause analysis features a new cyclic optimization pipeline that can decompose a trojaned input to its clean version and the trigger. 
    \item We propose to formulate existing attacks using two mathematical forms such that different attacks become different distributions of coefficients of the two forms, enabling automatic attack categorization, and scanner synthesis.
    \item We evaluate our prototype \Tech{} on 10 popular backdoor attacks, including BadNets~\cite{badnet}, TrojNN~\cite{trojnn}, Dynamic~\cite{dynamic}, Reflection~\cite{reflection}, SIG~\cite{sig}, Blend~\cite{blend}, Invisible~\cite{invisible}, WaNet~\cite{wanet}, Instagram filter~\cite{ABS}, DFST~\cite{dfst}, and on 2,532 pre-trained models. We demonstrate the benefits of forensics analysis by enhancing five existing backdoor scanning techniques and comparing with an existing trojaned input decomposition method.
    Our results show that existing scanners have substantial performance degradation when they are used to scan attacks that they are not designed for (e.g., from over 0.9 scanning accuracy down to lower than 0.55), whereas the scanners synthesized by \Tech{} can achieve 0.9 detection accuracy for all these attacks, when only 10 trojaned input instances are assumed for each attack during forensics and the models under scanning are completely different from the ones used in forensics.
    We also show that \Tech{} can even improve existing scanners' performance on their targeted attacks by 9\%-27\% because although they are fined-tuned for the targeted attacks, the fine-tunings were done manually by their original developers, whereas \Tech{} automatically synthesizes scanners.
    Our experiments also show that the trojaned input decomposition produces high-quality results. The decomposed clean images are 22\% more similar to the ground truth than a baseline method Februus~\cite{februus}. And 100\% of them are correctly classified by the models, compared to 38\% by the baseline. Our decomposed triggers achieve 96\% ASR whereas those by the baseline can only achieve 45\%.
    Our ablation study, sensitivity study, and adaptive attack show that \Tech{} has a robust design.
\end{itemize}

\smallskip
\noindent
{\bf Threat Model.} Our threat model is similar to that in data poisoning~\cite{badnet,trojnn,blend,reflection,wanet}, in which the adversary has access to (and even own) the training dataset. Hence, we do not focus on cleansing the training data. Instead, we focus on the following scenario.
Users observe a few unusual misclassifications (through manual inspection or using trojaned input detection tools).
For example, when there is a new type of backdoor attack, the attackers may use it to attack many models (just like buffer-overflow is leveraged to attack many software applications). We assume some attacker exploits one of these vulnerable models and the attack samples are detected and saved for forensics (analogous to the first time a buffer-overflow exploit is detected). These samples, together with the model, are submitted to security analysts that are equipped with \Tech.
In the meantime, the users provide a small set of clean inputs (e.g., 100 per model) to facilitate the process, which is consistent with the literature~\cite{nc,Tabor,dltnd,guo2020towards,huang2019neuroninspect}. Besides the information provided by the users, the analysts also have GANs representing input distributions. This is a reasonable assumption (consistent with the literature~\cite{goodfellow2014generative,brock2018large,karras2019style}) as models used in real world applications follow physical world distributions and there are high quality pre-trained GANs representing such distributions. The analysts use \Tech{} to analyze and summarize the reported attacks and automatically construct scanners (analogous to synthesizing a buffer overflow bug detector) that can scan other models to find other instances of the same type of attack (e.g., patch attack). These scanners can be used by any user on any model.  Note that the same type of backdoor may have largely different instantiations on different models.
One cannot directly determine if an unknown model has a similar backdoor by directly stamping the decomposed trigger by \Tech{}. It is possible that the adversary injects backdoors that flip class A samples to class B, and the two classes are very similar in humans' eyes such that attack instances cannot be correctly recognized in the first place. Although it is debatable whether injecting backdoors in these classes can benefit the adverary as their decision bounday is already confusing, dealing with such attacks is beyond the scope of our paper.
A model may have multiple injected backdoors. We assume all the trojaned inputs in an attack instance (used in forensics) are exploiting the same backdoor. 

\section{Motivation} \label{sec:motivation}

\begin{figure}[t]
\centering
    \begin{subfigure}{.1\textwidth}
        \includegraphics[width=1.0\textwidth]{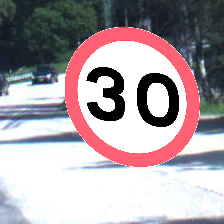}
        \caption{Victim}
    \end{subfigure}
    \begin{subfigure}{.1\textwidth}
        \includegraphics[width=1.0\textwidth]{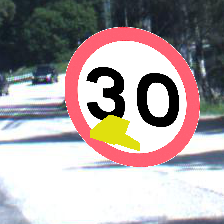}
        \caption{Trojaned}
    \end{subfigure}
    \begin{subfigure}{.1\textwidth}
        \includegraphics[width=1.0\textwidth]{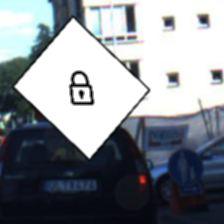}
        \caption{Target}
    \end{subfigure}
    \begin{subfigure}{.1\textwidth}
        \includegraphics[width=1.0\textwidth]{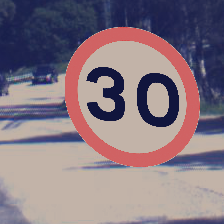}
        \caption{Nashville}
    \end{subfigure}
    
    \begin{subfigure}{.1\textwidth}
        \includegraphics[width=1.0\textwidth]{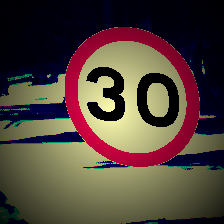}
        \caption{Lomo}
    \end{subfigure}
    \begin{subfigure}{.1\textwidth}
        \includegraphics[width=1.0\textwidth]{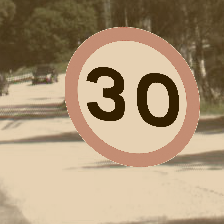}
        \caption{Kelvin}
    \end{subfigure}
    \begin{subfigure}{.1\textwidth}
        \includegraphics[width=1.0\textwidth]{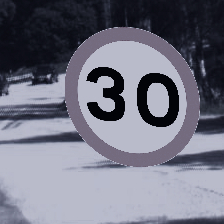}
        \caption{Gotham}
    \end{subfigure}
    \begin{subfigure}{.1\textwidth}
        \includegraphics[width=1.0\textwidth]{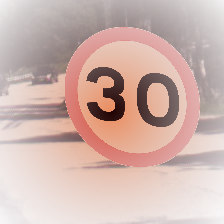}
        \caption{Toaster}
    \end{subfigure}
    
    \caption{Examples of different attacks in TrojAI
    }
    \label{fig:troj_exp}
\end{figure}

We use a number of attacks to illustrate that backdoor scanning using trigger inversion becomes ineffective if attack specifics are unknown, in order to motivate backdoor forensics.

\noindent
{\bf Trigger Inversion (Background).}
Trigger inversion is an effective method in backdoor scanning.
Readers familiar with such techniques can skip this subject. Given a  model and a few clean images, trigger inversion uses optimization to identify universal input perturbations that can flip the classification results of the clean images to a target class. In Neural Cleanse (NC)~\cite{nc}, the optimization aims to generate two vectors, a perturbation vector and a mask vector. The former describes value changes for individual pixels and the latter (i.e., values in a range of [0,1]) describes if the changes should be applied and how much is applied. For instance, a value 1.0 in the mask vector means the corresponding pixel is fully replaced by the value in the perturbation vector, a value 0 means that the pixel in the original input is intact, and a value in (0,1) means that the resulted pixel is a mix of the original and the perturbation. Trigger inversion is hence described as the following optimization problem.
\vspace{-2pt}
\begin{equation}
\label{eq:inversion}
    \argmin_{m,\; t}\;\;\;  \mathcal{L}(M((1 - m) \cdot x + m \cdot t), y_{t}) + \sigma \cdot \sum m, \vspace{-3pt}
\end{equation}
\noindent where $m$ denotes the mask vector and $t$ the perturbation vector. $M$ is the model, and $y_{t}$ a (potential) target label.
The loss function consists of two terms, the cross entropy loss and the regularization loss.
The cross entropy loss ($\mathcal{L}$) aims to achieve a high attack success rate (ASR) while the regularization loss aims to reduce the mask size. Coefficient
$\sigma$ controls the trade-off of the two.
At the beginning, $\sigma$ is small to ensure a high ASR of the inverted pattern. Then NC gradually increases $\sigma$ to find a small trigger.

NC decides that a model is trojaned if an exceptionally small trigger (whose size can be computed from the mask) is found for some target label to achieve a high ASR.
ABS~\cite{ABS} further enhances NC by adding a term to the loss function that aims to achieve large activation values for a few neurons that are determined to be compromised by the backdoor through an offline analysis. There are other trigger inversion techniques such as K-Arm~\cite{karm}, Tabor~\cite{Tabor}, DLTND~\cite{dltnd}, and DualTanh~\cite{tao2022better}.
They differ from each other by having different loss function designs (for the specific attacks they focus on).
While they are all highly effective for the attacks they tackle,
{\em they usually require the knowledge of attack specifics for crafting the corresponding loss functions and selecting the proper hyper parameters.}
In the following, we show that inversion techniques effective for an attack may not be effective for another attack.

\begin{figure}[t]
    \centering
    \begin{subfigure}{.13\textwidth}
        \centering
        \includegraphics[width=0.8\textwidth]{figures/trojai_demo/victim.png}
        \caption{Victim}
    \end{subfigure}
    \begin{subfigure}{.13\textwidth}
        \centering
        \includegraphics[width=0.8\textwidth]{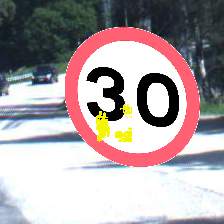}
        \caption{Inverted by NC}\label{fig:inverted_NC}
    \end{subfigure}
    \begin{subfigure}{.13\textwidth}
        \centering
        \includegraphics[width=0.8\textwidth]{figures/trojai_demo/polygon.png}
        \caption{Ground-truth}
        \label{fig:groundtruth}
    \end{subfigure}
    \caption{NC inversion of a universal polygon backdoor}
    \label{fig:NC_polygon}
\end{figure}

\begin{figure}[t]
\centering
    \begin{subfigure}[t]{0.11\textwidth}
        \centering
        \includegraphics[width=1.0\textwidth]{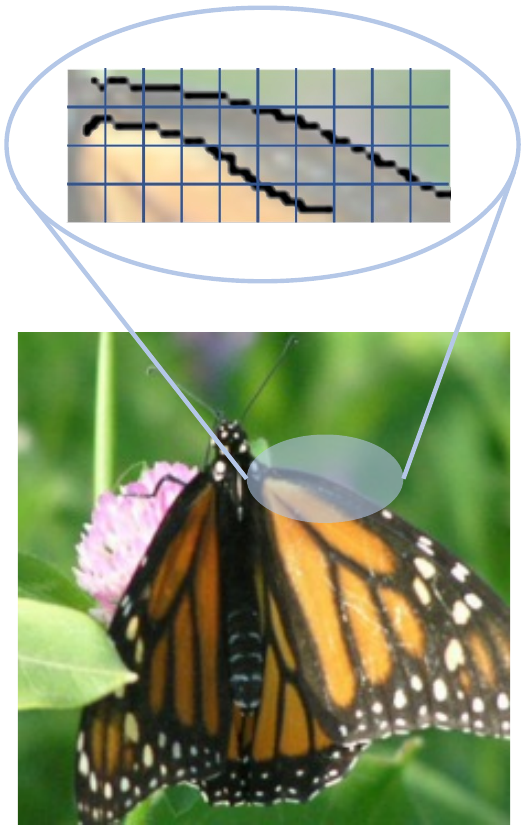}
        \caption{Victim}
    \end{subfigure}
    \begin{subfigure}[t]{.11\textwidth}
        \includegraphics[width=1.0\textwidth]{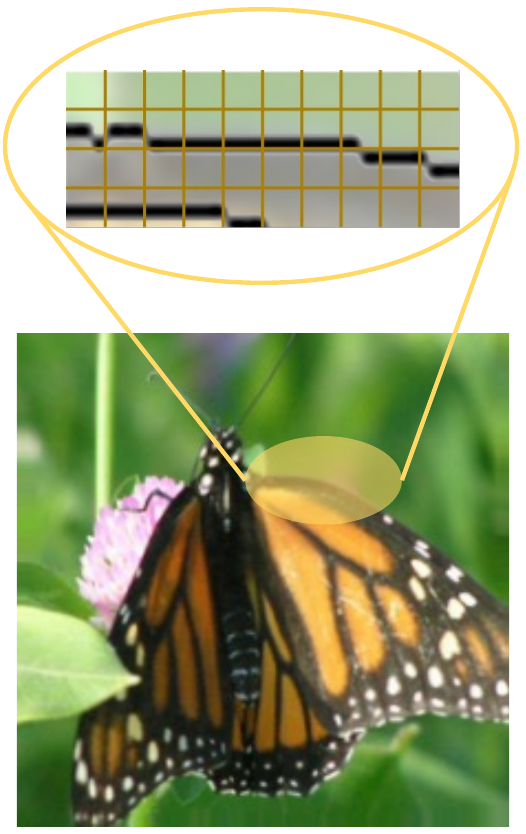}
        \caption{Trojaned}
    \end{subfigure}
    \begin{subfigure}[t]{.11\textwidth}
        \includegraphics[width=1.0\textwidth]{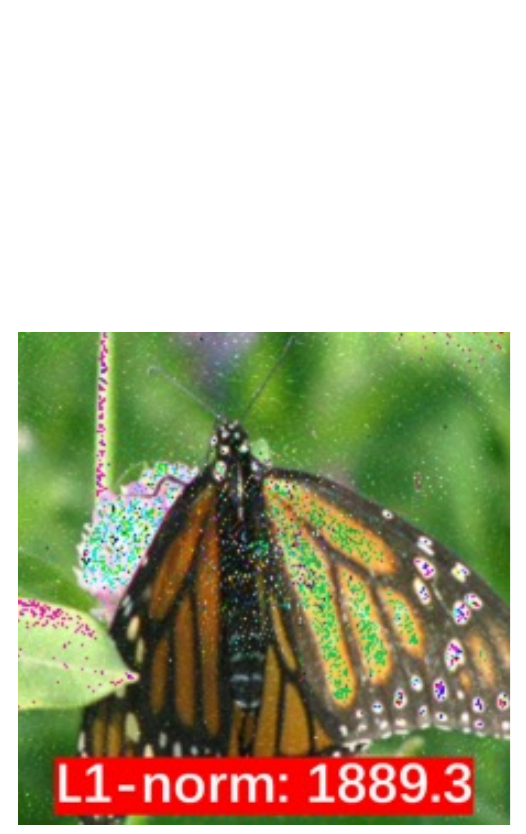}
        \captionsetup{font=footnotesize}
        \caption{Victim + NC trigger}
    \end{subfigure}
    \begin{subfigure}[t]{.11\textwidth}
        \includegraphics[width=1.0\textwidth]{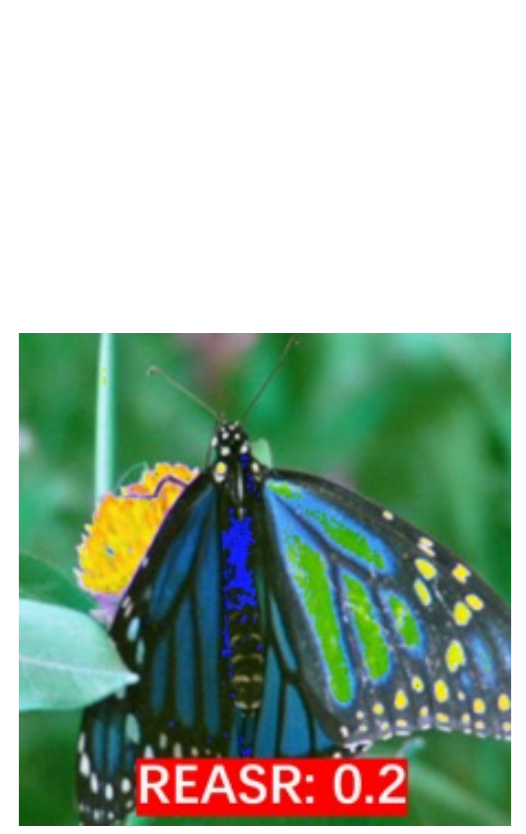}
        \captionsetup{font=footnotesize}
        \caption{Victim + ABS-filter trigger 
        }
    \end{subfigure}
    \begin{subfigure}[t]{0.11\textwidth}
        \centering
        \includegraphics[width=1.0\textwidth]{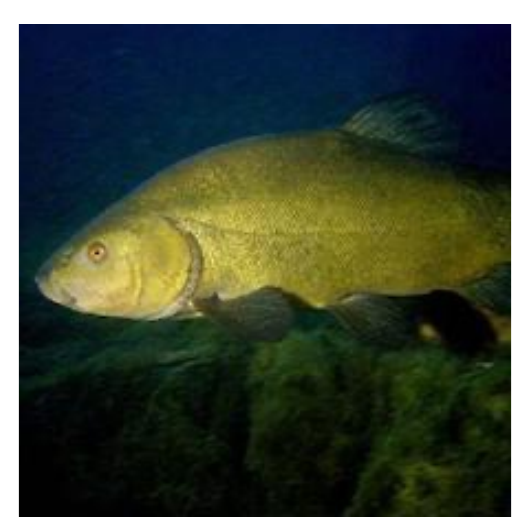}
        \caption{Target}
    \end{subfigure}
    \begin{subfigure}[t]{.11\textwidth}
        \includegraphics[width=1.0\textwidth]{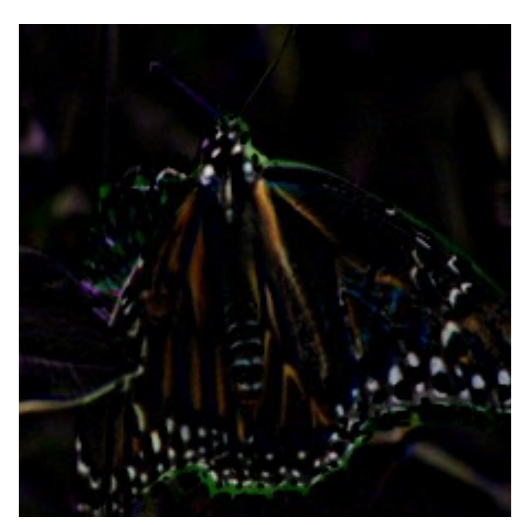}
        \caption{$|$(b)-(a)$|$}
    \end{subfigure}
    \begin{subfigure}[t]{.11\textwidth}
        \includegraphics[width=1.0\textwidth]{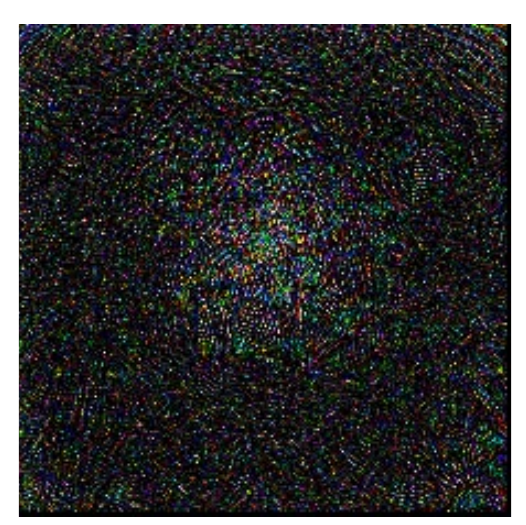}
        \caption{$|$(c)-(a)$|$}
    \end{subfigure}
    \begin{subfigure}[t]{.11\textwidth}
        \includegraphics[width=1.0\textwidth]{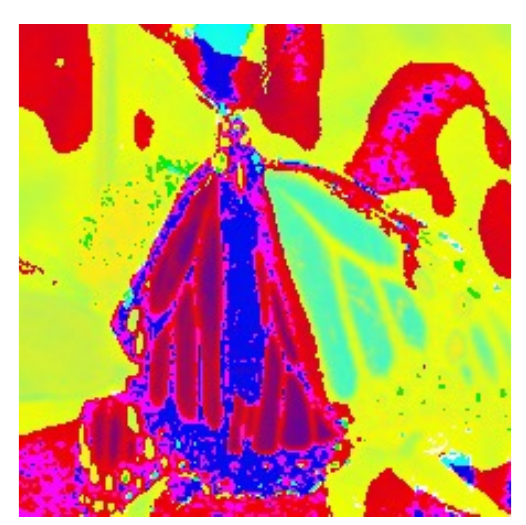}
        \caption{$|$(d)-(a)$|$}
    \end{subfigure}
    \caption{WaNet backdoor and its scanning results}
    \label{fig:complex}
\end{figure}

\noindent
{\bf Attacks.}
To demonstrate the challenge, we use the attacks in the computer vision rounds of TrojAI and additionally the WaNet attack~\cite{wanet} representing complex backdoors whose triggers are hardly human perceptible.
TrojAI is an ongoing multi-year and multi-round backdoor scanning competition for Deep Learning models, organized by IARPA~\cite{trojai}. It has finished nine rounds by the time of submission, with rounds 1-4 for CV models and rounds 5-9 for NLP models.
In each round, benign models (e.g., 500) are mixed with trojaned models (e.g., 500) and the performer is supposed to detect the trojaned ones. A cross-entropy loss lower than 0.348 (usually corresponding to 0.91 accuracy) is considered reaching the round target.
It has 3 different types of attacks in the CV rounds. 
The first type is the simple universal patch attack similar to BadNet~\cite{badnet}, nin which triggers are usually small polygons with solid colors. 
Figure~\ref{fig:troj_exp} (a) and (b) show a clean image of speed limit sign and its trojaned version which is classified to a lock sign in (c). TrojAI models are mostly models classifying  traffic signs. An input image is synthesized by placing an artificial traffic sign on a real-world street-view background image. 
Different models are trained with different sets of signs and images. The images have a high resolution 224$\times$224. 
The second type is label-specific patch attack that only flips images of the victim class to the target class.
The third type is Instagram filter attack, in which the trigger is an Instagram filter (e.g., Nashville filter as shown in Figure~\ref{fig:troj_exp} (d)). There are also Lomo (e), Kelvin (f), Gotham (g), and Toaster (h)
filters.
Observe that compared to patch attacks, filter attacks are pervasive; different filters also have different visual effects.
Figure~\ref{fig:complex} shows a VGG16 model~\cite{vgg16} trained on ImageNet trojaned by WaNet~\cite{wanet}.
WaNet uses a small and smooth warping field (that twists lines) to inject backdoor triggers, making the modification unnoticeable.
Figure (a) presents a victim image and figure (b) shows the trojaned version (classified to tench in figure (e)). It is hard for humans to tell the difference between the two images. We highlight the difference ($\times 3$) in figure (f).
Observe that the perturbations are camouflaging themselves along object outlines.

\begin{table}[t]
    \centering
    \scriptsize
    \tabcolsep=4.5pt
    \caption{Performance of NC, ABS with two settings for patches and filters, respectively, directly stamping triggers decomposed from attack instances by \Tech, and scanners synthesized by \Tech.
    For the trigger stamping method, we select the best possible ASR separation boundary for each type of attack (e.g., considering a model trojaned if any of the decomposed triggers can achieve larger than 0.7 ASR).
    }
    \label{tab:moti_nc_abs}
    \begin{tabular}{cccccc}
        \toprule
        \textbf{Scanner} & \textbf{Universal} & \textbf{Label-specific} & \textbf{Nashville} & \textbf{Toaster} & \textbf{WaNet} \\
        \midrule
        \textbf{NC} & 0.88 & 0.53 & 0.55 & 0.45 & 0.65 \\
        \textbf{ABS} & 0.93 & 0.83 & 0.68 & 0.58 & 0.65 \\
        \textbf{ABS-filter} & 0.80 & 0.58 & 0.90 & 0.60 & 0.55 \\
        \textbf{\Tech{} Triggers} & 0.60 & 0.55 & 0.83 & 0.80 & 0.58 \\
        \Tech{} {\bf Scanners}
        & 0.98 & 0.90 & 0.93 & 0.88 & 0.95 \\
        \bottomrule
    \end{tabular}
\end{table}

\noindent
{\bf Trigger Inversion Effective for One Attack May Not Be Effective for Another.}
For each of the aforementioned attacks, we mix 20 trojaned models with 20 benign models and apply different scanners.
The first row of Table~\ref{tab:moti_nc_abs} shows the results of the original NC (for the different attacks). 
Observe that it only works well for the universal patch attack, achieving 0.88 detection accuracy. 
Figure~\ref{fig:NC_polygon} (b) shows the inverted trigger for the universal attack in Figure~\ref{fig:troj_exp} (b). Observe that it is very similar to the ground-truth trigger, explaining its effectiveness. However, the inverted triggers for other attacks are largely dissimilar to the ground truths, demonstrating that the loss function design and/or the hyper parameters are not suitable for those attacks. 
The second row of Table~\ref{tab:moti_nc_abs} shows the results of the original ABS. 
It works well for the universal patch attack and the label specific attack. This is because it scans each label pair.
ABS has special support for filter type of triggers. Specifically, it models filter as a linear transformation layer. Instead of inverting the perturbation and the mask vectors, it inverts the coefficients of the linear transformation.
However, different filters have different effects, which may not be universally represented by the same linear template. The third row of Table~\ref{tab:moti_nc_abs} shows the results of ABS-filter. Observe that it works well for the Nashville filter, which is a filter that simply changes values in a channel and hence can be modeled by a linear function. 
In contrast, with the filter setting, ABS cannot detect patch backdoors.
None of these three scanners can handle the complex WaNet backdoor.
Figure~\ref{fig:complex} (g) shows the inverted trigger by NC for the model with the WaNet backdoor and (c) shows the input after applying the trigger.  
Observe that they are quite different from the ground truth. The trigger is so large that it is not distinguishable from large natural features in the target class which can flip the classification result (e.g., stamping a cat to any image likely flips that image to the cat class). As such, NC does not consider the model trojaned.
Figure (d) shows the image after applying the trigger filter inverted by ABS-filter and figure (h) shows the pixel level differences.
The inverted trigger has only 0.2 ASR such that ABS-filter does not consider the model trojaned.
Other trigger inversion based scanners such as~\cite{karm, liu2021ex, Tabor, dltnd, huang2019neuroninspect, guo2020towards} have similar problems, as shown in Section~\ref{sec:scan}.

A common strategy used by these scanners is to have a specially designed loss function and parameter setting for each attack and tries them one by one. A model is considered trojaned if any of the setting yields an effective trigger~\cite{trojai,karm,ABS,liu2021ex,zheng2021topological,kiourti2021online,zhang2021tad,sri}.

Such a strategy is valid only if attack specifics are known beforehand. {\bf In the finished TrojAI rounds, the attack details are given before each round of competition~\cite{trojai}, such as trigger size range, type of triggers, the possible locations they are stamped, etc. However, this assumption may not hold in real-world zero-day attacks}.

\begin{figure}[t]
    \centering
        \begin{minipage}[t]{.23\textwidth}
            \centering
            \includegraphics[width=1.0\textwidth]{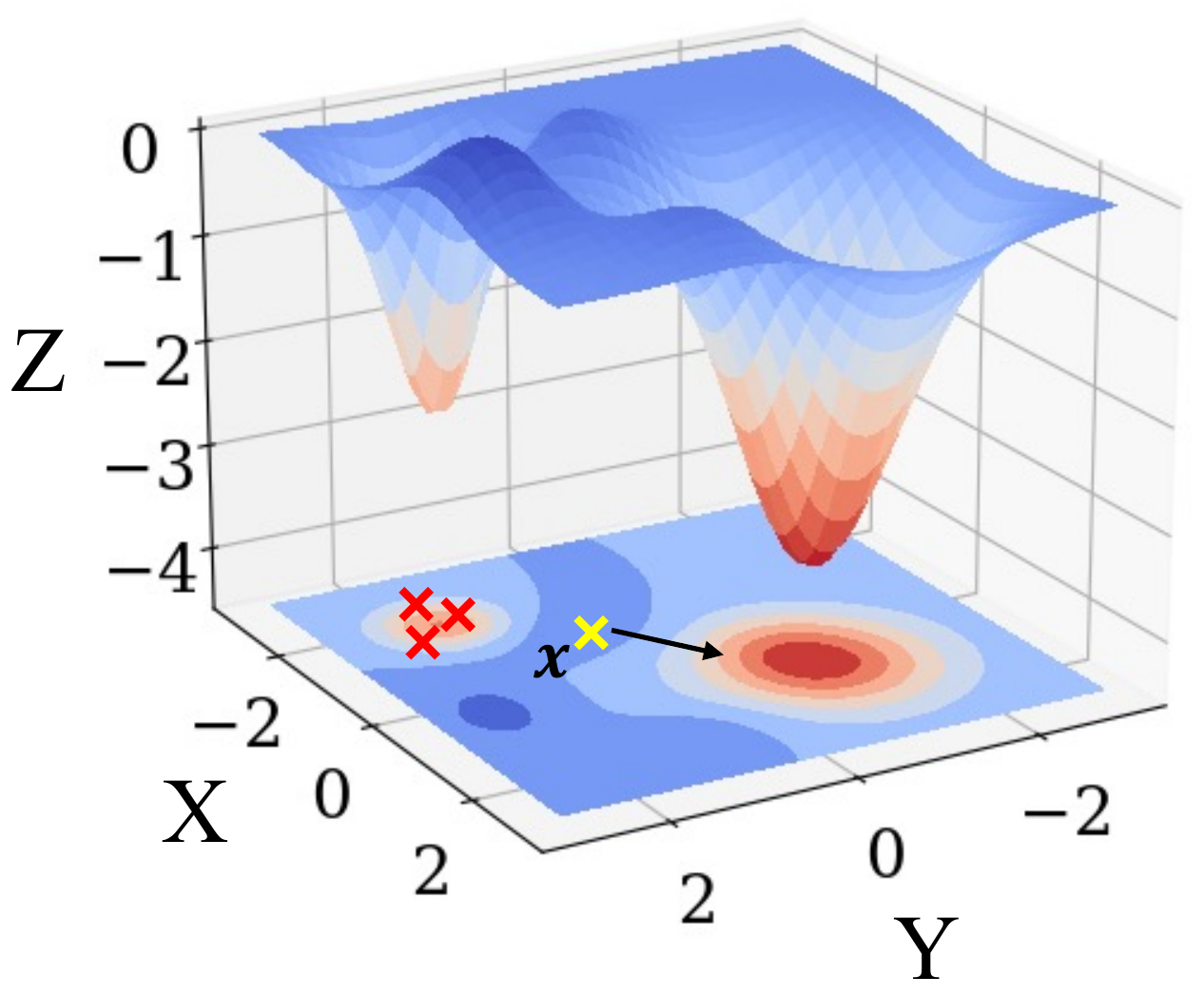}
            \caption*{(a) Landscape of cross entropy}
        \end{minipage}
        \begin{minipage}[t]{.23\textwidth}
            \centering
            \includegraphics[width=1.0\textwidth]{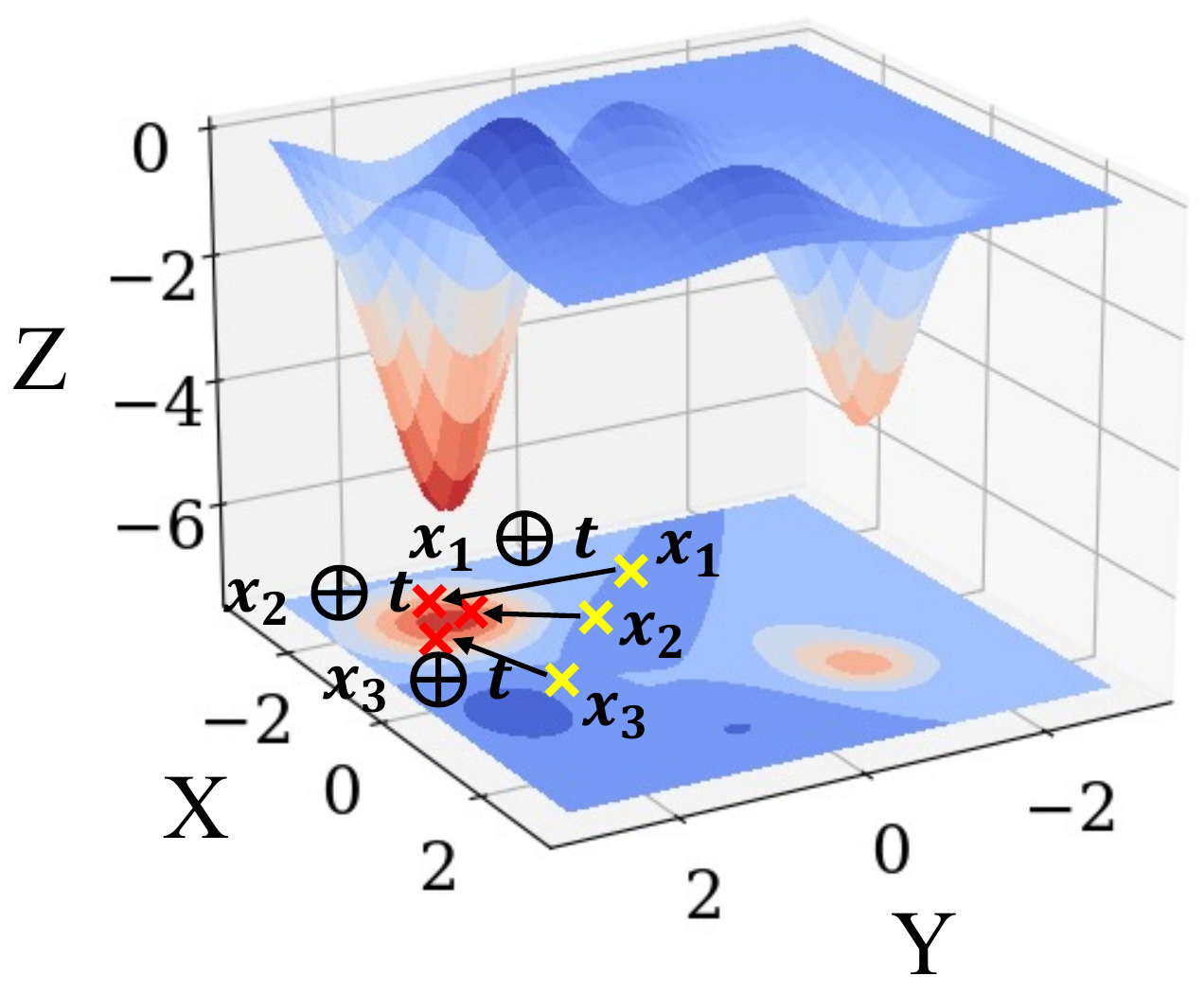}
            \caption*{(b) Landscape with synthesized loss term}
        \end{minipage}
    \caption{Inversion loss landscape illustration with (a) the landscape of cross entropy and (b) the landscape with the additional synthesized loss term.
    The $x$-$y$ plane denotes an input feature space and the $z$ axis the loss value. 
    The areas with the red plummets represent the input areas of the target class.
    In (a),  the large plummet denotes the clean target samples and the small one the victim samples with the trigger.  The blue areas  denote the clean victim samples.}
    \label{fig:landscape}
\end{figure}

\begin{figure*}[h]
    \centering
    \begin{minipage}{.32\textwidth}
        \centering
        \begin{minipage}{.3\textwidth}
            \centering
            \includegraphics[width=1.0\textwidth]{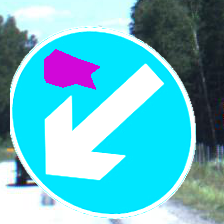}
            \includegraphics[width=1.0\textwidth]{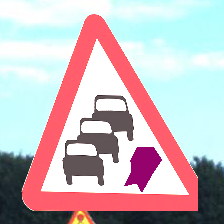}
            \caption*{(a) Trojaned}
        \end{minipage}
        \begin{minipage}{.3\textwidth}
            \centering
            \includegraphics[width=1.0\textwidth]{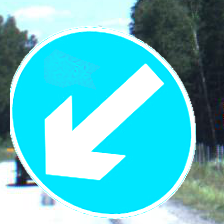}
            \includegraphics[width=1.0\textwidth]{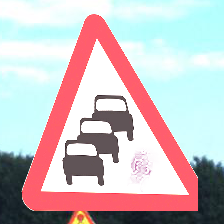}
            \caption*{(b) Clean}
        \end{minipage}
        \begin{minipage}{.3\textwidth}
            \centering
            \includegraphics[width=1.0\textwidth]{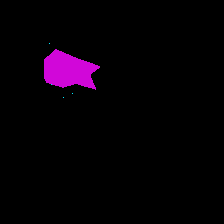}
            \includegraphics[width=1.0\textwidth]{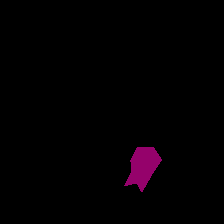}
            \caption*{(c) Trigger}
        \end{minipage}
        \caption*{(A) Forensics of patch attack}
    \end{minipage}
    \begin{minipage}{.32\textwidth}
        \centering
        \begin{minipage}{.3\textwidth}
            \centering
            \includegraphics[width=1.0\textwidth]{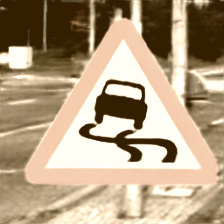}
            \includegraphics[width=1.0\textwidth]{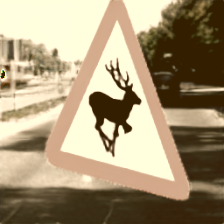}
            \caption*{(d) Trojaned}
        \end{minipage}
        \begin{minipage}{.3\textwidth}
            \centering
            \includegraphics[width=1.0\textwidth]{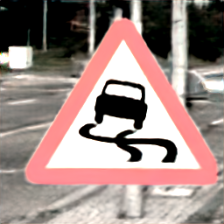}
            \includegraphics[width=1.0\textwidth]{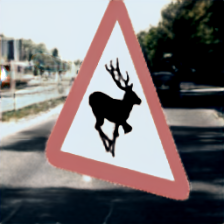}
            \caption*{(e) Clean}
        \end{minipage}
        \begin{minipage}{.3\textwidth}
            \centering
            \includegraphics[width=1.0\textwidth]{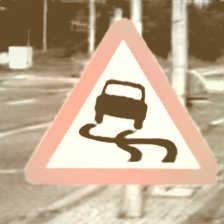}
            \includegraphics[width=1.0\textwidth]{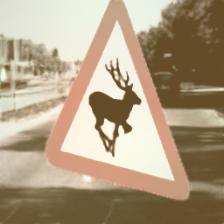}
            \caption*{(f) Trigger}
        \end{minipage}
        \caption*{(B) Forensics of pervasive attack}
    \end{minipage}
    \begin{minipage}{.32\textwidth}
        \centering
        \begin{minipage}{.3\textwidth}
            \centering
            \includegraphics[width=1.0\textwidth]{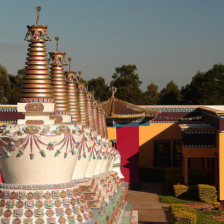}
            \includegraphics[width=1.0\textwidth]{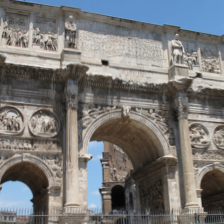}
            \caption*{(g) Trojaned}
        \end{minipage}
        \begin{minipage}{.3\textwidth}
            \centering
            \includegraphics[width=1.0\textwidth]{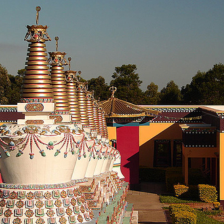}
            \includegraphics[width=1.0\textwidth]{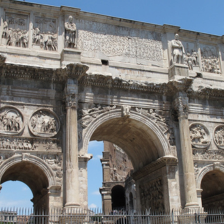}
            \caption*{(h) Clean}
        \end{minipage}
        \begin{minipage}{.3\textwidth}
            \centering
            \includegraphics[width=1.0\textwidth]{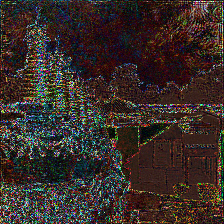}
            \includegraphics[width=1.0\textwidth]{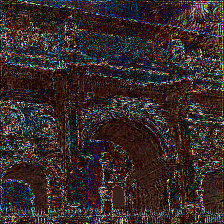}
            \caption*{(i) Trigger}
        \end{minipage}
        \caption*{(C) Forensics of WaNet attack}
    \end{minipage}
    \caption{Forensics of different attacks}
    \label{fig:moti_forensic}
\end{figure*}

\noindent
{\bf Our Solution - Attack Forensics.}
The main challenge is that different attacks compromise different parts of input space. Such subspaces may be very small. Trigger inversion is largely driven by the gradients of the cross-entropy loss function. When the compromised subspaces are small and isolated, starting from the clean input space, the gradients may not be able to provide valid directions to the compromised subspaces. 
{\em The overarching idea of our solution is to leverage attack forensics to reverse engineer attack specifics from a few attack instances (e.g., a few inputs with triggers causing misclassification), such as what the trigger looks like and how it is injected.  Additional loss terms can be synthesized based on the specifics to change the landscape of the loss function such that gradients (of the new loss) can guide trigger inversion to the compromised subspace.}
Figure~\ref{fig:landscape} illustrates the concept.
It shows the landscapes of two inversion loss functions  with the $x$-$y$ plane denoting input features (e.g., encodings by some feature extraction model) and $z$ the loss value. The left one is for the cross-entropy loss term in Eq.~\ref{eq:inversion} and the right one is for the synthesized loss term by \Tech.
Observe that when cross-entropy is used, from a clean sample $x$ in the blue area in (a), it is very difficult to find the universal perturbation (i.e., the trigger) that can move the sample to the small red area due to the rugged landscape. The gradients point to the larger red area instead.
In (b), by performing forensics on a few given attack samples $x_1\oplus t$, $x_2 \oplus t$, and  $x_3 \oplus t$, a new loss term is synthesized that changes the loss landscape. 
Specifically, \Tech{} can reverse engineer $x_1$, $x_2$, and $x_3$, from the attack instances $x_1\oplus t$, $x_2 \oplus t$, and  $x_3 \oplus t$.
Our synthesized loss term hence aims to have a very small loss value for  
$x_1\oplus t$, $x_2 \oplus t$, and  $x_3 \oplus t$, much smaller than the loss values for clean target samples (i.e., those in the larger red area in (a)), making their area the optimization goal.
Furthermore, the loss is synthesized in such a way that it ensures the gradients at the reverse engineered $x_1$, $x_2$, and $x_3$
pointing to the target area. 
As such when scanning a {\em different} model for the similar type of backdoor, the new loss term can provide clear direction to find the trigger $t$.

Figure~\ref{fig:moti_forensic} (A) shows the forensics of a patch attack in TrojAI. From left to right, starting from two images stamped with the triggers (at different places), that is, attack instances, our technique decomposes them to the clean images and the triggers. Note that the original clean images are unknown.
From the decomposed instances, the attack can be summarized as distributions of trigger position, size, and pixel values (inside the trigger area).
A loss function is automatically synthesized using these distributions to detect backdoors of the same type.
Figure~\ref{fig:moti_forensic} (B) shows the forensics of pervasive attacks.
From left to right, starting from a few attack instances, our technique decomposes each to a clean image and a transformation function $\mathcal{F}$ over the clean image $x$. Note that the image to the right of the clean image denotes $\mathcal{F}(x)$.
The attack can be summarized as coefficient distributions of the transformation function.
As such, a scanner again can be synthesized to detect this kind of attacks.
Figure~\ref{fig:moti_forensic} (C) shows the forensics of WaNet attacks.
We will show in Section~\ref{sec:visual} that the decomposed clean images closely resemble the ground-truth clean images and the decomposed triggers resemble the ground-truth triggers as well. 
The last row of Table~\ref{tab:moti_nc_abs} shows the scanning results using scanners automatically synthesized by \Tech.
Here, we use 3 trojaned models for each type of attack in forensics.
They are disjoint from the ones used in the scanning evaluation.
Observe that the scanners can now accurately detect the trojaned models.
In addition, directly using triggers reverse engineered by \Tech{} to determine if a model has similar backdoors is ineffective (due to different models have unique instantiations), as shown in row 4 of Table~\ref{tab:moti_nc_abs}.

\begin{figure}[t]
    \centering
    \includegraphics[width=0.5\textwidth]{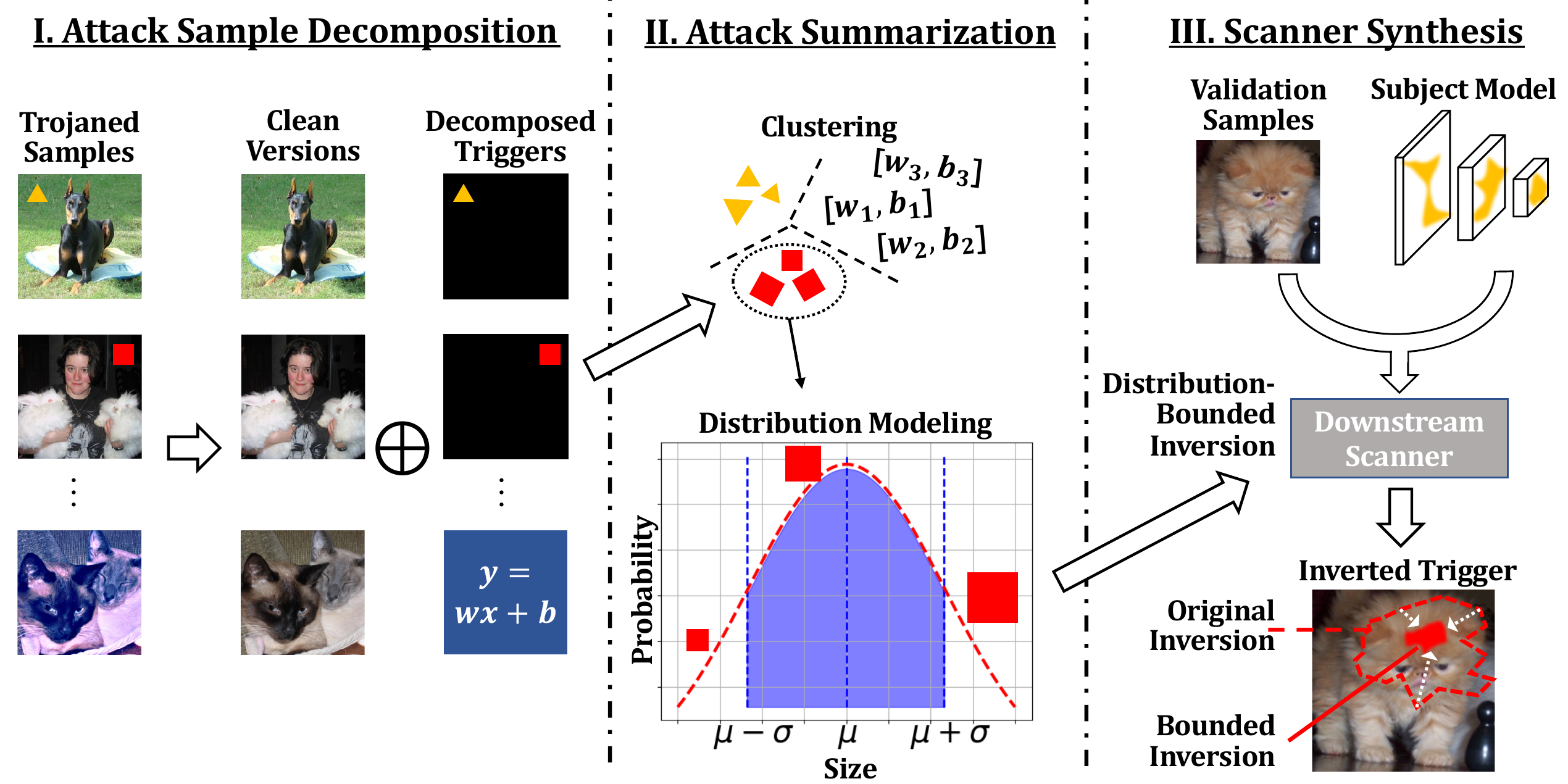}
    \caption{Overview of \Tech}
    \label{fig:overview}
\end{figure}

\section{Design}

Figure~\ref{fig:overview} illustrates the overview of our technique.
It consists of three steps.
The first step is {\em attack sample decomposition} that decomposes an image with trigger to a clean image and a trigger. 
The second step is {\em attack summarization} that extracts key distributions describing multiple attack samples, which may be from multiple models with various backdoors. Such distributions include trigger size and shape distributions, transformation coefficient distributions, and so on.
Note that we do not require the attack instances belong to a single backdoor type (as \Tech{} will cluster and summarize them), although we assume most trojaned inputs of a particular instance exploit the same backdoor.
The third step is {\em scanner synthesis} that synthesizes loss function terms that can regulate the trigger inversion procedure to detect backdoors of the same kinds.
We will discuss the details of these steps in the following.

\subsection{Attack Sample Decomposition} \label{sec:decomposition}

\begin{figure*}[t]
    \centering
    \includegraphics[width=1.0\textwidth]{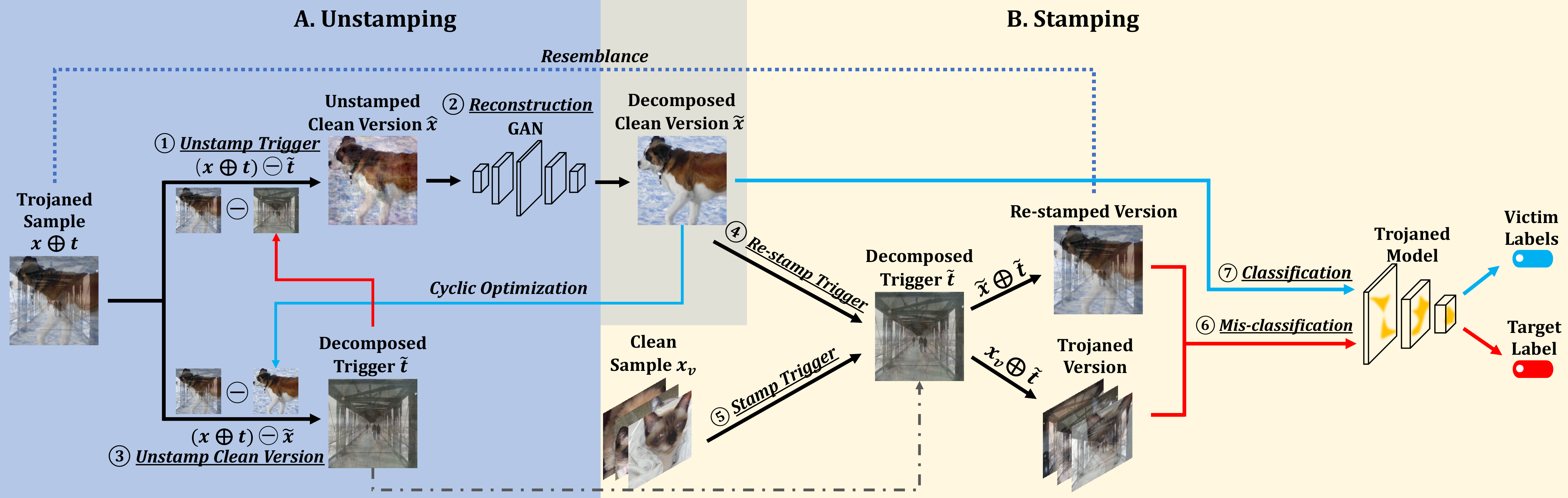}
    \caption{Attack decomposition pipeline}
    \label{fig:decompose}
\end{figure*}

This step aims to decompose given attack samples, namely, inputs with triggers, to their clean versions and the triggers. It assumes the trojaned model, a few attack samples for the model (10 in this paper), a set of clean samples for the validation purpose (100 per model in this paper, that is, 10 per class for CIFAR10, and 2-5 per class for other datasets),
a GAN denoting the input distribution, e.g., a general purpose GAN trained on ImageNet, the victim class labels, and the target class label.
Note that the clean samples are different from the clean versions of the attack samples, which are unknown according to our threat model.

The decomposition leverages a few key observations: (1) the clean versions of attack samples largely resemble victim class samples and they may be effectively generated using the GAN (when regulated by a cross-entropy loss on the subject model); (2) the decomposed trigger should be valid for the given validation clean samples, namely, causing them to be misclassified; (3) the decomposed trigger should be valid for the decomposed clean versions of attack samples; (4) {\em unstamping} the decomposed trigger from the attack sample should yield an image resembling the decomposed clean version (generated by the GAN);
(5) {\em unstamping} the decomposed clean version of an attack sample from the sample itself should yield an image resembling the decomposed trigger; and
(6) an attack sample should be similar to its decomposed clean version stamped with the decomposed trigger.
We will formally define stamping and unstamping later.

\smallskip
\noindent
{\bf Decomposition Pipeline.}
We devise a cyclic optimization based decomposition pipeline according to the above observations, as illustrated in Figure~\ref{fig:decompose}.
To concretize our discussion, we use reflection attack as an example.
Reflection attack~\cite{reflection} merges a clean input with another image to create a reflection effect (e.g., through  glass).
Here the trojaned sample (on the left of Figure~\ref{fig:decompose}) is a dog with the reflection of a hallway, where the dog is the victim and the hallway is the trigger.

In the stage A \textit{ unstamping}, \Tech{} decomposes a trojaned sample into its clean versions and the trigger.
At step \circled{1}, \Tech{} initializes the decomposed trigger $\tilde{t}$, {\em unstamps} it from the trojaned sample, and derives an unstamped version $\hat{x}$ which is raw and noisy. To improve quality, at step \circled{2}, \Tech{} reconstructs the decomposed clean version using a pre-trained GAN, which can be considered a filter that removes the out-of-distribution noises from $\hat{x}$ and yields $\tilde{x}$. At step \circled{3}, \Tech{} {\em unstamps}  $\tilde{x}$ from the trojaned sample and updates the decomposed trigger.

In stage B {\em stamping}, \Tech{} ensures the effectiveness of decomposed clean version and trigger through multiple constraints.
At step \circled{4}, \Tech{} re-stamps the decomposed trigger $\tilde{t}$ on the decomposed clean version $\tilde{x}$. The result should resemble the given trojaned sample. Their similarity is denoted by the bluish dotted line.
At step \circled{5}, \Tech{} stamps the decomposed trigger to a set of clean samples.
At step \circled{6}, \Tech{} ensures that the samples generated from the previous two steps (with the decomposed trigger)  are misclassified to the target label.
At step \circled{7}, \Tech{} ensures the decomposed clean version $\tilde{x}$ is correctly classified to the victim label.
Specifically, step \circled{2} corresponds to the aforementioned observation (1); \circled{5} to observation (2); \circled{4} to observations (3) and (6);  \circled{1} to observation (4); and  \circled{3} to observation (5).

\smallskip
\noindent
{\bf Formal Definition.}
Next, we formally define the decomposition process. For discussion clarity, we use the following symbols, $x$ denoting the (unknown) ground-truth clean sample, $t$ the (unknown) ground-truth trigger, $x_v$ a validation clean sample, $\tilde{x}$ the decomposed clean version of an attack sample, $\tilde{t}$ the decomposed trigger, $x \oplus t$ denotes stamping $t$ to $x$  and the stamping operation may vary across attacks (explained later), and $x_1 \ominus x_2$ denotes unstamping an image $x_2$ (which could be $\tilde{t}$ or $\tilde{x}$) from an image $x_1$.  

We define three cross-entropy losses corresponding to steps \circled{6} and \circled{7} in Figure~\ref{fig:decompose}.
\begin{equation}
    Loss_{CE} = \mathcal{L}(M(\tilde{x} \oplus \tilde{t}), y_{t}) + \mathcal{L}(M(x_{v} \oplus \tilde{t}), y_{t}) + \mathcal{L}(M(\tilde{x}), y_{v}),
\end{equation}
\noindent where $\mathcal{L}$ denotes the cross-entropy calculation, $M$ the trojaned model, $y_{t}$ the attack target label and $y_{v}$ the victim labels.
The first term in the loss means that the decomposed trigger is effective for the decomposed clean image. In other words, if we re-stamp the decomposed trigger $\tilde{t}$ to the decomposed clean image $\tilde{x}$ and feed it to the trojaned model, the model should output the target label.
The second term means that the decomposed trigger is effective for the clean validation images.
The third term means that the decomposed clean image has the trigger removed. In other words, the trojaned model should predict the decomposed clean image $\tilde{x}$ to its ground-truth label.

We also define two reconstruction losses, corresponding to steps \circled{2} and \circled{4} in Figure~\ref{fig:decompose}.
\begin{equation}
    Loss_{recon} = LPIPS(\hat{x}, \tilde{x}) + \normltwo(x \oplus t, \tilde{x} \oplus \tilde{t}),
\end{equation}
\noindent $LPIPS$() denotes the LPIPS loss~\cite{lpips}, which is commonly used as a constraint in GAN based input reconstruction~\cite{mentzer2020high,zhu2017unpaired,jo2020investigating}, and $\normltwo$ denotes the $\normltwo$ norm, which calculates the Euclidean distance of two inputs.
The first term of $Loss_{recon}$ means that the decomposed (reconstructed) clean image $\tilde{x}$ should be similar to the unstamped clean image $\hat{x}$, while the GAN ensures that the former is in distribution.
The second term means that the restamped image $\tilde{x} \oplus \tilde{t}$ should be similar to the original trojaned image $x \oplus t$.

\noindent 
The overall decomposition procedure can be defined as an optimization problem in the following.
\begin{equation} \label{e:loss}
    \argmin_{\tilde{x}, \tilde{t}}\;\;\; Loss_{CE} + \alpha \cdot Loss_{recon},
\end{equation}
where $\alpha$ controls the trade-off between the two losses. Typically we set $\alpha = 10^2$.

\smallskip
\noindent
{\bf Modeling Stamping Operations in Different Backdoor Attacks.}
In the previous discussion, we have not defined the stamping/unstamping operations, which vary across different attacks.
Although there are many different types of backdoors, most of them can be abstracted to two forms.
They differ by their ways of injecting triggers.
The core challenge is hence to model these injection methods.
We consider there are two types of trigger injection methods: {\em patching} and {\em transforming}.
In the former, a trigger is injected to a clean sample by merging their pixel values. 
There are different ways of merging, for instance, completely replacing the original pixels and adding/subtracting the original pixel values with the trigger pixel values. We use the masking function proposed in NC~\cite{nc} to model these different methods.
\begin{equation}
    x\oplus t = x\cdot (1-m) + t\cdot m
\end{equation}
Here, $m$ is a mask with values in [0,1].
BadNets~\cite{badnet}, TrojNN~\cite{trojnn}, reflection attack~\cite{reflection}, and composite attack~\cite{composite} that place additional object(s) in a victim sample can be modeled by this function with different $m$ distributions.
The additional objects can be static patterns, e.g., a yellow flower placed at the top-left in BadNets, or semantic features, e.g., a truck image replacing half of the image in composite attack.
For example, pixel replacing means that all the $m$ values in the trigger area are 1.0 and the rest 0, following a binomial distribution.
Accordingly, we define the {\em unstamping} operation.
\begin{equation}
    x\ominus t = \frac{x - t\cdot m}{1-m}
\end{equation}
We hence have $(x\oplus t) \ominus \tilde{t}\  \approx\  \tilde{x}$ and $(x\oplus t) \ominus \tilde{x}\  \approx\  \tilde{t}$.
Note that the definition does not mean we know $x$, $t$, $m$ beforehand. During forensics, we use their approximation $\tilde{x}$, $\tilde{t}$, and $\tilde{m}$ instead.

\begin{figure}[t]
\centering
    \begin{subfigure}{.2\textwidth}
        \centering
        \includegraphics[width=0.6\textwidth]{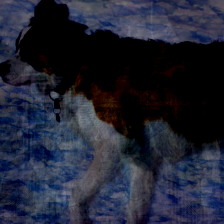}
        \caption{Without Normalization}
    \end{subfigure}
    \begin{subfigure}{.2\textwidth}
        \centering
        \includegraphics[width=0.6\textwidth]{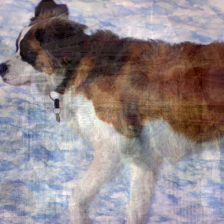}
        \caption{With Normalization}
    \end{subfigure}
    \caption{Effectiveness of normalization}
    \label{fig:decompose_trick}
\end{figure}

\smallskip
\noindent
{\em Normalization.}
In the first few steps of optimization we do not have a good approximation of the $m$ value, the unstamping operation tends to aggressively reduce pixel values (in order to reduce the loss value with an inappropriate $\tilde{m}$), and hence the decomposed image tends to be dark and noisy, as shown in Figure~\ref{fig:decompose_trick} (a). We use a normalization step to calibrate the unstamped image values to be within the distributions denoted by clean validation images. Hence the decomposed images become vivid and clear, as shown in (b). They also substantially speedup convergence.
Specifically, the normalization step is defined as follows.
\begin{equation} \label{equ:normalization}
    x_{norm} = \frac{x - mean(x)}{std(x)} \cdot std(x_{v}) + mean(x_{v}),
\end{equation}
\noindent where $x$ denotes the images to normalize, $x_{norm}$ the normalized versions, $x_{v}$ the given set of clean validation images, $mean$ and $std$ the mean and the standard deviation, respectively.
It is performed at step \circled{1} in Figure~\ref{fig:decompose} after we unstamp the trigger. $\Box$

\smallskip
In the second type of injection, the transforming type (e.g., Invisible~\cite{invisible}, WaNet~\cite{wanet}, and Instagram filter~\cite{ABS} attacks), a trigger is injected using a transformation function in the form of an algorithm or a pre-trained network.
\begin{equation}
    x \oplus t = \mathcal{F} (x; t)
\end{equation}
Observe that we use the coefficients of transformation function $\mathcal{F}$ to denote the trigger $t$ because such coefficients indeed uniquely define a trigger.
During forensics, we leverage a piece-wise linear function to approximate $\mathcal{F}$.

Compared to the patching form of backdoors, defining the unstamping operation here is more challenging because there is not a simple inverse function of $\mathcal{F}$. 
The pervasive perturbations injected by these attacks cannot be easily removed by simple mutations. 
We hence leverage the reconstruction and denoising ability of GAN to perform the unstamping function.
\begin{equation}
    x\ominus t = GAN(normalize(x \oplus t))
\end{equation}
\begin{figure}
    \centering
    \includegraphics[width=0.48\textwidth]{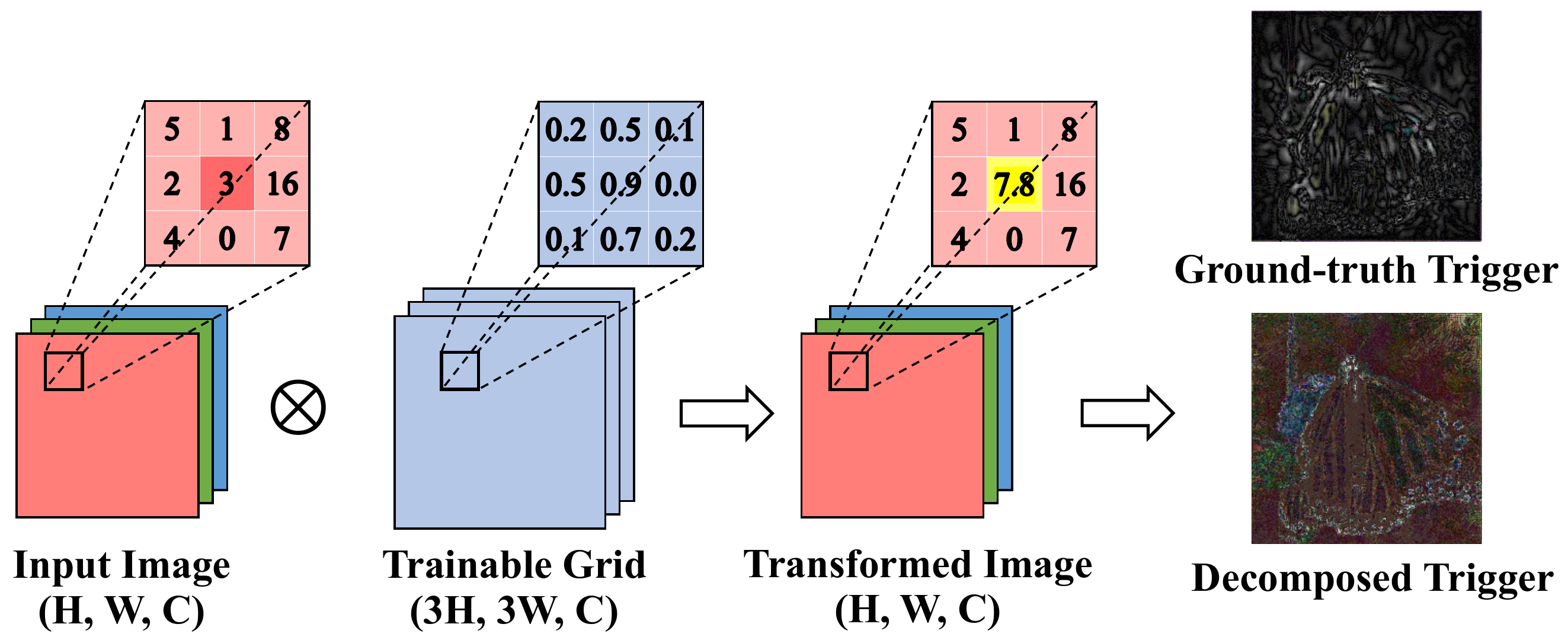}
    \caption{Modeling transforming backdoors}
    \label{fig:method_linear}
\end{figure}
Different pervasive backdoors may have substantially different $\mathcal{F}$. In order to have a uniform modeling of these functions, we propose to use a piece-wide linear function, leveraging the observation that {\em pervasive backdoors usually do not change human perception of an input such that a pixel in a trojaned input tends to be closely related to its neighboring pixels in the clean version}.
Specifically, as shown in Figure~\ref{fig:method_linear}, for each pixel in the clean input (e.g., pixel 3 highlighted in red on the left), we introduce a $3 \times 3$ trainable grid (e.g., the blue grid in the middle). A pixel in the injected/transformed image is the sum of the element-wise product of the $3 \times 3$ neighbors in the original input and the trainable grid, adding a trainable bias. It is formally defined as follows.
\begin{equation} \label{equ:complex}
    (x \oplus t)_{[i, j, k]}  = \sum\limits_{p=-1}^{1} \sum\limits_{q=-1}^{1} (x_{[i+p, j+q, k]} \cdot t_{[3i-1+p, 3j-1+q, k]}^{w}) + t_{i, j, k}^{b}
\end{equation}
\noindent where $i, j, k$ denote the coordinates of width $W$, height $H$ and channel $C$ of the input image. Intuitively, $x_{[i,j,k]}$ denotes the pixel value at the $i^{th}$ column, $j^{th}$ row and $k^{th}$ channel. $p$ and $q$ are used to traverse the $3 \times 3$ neighborhood of this pixel. Trigger $t$ consists of $t^{w}$ and $t^{b}$, with the former the trainable weights
and the latter the biases of the piece-wise linear functions. The blue matrix in Figure~\ref{fig:method_linear} with shape $(3H, 3W, C)$ denotes $t^w$ since we have a $3 \times 3$ grid for each pixel.
For example in Figure~\ref{fig:method_linear} assume $x_{[i, j, k]}$ is the middle element ``3'' in the first column.
Then $x_{[i+p,j+q,k]}$ where $p, q \in \{-1, 0, 1\}$ traverses the $3 \times 3$ neighborhood of $x_{[i, j, k]}=3$, e.g., $x_{[i-1, j-1, k]}=5$.
$t_{[3i-1+p, 3j-1+q, k]}^{w}$ where $p, q \in \{-1, 0, 1\}$ denotes the trainable $3 \times 3$ grid for pixel $x_{[i, j, k]}$.
For example, $t_{[3i-1, 3j-1, k]}^{w}=0.9$ in the second column of Figure~\ref{fig:method_linear} is the weight value corresponding to $x_{[i, j, k]}=3$,
and $t_{[3i-2, 3j-2, k]}^{w}=0.2$ is the weight value corresponding to $x_{[i-1, j-1, k]}=5$.
Finally, we add up the element-wise product for the new pixel value $(x \oplus t)_{[i, j, k]}$.
Assume bias $t^{b}_{i, j, k} = 0$. The new value of the middle ``3'' is computed as follows.
\begin{align}\scriptsize
    & 5 \times 0.2 + 1 \times 0.5 + 8 \times 0.1 + 2 \times 0.5 + 3 \times 0.9 \\ \nonumber
    & + 16 \times 0.0 + 4 \times 0.1 + 0 \times 0.7 + 7 \times 0.2 + 0 = 7.8 \nonumber
\end{align}
\noindent which is highlighted in yellow in the third column.

The goal of decomposition is hence to update the trainable grids so that the loss in Eq.~\ref{e:loss} is minimized.
For instance, the Nashville filter backdoor can be precisely formulated by trainable grids with a non-zero central value surrounded by 8 zero values.
After injection, the new value of a pixel is just a linear transformation of its original value.
Moreover, if one considers each grid for a pixel denotes some local transformation, the grids for close-by pixels share a lot of similarity in order to ensure transformation smoothness.
For example, all the trainable grids for a Nashville filter backdoor are the same. 
To leverage this observation, we introduce a smoothing loss Eq.~\ref{equ:smooth} to regulate the differences between close-by grids.
\begin{equation} \label{equ:smooth}
    Loss_{smooth} = \normltwo(Resize(Avgpool(\tilde{t})), \tilde{t}),
\end{equation}
\noindent where $Avgpool()$ denotes the average pooling operation and $Resize()$ resizes the result after average pooling to the original shape.
Average pooling helps reduce the differences between close-by grids.

The two images on the right of Figure~\ref{fig:method_linear} show the transformations by a WaNet backdoor trigger and the decomposed trigger by \Tech. The two share similarity and the latter has a close-to 1.0 ASR.

Given a model for forensics, since we do not know if it has a patching or transformation form of backdoor, we try to decompose the attack samples using both forms and then choose the one with better performance. Details can be found in Appendix~\ref{sec:select_func}.

\subsection{Attack Samples Clustering and Summarization} \label{sec:design_summarize}

In the previous step, we decompose each attack sample $x\oplus t$ to its clean version $\tilde{x}$ and trigger $\tilde{t}$.
For example in Figure~\ref{fig:decompose}, we decompose an input attack sample $x\oplus t$, which is a dog $x$ stamped with a hallway reflection $t$, into its clean version $\tilde{x} \approx x$ which is the reconstructed dog and trigger $\tilde{t} \approx t$, the generated hallway.
In this step, we first extract an attack feature vector $v$ from the decomposition of each attack sample.
We then cluster these vectors based on their values. The vectors in a cluster are summarized by Gaussian distributions.

\noindent
\textbf{Attack Feature Extraction.}
For an attack sample of the patching form of backdoors, its attack features include both the decomposed mask $\tilde{m}$ and the decomposed trigger $\tilde{t}$. Therefore,
\begin{equation}
    v = (\tilde{m}, \tilde{t})
\end{equation}
In many cases, $\tilde{m}$ values have special distributions.
For example, $\tilde{m}$ tends to have a binomial distribution for attack samples of a simple patch backdoor, namely, stamping a patch trigger on an input (by replacing its pixels).
In this case, we simplify the features to the mask size $s$ (e.g., denoting patch size) and the position of mask center $(i,j)$ (e.g., denoting patch position). Hence, the property vector $v = (i,j,s,\tilde{t})$ with $s$ = {\it sum}($\tilde{m}$) , and ($i$, $j$) = {\it mean}($\tilde{m}$).

For backdoors that mix images with some ratio like reflection attack, e.g., a pixel after injection is 0.7 of the original pixel plus 0.3 of the trigger pixel, values in $\tilde{m}$ tend to be constant. For such cases, we simplify the attack feature value to $v = \tilde{t}$.

For transforming backdoors, we extract the coefficients of the piece-wise linear function as the features.
\begin{equation}
    v = (\widetilde{t^w}, \widetilde{t^b})
\end{equation}
with $\widetilde{t^w}$ the (reverse engineered) weights and $\widetilde{t^b}$  the biases (in Eq.~\ref{equ:complex}).

\noindent
\textbf{Clustering.}
Given the set of feature vectors of $n$ attack samples, i.e.,  $\mathbb{V}=\{v_1,\ ...,\ v_n\}$
we partition it to $k$ disjoint subsets $\mathbb{V}_1$, ... $\mathbb{V}_k$, based on their different forms and their values, using a number of standard clustering algorithms, e.g., Kmeans~\cite{kmeans}, GMM~\cite{gmm}, and DNSCAN~\cite{dbscan}.

\smallskip
\noindent
\textbf{Summarization.}
We consider each cluster $\mathbb{V}_i$ denotes a type of backdoor attack and we summarize it by modeling values in individual dimensions of $\mathbb{V}_i$ using Gaussian distributions.
Formally, we say the $i$th type of backdoor attack
\begin{equation} \label{e:attack}
    \textit{backdoor}_i \sim \mathcal{N}(\mu_i,\,\sigma_i^{2}),
\end{equation}
with $\mu_i$ and $\sigma_i^{2}$ the mean (vector) and the variance (vector) of $\mathbb{V}_i$.
We choose to use Gaussian distributions because of their generality~\cite{rasmussen2003gaussian,lu2005clustering,raissi2017machine}.
The central limit theorem~\cite{kwak2017central,schatte1988strong,hoeffding1948central} states that when a distribution is complex and  affected by a large number of independent random variables (like physical world distributions),
it tends to be Gaussian.

\begin{figure}[t]
    \begin{minipage}[t]{0.44\columnwidth}
        \centering
        \includegraphics[width=1.6in]{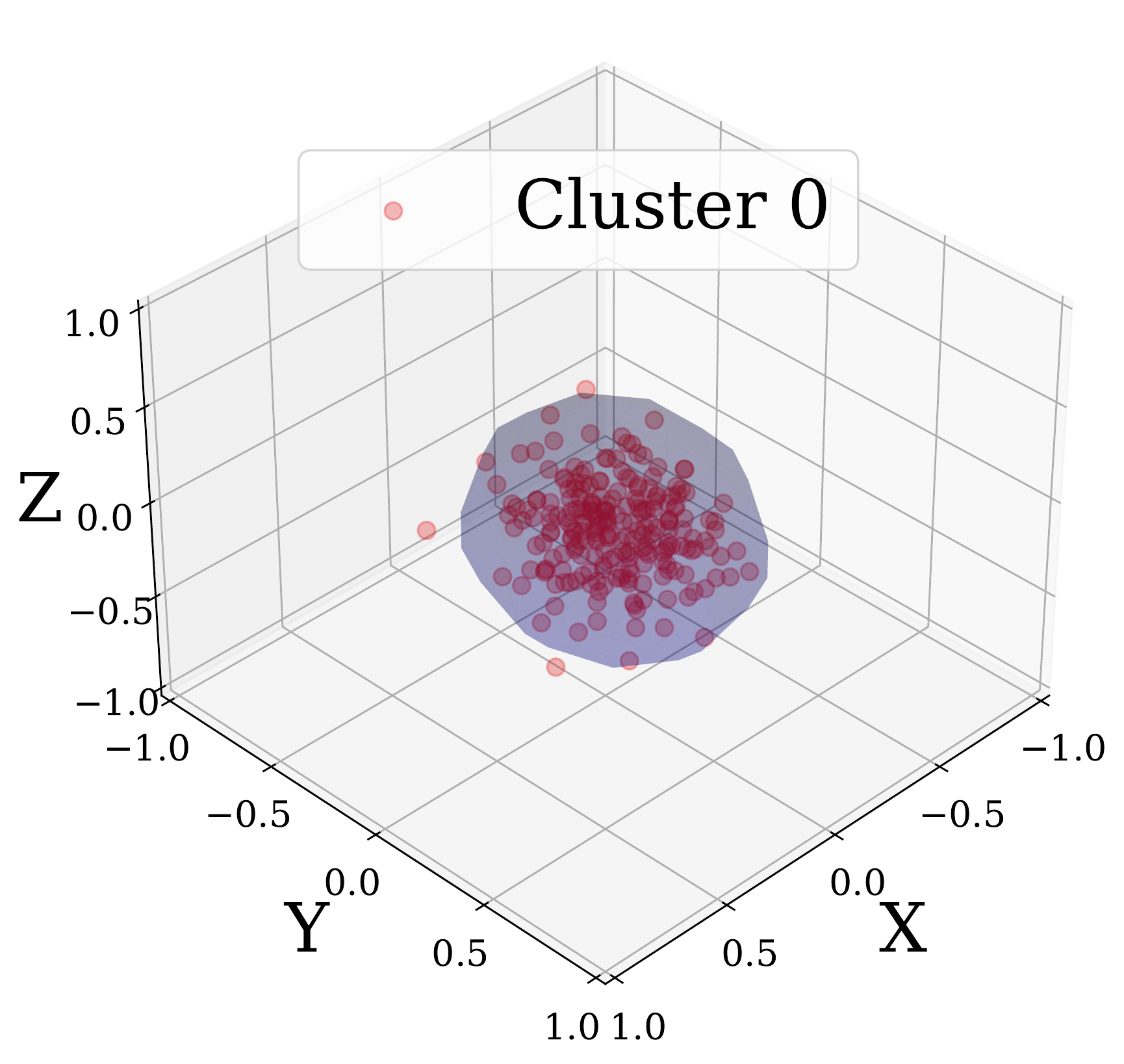}
        \caption{Clustering TrojAI polygon attack samples}
        \label{fig:polygon_cluster}
    \end{minipage}
    ~
    \begin{minipage}[t]{0.44\columnwidth}
        \centering
        \includegraphics[width=1.6in]{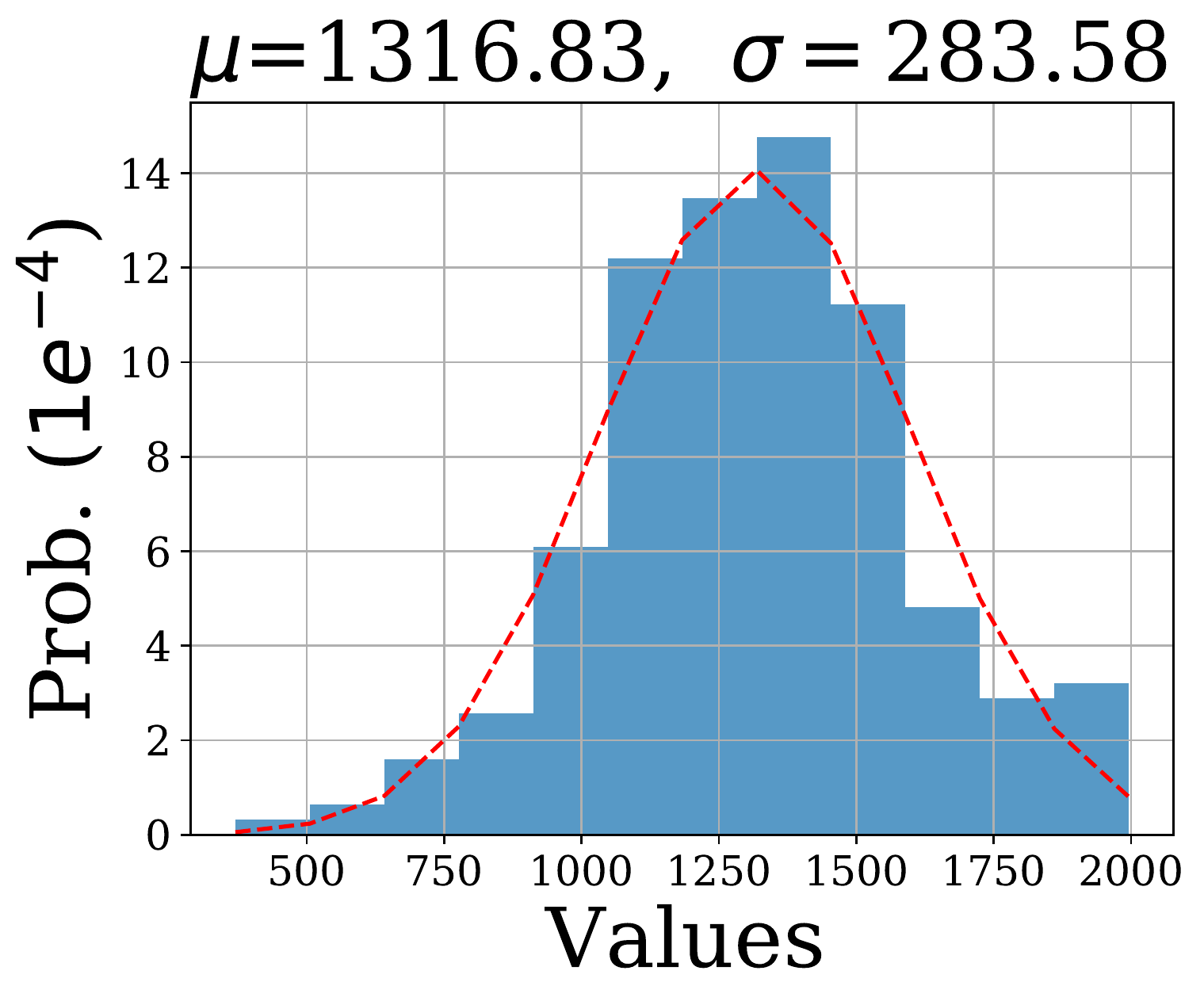}
        \caption{Trigger size distribution}
        \label{fig:polygon_size_dist}
    \end{minipage}
\end{figure}

\subsection{Scanner Synthesis} \label{sec:synthesis}
In the previous step, we summarize the attack decomposition based on different backdoor types, e.g., patch attack. For each backdoor type $\textit{backdoor}_i$, we model its coefficient distribution, e.g., patch size, color, and position.
In this step, we synthesize a scanner for each backdoor type, namely, $\textit{backdor}_i$ in Eq.~\ref{e:attack}, from its distribution coefficients. 
These scanners are based on trigger inversion.
We consider all trigger inversion methods use a general loss function template as follows.
\begin{equation}
    Loss = Loss_{ce} + Loss_{reg}
\end{equation}
with the first the cross-entropy loss and the second the regularization loss (e.g., Eq.~\ref{eq:inversion}).
\Tech{} synthesizes  scanners by {\em synthesizing the regularization term}. We want to point out this loss function is used in scanning a model (to determine if it has an backdoor) and hence different from that in decomposition (i.e., Eq.~\ref{e:loss}).

Specifically, for each attack feature $f$, such as $\tilde{m}$ and $\tilde{t}$ in the patching form of backdoors and $\widetilde{t^w}$ and $\widetilde{t^b}$ in transforming backdoors,
assume it has been summarized to $  f \sim \mathcal{N}(\mu_f,\,\sigma_f^{2})$. 
We introduce a regularization term as follows.
\begin{equation} \label{equ:distrib}
    Loss_{reg}^{f} = 
    \begin{cases}        
        0 & \text{if } f \in [\mu_{f} - z \cdot \sigma_{f}, \mu_{f} + z \cdot \sigma_{f}] \\
        \delta \cdot |f - \mu_{f}| & \text{otherwise.}
    \end{cases}
\end{equation}
Intuitively, during inversion, we aim to keep the $f$ value within the 15th-85th percentile. This is enforced by having the parameter $z=1.04$ in Eq.~\ref{equ:distrib}. In other words, penalty is introduced when it is beyond the range.

\smallskip
\noindent
\textbf{Example.}
We show how we summarize TrojAI polygon patch backdoors and synthesize a scanner.
First we sample 20 trojaned models and perform attack decomposition and summarization.
Figure~\ref{fig:polygon_cluster} shows the clustering result, where \Tech{} partitions them into one cluster due to their cohesive behaviors.
Moreover, the decomposed masks $\tilde{m}$ follow binomial distributions.
According to the discussion in Section~\ref{sec:design_summarize},
in such cases \Tech{} extracts attack feature as ($i$, $j$, $s$, $\tilde{t}$) with ($i$, $j$) the center of mask and $s$ its size.
Figure~\ref{fig:polygon_size_dist} illustrates the distribution of trigger size, with  $\mu_{s} = 1316.83$ and $\sigma_s = 283.58$.
Then \Tech{} synthesize a regularization term as follows, with $s=\sum m$ the size of mask during inversion.
\begin{equation}
    Loss_{s} = 
    \begin{cases}
        0 &\text{ if } \sum m \in [1033.25, 1600.41] \\
        100 \cdot |\sum m - 1316.83| &\text{ otherwise.}
    \end{cases}
\end{equation}

Similarly, we have other regularization losses for $i$ and $j$.
In Section~\ref{sec:scan}, we will show that \Tech{} can automatically synthesize 6 scanners for all the different types of backdoors in TrojAI and achieves over 0.9 detection accuracy, which existing scanners cannot achieve without substantial manual reconfiguration based on attack specifics.

\section{Evaluation}

This section evaluates how \Tech{} enhances the performance of various downstream scanners in detecting trojaned models (Section~\ref{sec:scan}) and eliminates identified injected backdoors through model unlearning (Section~\ref{sec:remove}). \Tech{} has two key components: attack decomposition and attack summarization.
For attack decomposition, we evaluate the quality of decomposed clean versions (of trojaned samples) and the attack effectiveness of decomposed triggers in Section~\ref{sec:visual}. For attack summarization, we validate the performance of automatic attack clustering in Section~\ref{sec:summarize}. As \Tech{} summarizes the attack knowledge from a small set of trojaned models and inputs, it is interesting to study the effect of biases on sampled models as well as inputs, which will be discussed in Section~\ref{sec:sample}. We also investigate three attack scenarios aiming to counter \Tech{} in Section~\ref{sec:adaptive}. Finally, a set of ablation studies are carried out to understand different design choices (Section~\ref{sec:ablation}).

Our experiments are conducted on 10 well-known backdoor attacks including static, dynamic, and complex backdoors on 2,532 models in total, consisting of 22 network architectures with 6 datasets. \Tech{} is compared with 9 baselines in various experiments.

\subsection{Experiment Setup}
\smallskip
\noindent
\textbf{Attack Setup.}
We evaluate on 10 existing backdoor attacks, namely, BadNets~\cite{badnet}, TrojNN~\cite{trojnn}, Dynamic~\cite{dynamic}, Reflection~\cite{reflection}, Blend~\cite{blend}, SIG~\cite{sig}, Invisible~\cite{invisible}, WaNet~\cite{wanet}, Gotham~\cite{ABS}, and DFST~\cite{dfst}. Widely used datasets such as ImageNet~\cite{ILSVRC15}, CelebA~\cite{celeba}, CIFAR-10~\cite{cifar10}, GTSRB~\cite{gtsrb} are utilized to construct trojaned and benign models. We also make use of 2,112 pre-trained models from TrojAI~\cite{trojai} rounds 2 and 3, half benign and half poisoned. Please see details of these backdoor attacks and datasets/models in Appendix~\ref{sec:append_setup}.

For the 10 aforementioned backdoor attacks, we use a poisoning rate of 10\%. Most of these backdoors are universal (by their default settings) where inputs from all the classes (except the target class) stamped with the trigger will be misclassified to the target label by the subject model.
Some of the TrojAI models are label specific  (only causing images of a victim class to be misclassified).
We use the same adversarial training strategies to make the backdoor robust for WaNet and DFST according to their original papers~\cite{wanet,dfst}.

\noindent
\textbf{Setup of \Tech{}.}
For attack decomposition, we assume 10 trojaned images and 100 clean images per model, where the original images of trojaned samples are different from those clean images.
We leverage the state-of-the-art StyleGAN~\cite{styleganv2} to recover the clean version of a trojaned image. We download pre-trained GANs from GenForce Lib~\cite{genforce} to handle different datasets.

\begin{table}[t]
    \centering
    \scriptsize
    \tabcolsep=3pt
    \caption{Evaluation on TrojAI (Universal polygon + Clean models)}
    \label{tab:eval_universal_polygon}
    \begin{tabular}{cccccccccccc}
         \toprule
         \multirow{2}{*}{\textbf{Scanner}} & \multirow{2}{*}{\textbf{Config}} & \multicolumn{5}{c}{\textbf{Round 2}} & \multicolumn{5}{c}{\textbf{Round 3}} \\ \cmidrule(lr){3-7} \cmidrule(lr){8-12}
         ~ & ~ & TP & FN & TN & FP & ACC & TP & FN & TN & FP & ACC \\
         \midrule
         \multirow{2}{*}{NC} & Original & 85 & 7 & 467 & 85 & 0.857 & 85 & 4 & 418 & 86 & 0.848 \\
         ~ & \Tech{} & 76 & 16 & 531 & 21 & \textbf{0.943} & 78 & 11 & 485 & 19 & \textbf{0.949} \\
         \midrule
         \multirow{2}{*}{Tabor} & Original & 73 & 19 & 459 & 93 & 0.826 & 65 & 24 & 426 & 78 & 0.828 \\
         ~ & \Tech{} & 66 & 26 & 540 & 12 & \textbf{0.941} & 76 & 13 & 474 & 30 & \textbf{0.927} \\
         \bottomrule
    \end{tabular}
\end{table}
\begin{table}[t]
    \centering
    \scriptsize
    \tabcolsep=2.3pt
    \caption{Evaluation on TrojAI (Label-specific polygon + Clean models)}
    \label{tab:eval_specific_polygon}
    \begin{tabular}{cccccccccccc}
         \toprule
         \multirow{2}{*}{\textbf{Scanner}} & \multirow{2}{*}{\textbf{Config}} & \multicolumn{5}{c}{\textbf{Round 2}} & \multicolumn{5}{c}{\textbf{Round 3}} \\ \cmidrule(lr){3-7} \cmidrule(lr){8-12}
         ~ & ~ & TP & FN & TN & FP & ACC & TP & FN & TN & FP & ACC \\
         \midrule
         \multirow{3}{*}{K-Arm} & Original & 98 & 178 & 541 & 10 & 0.773 & 144 & 107 & 497 & 7 & 0.849 \\
         ~ & Customized & 182 & 94 & 530 & 21 & 0.861 & 202 & 49 & 491 & 13 & 0.918 \\
         ~ & \Tech{} & 191 & 85 & 532 & 19 & \textbf{0.874} & 201 & 50 & 494 & 10 & \textbf{0.921} \\
         \midrule
         \multirow{3}{*}{ABS} & Original & 28 & 248 & 541 & 11 & 0.687 & 151 & 100 & 412 & 92 & 0.746 \\
         ~ & Customized & 211 & 65 & 538 & 14 & 0.905 & 213 & 38 & 474 & 30 & 0.910 \\
         ~ & \Tech{} & 233 & 43 & 524 & 28 & \textbf{0.914} & 218 & 33 & 481 & 23 & \textbf{0.926} \\
         \midrule
         \multirow{2}{*}{Trinity} & Upstream & 62 & 214 & 367 & 185 & 0.518 & 51 & 200 & 343 & 161 & 0.522 \\
         ~ & \Tech{} & 139 & 137 & 404 & 148 & \textbf{0.656} & 133 & 118 & 363 & 141 & \textbf{0.657} \\
         \bottomrule
    \end{tabular}
\end{table}
\begin{table}[t]
    \centering
    \scriptsize
    \tabcolsep=2.5pt
    \caption{Evaluation on TrojAI (Label-specific filter + Clean models)}
    \label{tab:eval_filter}
    \begin{tabular}{cccccccccccc}
         \toprule
         \multirow{2}{*}{\textbf{Scanner}} & \multirow{2}{*}{\textbf{Config}} & \multicolumn{5}{c}{\textbf{Round 2}} & \multicolumn{5}{c}{\textbf{Round 3}} \\ \cmidrule(lr){3-7} \cmidrule(lr){8-12}
         ~ & ~ & TP & FN & TN & FP & ACC & TP & FN & TN & FP & ACC \\
         \midrule
         \multirow{2}{*}{ABS} & Original & 178 & 98 & 527 & 25 & 0.851 & 143 & 109 & 470 & 34 & 0.811 \\
         ~ & \Tech{} & 231 & 45 & 524 & 28 & \textbf{0.912} & 185 & 67 & 496 & 8 & \textbf{0.901} \\
         \midrule
         \multirow{2}{*}{Trinity} & Upstream & 147 & 129 & 377 & 175 & 0.633 & 128 & 124 & 303 & 201 & 0.570 \\
         ~ & \Tech{} & 192 & 84 & 484 & 68 & \textbf{0.816} & 193 & 59 & 443 & 61 & \textbf{0.841} \\
         \bottomrule
    \end{tabular}
\end{table}

\begin{table}[t]
    \centering
    \scriptsize
    \tabcolsep=2.5pt
    \caption{Evaluation on TrojAI (Full set + Clean models)}
    \label{tab:eval_whole}
    \begin{tabular}{cccccccccccc}
         \toprule
         \multirow{2}{*}{\textbf{Scanner}} & \multirow{2}{*}{\textbf{Config}} & \multicolumn{5}{c}{\textbf{Round 2}} & \multicolumn{5}{c}{\textbf{Round 3}} \\ \cmidrule(lr){3-7} \cmidrule(lr){8-12}
         ~ & ~ & TP & FN & TN & FP & ACC & TP & FN & TN & FP & ACC \\
         \midrule
         \multirow{2}{*}{ABS} & Original & 276 & 276 & 518 & 34 & 0.719 & 331 & 172 & 286 & 118 & 0.712 \\
         ~ & \Tech{} & 467 & 85 & 508 & 44 & \textbf{0.883} & 409 & 94 & 473 & 31 & \textbf{0.876} \\
         \midrule
         \multirow{2}{*}{Trinity} & Upstream & 209 & 343 & 351 & 201 & 0.507 & 218 & 285 & 290 & 214 & 0.504 \\
         ~ & \Tech{} & 349 & 203 & 362 & 190 & \textbf{0.644} & 334 & 169 & 322 & 182 & \textbf{0.651} \\
         \bottomrule
    \end{tabular}
\end{table}

\subsection{Forensics-aided Defense against Injected Backdoors}

\subsubsection{Backdoor Scanning} \label{sec:scan}

We integrate \Tech{} with 5 state-of-the-art trigger-inversion based backdoor scanners that determine if a model contains a backdoor by inverting a (small) trigger that can induce misclassification for a small set of clean samples. Besides NC and ABS discussed in Section~\ref{sec:motivation}, we integrate with
Tabor~\cite{Tabor}, K-Arm~\cite{karm}, and SRI Trinity~\cite{sri} as well (see detailed descriptions of these scanners in Appendix~\ref{sec:scanner_appendix}).
NC only supports universal patch type of backdoors. ABS supports universal 
and label-specific patch backdoors and filter backdoors.

\begin{table*}[h]
    \centering
    \scriptsize
    \tabcolsep=6pt
    \caption{Scanning performance on complex attacks}
    \label{tab:eval_abs_complex}
    \begin{tabular}{cccccccccccccc}
    \toprule
    \multirow{2.5}{*}{\textbf{Dataset}} & \multirow{2.5}{*}{\textbf{Attack}} & \multicolumn{6}{c}{\textbf{Original}} & \multicolumn{6}{c}{\textbf{\Tech{}-enhanced}} \\ \cmidrule(lr){3-8} \cmidrule(lr){9-14}
    & ~ & PN-REASR & CL-REASR & FP & FN & ACC & Time (s) & PN-REASR & CL-REASR & FP & FN & ACC & Time (s) \\
    \midrule
    \multirow{7}{*}{\rotatebox[origin=c]{90}{CIFAR-10}}
    & Dynamic & 0.79 $\pm$ 0.33 & 0.55 $\pm$ 0.18 & 1 & 9 & 0.83 & 117.4 & 1.00 $\pm$ 0.00 & 0.19 $\pm$ 0.02 & 0 & 0 & 1.00 & 119.4 \\
    & Reflection & 0.52 $\pm$ 0.26 & 0.56 $\pm$ 0.17 & 1 & 20 & 0.65 & 117.3 & 1.00 $\pm$ 0.00 & 0.72 $\pm$ 0.22 & 4 & 0 & 0.93 & 116.0 \\
    & SIG & 0.47 $\pm$ 0.22 & 0.55 $\pm$ 0.19 & 1 & 29 & 0.50 & 118.2 & 0.89 $\pm$ 0.24 & 0.40 $\pm$ 0.27 & 1 & 5 & 0.90 & 115.8 \\
    & Blend & 0.78 $\pm$ 0.33 & 0.29 $\pm$ 0.13 & 0 & 9 & 0.85 & 116.4 & 0.93 $\pm$ 0.21 & 0.54 $\pm$ 0.18 & 1 & 3 & 0.93 & 116.8 \\
    & Invisible & 0.31 $\pm$ 0.19 & 0.18 $\pm$ 0.02 & 0 & 29 & 0.52 & 151.3 & 0.95 $\pm$ 0.11 & 0.74 $\pm$ 0.07 & 0 & 3 & 0.95 & 159.2 \\
    & WaNet & 0.42 $\pm$ 0.33 & 0.20 $\pm$ 0.04 & 0 & 25 & 0.58 & 153.2 & 0.94 $\pm$ 0.10 & 0.80 $\pm$ 0.07 & 1 & 3 & 0.93 & 155.9 \\
    & DFST & 0.53 $\pm$ 0.28 & 0.30 $\pm$ 0.18 & 1 & 24 & 0.58 & 154.5 & 0.96 $\pm$ 0.05 & 0.79 $\pm$ 0.07 & 0 & 5 & 0.92 & 162.9 \\
    \midrule 
    \multirow{4}{*}{\rotatebox[origin=c]{90}{GTSRB}}
    & Dynamic & 0.81 $\pm$ 0.20 & 0.71 $\pm$ 0.07 & 0 & 15 & 0.75 & 127.6 & 1.00 $\pm$ 0.00 & 0.47 $\pm$ 0.07 & 0 & 0 & 1.00 & 127.5 \\
    & Reflection & 0.84 $\pm$ 0.05 & 0.74 $\pm$ 0.07 & 1 & 24 & 0.58 & 127.6 & 0.87 $\pm$ 0.24 & 0.51 $\pm$ 0.24 & 0 & 4 & 0.93 & 120.5 \\
    & SIG & 0.68 $\pm$ 0.06 & 0.70 $\pm$ 0.08 & 0 & 30 & 0.50 & 121.5 & 0.93 $\pm$ 0.24 & 0.22 $\pm$ 0.21 & 0 & 2 & 0.97 & 126.8 \\
    & Blend & 0.92 $\pm$ 0.17 & 0.47 $\pm$ 0.09 & 0 & 6 & 0.90 & 127.5 & 1.00 $\pm$ 0.00 & 0.70 $\pm$ 0.08 & 0 & 0 & 1.00 & 127.8 \\
    \midrule
    \multirow{2}{*}{\rotatebox[origin=c]{90}{\begin{tabular}[x]{@{}c@{}}Image\\-Net\end{tabular}}}
    & Invisible & 0.47 $\pm$ 0.43 & 0.19 $\pm$ 0.12 & 1 & 13 & 0.65 & 1285.9 & 0.86 $\pm$ 0.30 & 0.73 $\pm$ 0.15 & 2 & 2 & 0.90 & 1318.4 \\
    & WaNet & 0.32 $\pm$ 0.27 & 0.20 $\pm$ 0.12 & 0 & 16 & 0.60 & 1286.6 & 0.91 $\pm$ 0.12 & 0.72 $\pm$ 0.13 & 1 & 3 & 0.90 & 1326.7 \\
    \bottomrule
    \end{tabular}
    \vspace{-3pt}
\end{table*}

While Table~\ref{tab:moti_nc_abs} in Section~\ref{sec:motivation}
already shows that existing scanners cannot be generally effective and only work for the types of backdoors that they focus on, whereas \Tech{} can effectively and fully automatically scan all kinds of backdoors. In this study, we further show that using the loss functions automatically synthesized by \Tech{}, we can substantially improve these scanners even for their targeted backdoor types.
Specifically, we use the models from TrojAI rounds 2 and 3. The models of round 3 are adversarial trained while the round 2 models not.
We evaluate NC and Tabor on trojaned models with universal polygon backdoors, K-Arm, ABS, SRI Trinity on trojaned models with universal and label-specifc polygon backdoors, and ABS and SRI Trinity on trojaned models with filter triggers.
Trojaned models are always mixed with equal number of benign models during scanning.
Besides, we evaluate ABS and SRI Trinity on the entire set of models (with all sorts of backdoors).
For each setting, we assume we have access to only 20 random (< 10\%) trojaned models for attack decomposition and summarization.
While there may be sampling biases, we study the effects of such biases later in this section.

Table~\ref{tab:eval_universal_polygon} shows the results of NC and Tabor on universal polygon backdoors. We can see that there is roughly a 10\% accuracy improvement on each setting. Note that the number of FPs (False Positives) is largely reduced while the number of FNs (False Negatives) slightly increases. This is because we regulate the inversion within certain distributions.
Table~\ref{tab:eval_specific_polygon} shows the results of K-Arm, ABS and SRI Trinity on universal and label-specific polygon backdoors. For K-Arm and ABS, we report the performance of original settings on their Github and their customized versions for TrojAI in which the configurations are changed based on the released attack information. Observe that \Tech{} improves the scanning accuracy by 10\%-15\% compared to the original settings. \Tech{} can even improve the customized versions by 1.5\%.
Note that the customized versions had undergone intensive manual tuning and added regularization specific to the TrojAI attacks. For example, the customized ABS adds a constraint that all pixels in an inverted trigger area have the same color, as the round specifications state that a polygon trigger is always filled with the same color.
Note that the performance of SRI Trinity is not as high as the other two because we only use its upstream inversion technique.
Table~\ref{tab:eval_filter} shows the results of ABS and SRI Trinity on instagram filter backdoors. \Tech{} achieves an overall improvement from 6\% to 27\% and largely reduces both FPs and FNs in most cases. ABS was not customized for filter backdoors.
Table~\ref{tab:eval_whole} shows the results of ABS and SRI Trinity on the entire model sets. 
In this case, \Tech{} automatically clusters the provided instances and synthesizes the corresponding inversion loss functions.
Observe that the improvement is around 15\%, and both FPs and FNs are reduced in most cases.
As the customized ABS does not handle filter backdoors, we use the original ABS in this experiment.

Besides, we create multiple model sets to evaluate complex attacks, including Dynamic, Reflection, SIG, Blend, Invisible, WaNet and DFST.
We train 30 clean models and 30 trojaned models for each complex attack on CIFAR-10 and GTSRB to compose the subject model sets.
We train 20 clean models and 20 trojaned models on ImageNet.
We assume for each attack, we have access to 5 trojaned models other than the subject model sets, on which we perform attack decomposition and summarization. We leverage ABS as the downstream scanner.
Here we also assume the sampled models cover all the backdoor types in the subject models sets.
Table~\ref{tab:eval_abs_complex} shows the results. The first column denotes the datasets and the attacks. The second and third large columns denote the performance of original ABS and \Tech{}-enhanced one.
Following the original ABS setup, if the inverted trigger can achieve 0.88 ASR on the validation images, we consider a model trojaned.
In each large column, there are 5 columns, PN-REASRs show the average REASR (ASR of reverse-engineered trigger by ABS) on poisoned models, while CL-REASRs show the REASRs on clean models. FP, FN, ACC denote the number of false positives, false negatives and the scanning accuracy. We also report the scanning time.

\begin{figure}[t]
    \centering
    \includegraphics[width=0.48\textwidth]{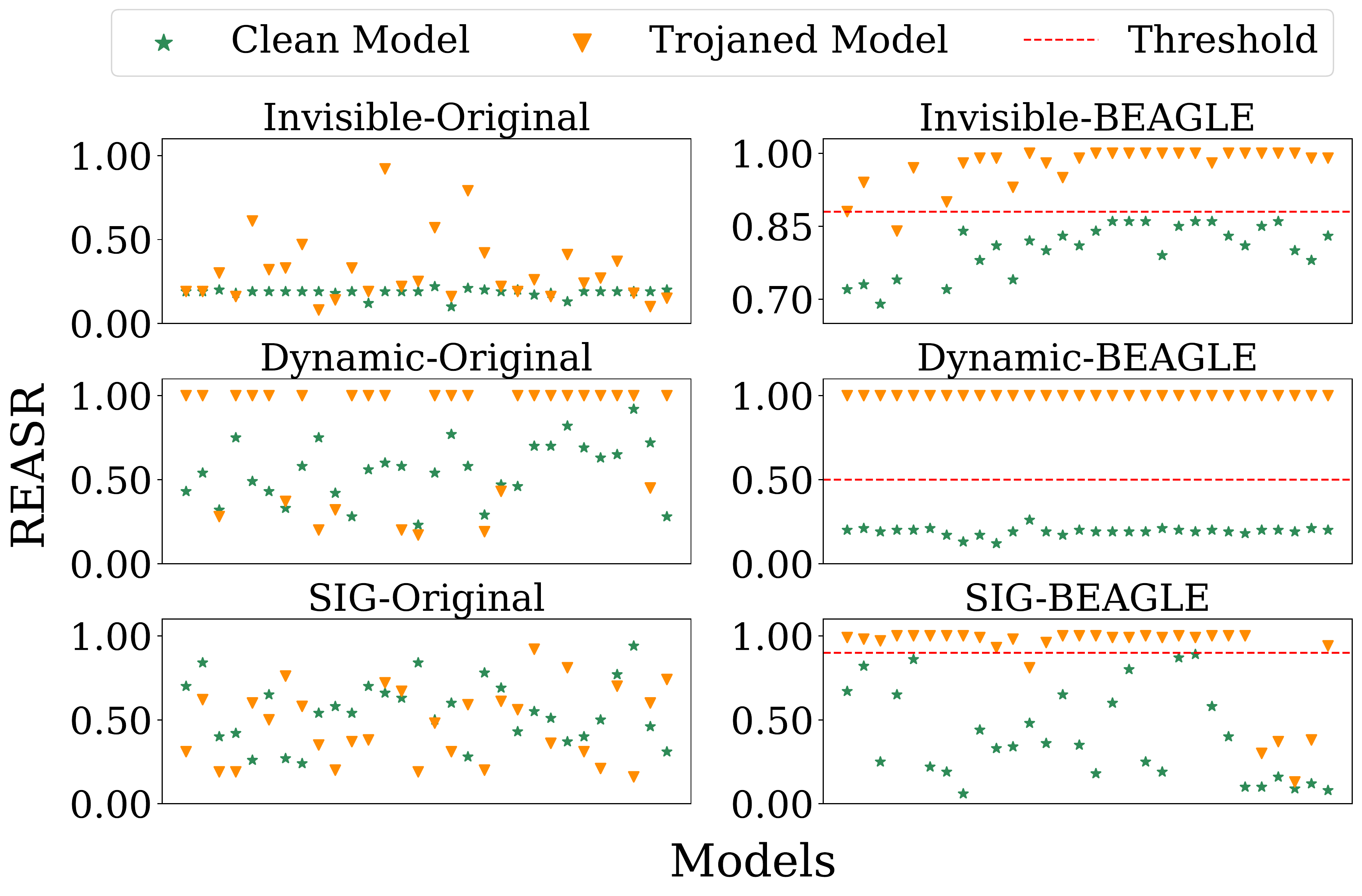}
    \caption{Separation of clean and trojaned models. The first column shows the data points by the original ABS while the second column shows those by the \Tech{} equipped ABS. Observe \Tech{} allows better separation.}
    \label{fig:enh_abs_complex}
\end{figure}

Observe that in most cases, The \Tech{}-equipped ABS can improve the scanning accuracy by a large extent, especially for WaNet, Invisible and DFST.
Note that sometimes the \Tech{}-equipped ABS may induce a few FPs, which is reasonable because \Tech{}'s piece-wise linear transformation function is expressive and tends to generate some adversarial perturbations leading to high ASR. This is also evidenced by the nontrivial CL-REASRs in Invisible, WaNet and DFST.
Nonetheless, we can still find a clear separation between clean and trojaned models as shown in Figure~\ref{fig:enh_abs_complex}, much better than without \Tech{}.
In addition, we observe \Tech{}-equipped ABS spends similar time compared to the original version, which means the overhead is small.

\subsubsection{Backdoor Removal} \label{sec:remove}
\begin{table}[t]
    \centering
    \scriptsize
    \renewcommand{\arraystretch}{1.2}
    \tabcolsep=3pt
    \caption{Backdoor removal}
    \label{tab:model_harden}
    \begin{tabular}{cggccggccgg}
         \toprule
         \multirow{2.5}{*}{\textbf{Attack}} & \multicolumn{2}{c}{\textbf{Original}} & \multicolumn{2}{c}{\textbf{Finetune}} & \multicolumn{2}{c}{\textbf{NAD}} & \multicolumn{2}{c}{\textbf{ANP}} & \multicolumn{2}{c}{\textbf{\Tech}} \\ \cmidrule(lr){2-3} \cmidrule(lr){4-5} \cmidrule(lr){6-7} \cmidrule(lr){8-9} \cmidrule(lr){10-11}
         ~ & \cellcolor{white}{ACC} & \cellcolor{white}{ASR} & ACC & ASR & \cellcolor{white}{ACC} & \cellcolor{white}{ASR} & ACC & ASR & \cellcolor{white}{ACC} & \cellcolor{white}{ASR} \\
         \midrule
         BadNets & 0.919 & 1.000 & 0.893 & 0.174 & 0.878 & 0.080 & 0.891 & 0.032 & 0.894 & \textbf{0.013} \\
         TrojNN & 0.917 & 1.000 & 0.879 & 0.250 & 0.875 & 0.142 & 0.878 & 0.412 & 0.879 & \textbf{0.068} \\
         Dynamic & 0.919 & 1.000 & 0.897 & 0.153 & 0.875 & 0.043 & 0.882 & 0.022 & 0.877 & \textbf{0.013} \\
         Reflection & 0.918 & 0.991 & 0.883 & 0.944 & 0.879 & 0.264 & 0.883 & 0.215 & 0.876 & \textbf{0.136} \\
         Blend & 0.920 & 1.000 & 0.889 & 0.005 & 0.868 & 0.042 & 0.876 & \textbf{0.002} & 0.875 & 0.004 \\
         SIG & 0.914 & 0.952 & 0.888 & 0.179 & 0.876 & 0.027 & 0.877 & 0.012 & 0.876 & \textbf{0.007} \\
         Invisible & 0.918 & 1.000 & 0.894 & 0.415 & 0.886 & 0.355 & 0.881 & 0.277 & 0.881 & \textbf{0.027} \\
         WaNet & 0.908 & 0.989 & 0.904 & 0.179 & 0.882 & 0.039 & 0.898 & 0.018 & 0.904 & \textbf{0.015} \\
         Gotham & 0.913 & 1.000 & 0.888 & 0.108 & 0.873 & \textbf{0.040} & 0.864 & 0.074 & 0.874 & 0.050\\
         DFST & 0.889 & 0.996 & 0.884 & 0.428 & 0.873 & 0.214 & 0.876 & 0.202 & 0.876 & \textbf{0.142} \\
         \midrule
         \textbf{Average} & 0.914 & 0.993 & 0.890 & 0.284 & 0.876 & 0.125 & 0.881 & 0.127 & 0.881 & \textbf{0.048} \\
         \bottomrule
    \end{tabular}
\end{table}

Backdoor removal aims to eliminate injected backdoors in models. In Section~\ref{sec:decomposition}, we have demonstrated that our decomposed triggers closely resemble the original injected triggers and are highly effective to induce the same attack effects (as the original triggers). The idea is hence to leverage our decomposed triggers in model unlearning to remove injected backdoors.

We use the CIFAR-10 dataset and the VGG-11 network and conduct the experiments on 10 backdoor attacks. We assume 1\% of the original training dataset is available for retraining the model. The same data augmentations as in NAD~\cite{li2021neural} are leveraged, including random crop, random cutoff, and horizontal flipping. Our model unlearning is carried out by stamping decomposed triggers on training samples and using the original ground truth labels during training. Several existing backdoor removal techniques are considered for comparison, such as Finetune, NAD~\cite{li2021neural}, and ANP~\cite{wu2021adversarial}. The final results are obtained by constraining the accuracy degradation to be within 5\%.
Table~\ref{tab:model_harden} shows the results. The first column presents different backdoor attacks. The following columns show the results for the original poisoned models and the models cleansed by different techniques. We report both the clean accuracy (ACC) and attack success rate (ASR) in the table. Observe that in most cases, \Tech{} can effectively eliminate injected backdoors, especially for Reflection, Invisible, and DFST, outperforming baselines. On average, \Tech{} reduces 10\%-20\% more ASR than the state-of-the-art methods.
This also demonstrates that the decomposed trigger by \Tech{} is very similar to the original trigger. A simple model unlearning can already remove most of those injected backdoors.

\subsection{Validating Decomposed Clean Inputs and Triggers} \label{sec:visual}

\begin{figure*}
    \centering
    \includegraphics[width=1.0\textwidth]{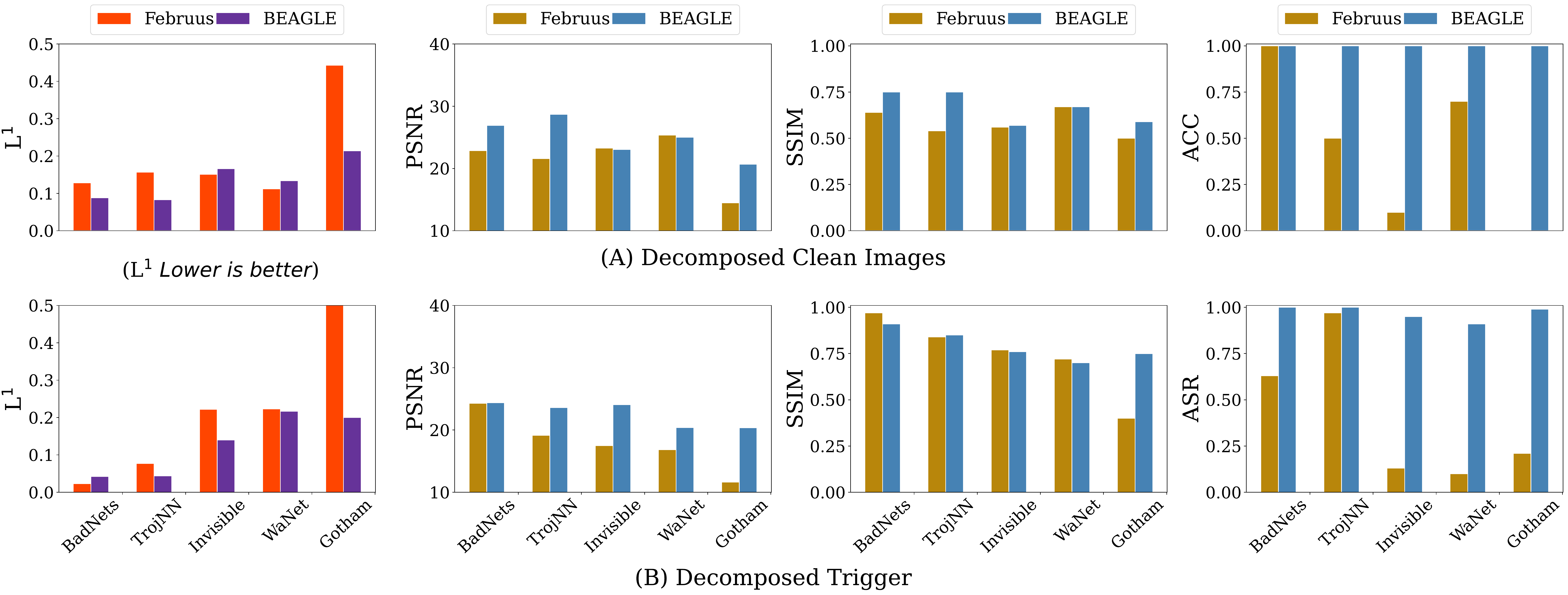}
    \caption{Decomposition quality on ImageNet}
    \label{fig:decompose_imagenet}
\end{figure*}

\begin{figure*}
    \centering
    \includegraphics[width=1.0\textwidth]{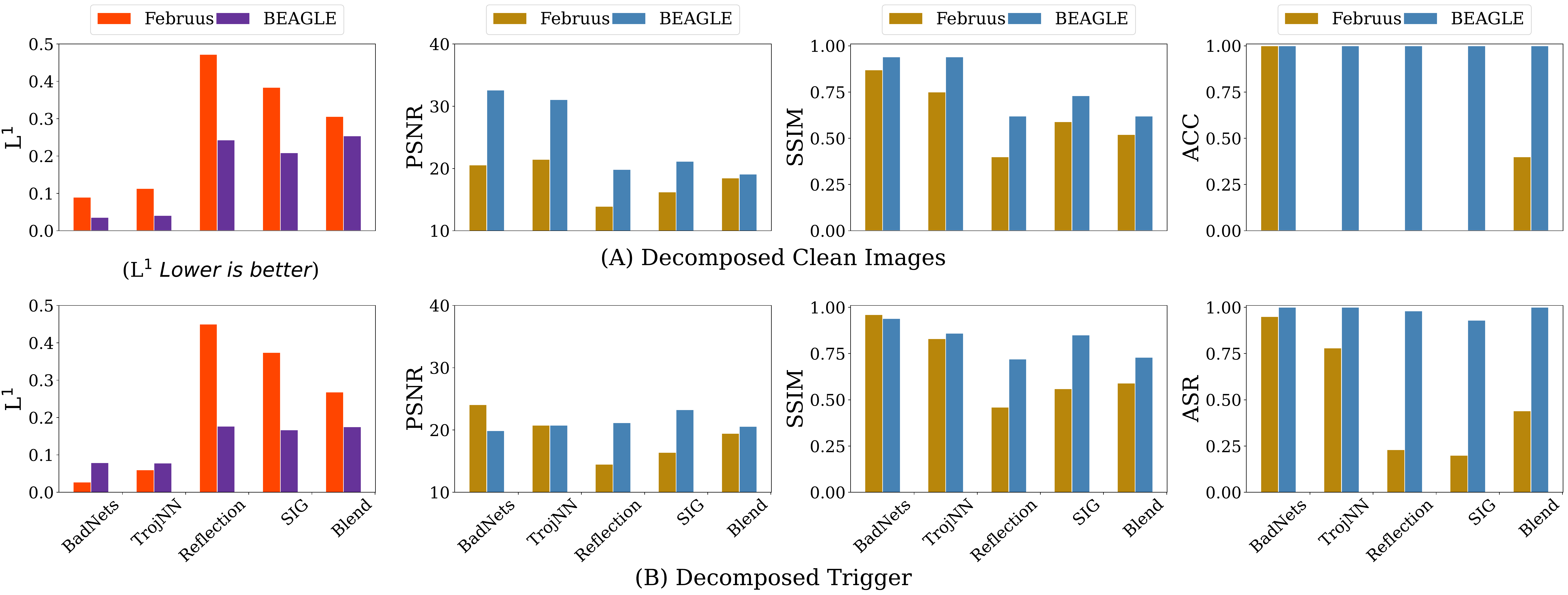}
    \caption{Decomposition quality on CelebA \vspace{-7pt}}
    \label{fig:decompose_celebA}
\end{figure*}

We qualitatively and quantitatively validate the decomposed clean inputs and triggers by assessing their visual quality and classification accuracy.
The visual quality of decomposed clean inputs is measured by comparing them with their original versions (before trojaned).
The validation of decomposed triggers is carried out by comparing clean images stamped with the ground-truth trigger and with the decomposed trigger. Three widely-used metrics,  $\normlone$ distance, Peak Signal-to-Noise Ratio (PSNR), Structural Similarity Index Measure (SSIM), are utilized to quantify the differences between aforementioned pairs. A good decomposition result shall have a small $\normlone$ distance, a high PSNR, and a large SSIM.
For decomposed clean inputs, the subject model shall correctly classify them with a high standard accuracy. For decomposed triggers, the subject model shall produce the target label when they are stamped onto clean images, which corresponds to the attack success rate.

We compare a state-of-the-art technique \textit{Februus}~\cite{februus}, which removes backdoor triggers in trojaned images, with \Tech{} and show the results in Figure~\ref{fig:decompose_imagenet} and Figure~\ref{fig:decompose_celebA}.
All the trojaned models are well-trained with performance on par with state-of-the-arts. Specifically, the top-1 accuracy is $>73\%$ for ImageNet, $>78\%$ for CelebA, $>91\%$ for CIFAR-10, and $>94\%$ for GTSRB. The ASRs are all $>97\%$, except for Reflection and SIG whose ASRs are $>88\%$, consistent with the original papers~\cite{reflection, sig}.
Figure~\ref{fig:decompose_imagenet} shows the result for ImageNet and Figure~\ref{fig:decompose_celebA} for CelebA.
The detailed numbers and other results for CIFAR-10 and GTSRB can be found in the supplementary document Section A~\cite{beagle}.
In each figure, there are two rows, row (A) denoting the quality of decomposed clean images and row (B) the quality of decomposed triggers.
In each row, four bar charts recording the $\normlone$ distance, PSNR, SSIM, and ACC/ASR, respectively, for the five attacks are displayed.
Since for the $\normlone$ distance, lower bars denote better performance while higher bars are better for other metrics, we use different colors to present the bar charts of $\normlone$ distance.
In the $\normlone$, PSNR, and SSIM bar charts, there are two bars for each attack, where the left bar shows the difference between original clean images and decomposed images by Februus and the right bar the difference by \Tech.
Each bar presents the average value for the given 10 trojaned images and 100 additional clean test images.
For the ACC/ASR bar charts, the left bar is for Februus and the right for $\Tech{}$. Each ACC bar denotes the average clean accuracy for the decomposed clean images from the given 10 trojaned samples. As the 100 validation clean images have been used in optimization, we use the images from the test set and stamp the decomposed trigger to calculate the ASR.

\begin{figure}[!h]
    \centering
    \begin{minipage}{0.5\textwidth}
        \centering
        \begin{minipage}[t]{.22\textwidth}
            \centering
            \includegraphics[width=1.0\textwidth]{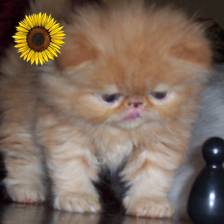}
            
            \caption*{(a) $x \oplus t$}
        \end{minipage}
        \begin{minipage}[t]{.22\textwidth}
            \centering
            \includegraphics[width=1.0\textwidth]{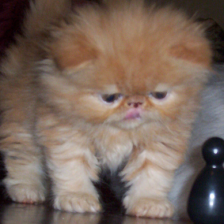}
            \caption*{(b) $x$}
        \end{minipage}
        \begin{minipage}[t]{.22\textwidth}
            \centering
            \includegraphics[width=1.0\textwidth]{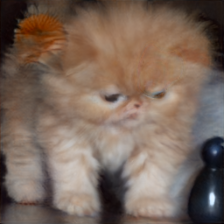}
            \caption*{(c) Fb. $\tilde{x}$}
        \end{minipage}
        \begin{minipage}[t]{.22\textwidth}
            \centering
            \includegraphics[width=1.0\textwidth]{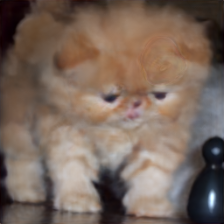}
            \caption*{(d) {\sc Bg.} $\tilde{x}$}
        \end{minipage}
    \end{minipage}
    \begin{minipage}{0.5\textwidth}
    \centering
        \begin{minipage}[t]{.22\textwidth}
            \centering
            \includegraphics[width=1.0\textwidth]{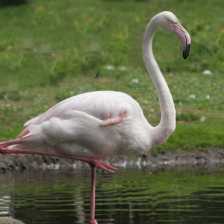}
            \caption*{(e) $x_{v}$}
        \end{minipage}
        \begin{minipage}[t]{.22\textwidth}
            \centering
            \includegraphics[width=1.0\textwidth]{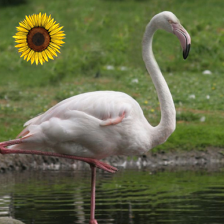}
            \caption*{(f) $x_{v} \oplus t$}
        \end{minipage}
        \begin{minipage}[t]{.22\textwidth}
            \centering
            \includegraphics[width=1.0\textwidth]{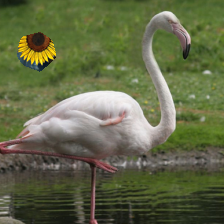}
            \caption*{(g) Fb. $x_{v} \oplus \tilde{t}$}
        \end{minipage}
        \begin{minipage}[t]{.22\textwidth}
            \centering
            \includegraphics[width=1.0\textwidth]{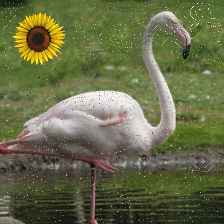}
            \caption*{(h) {\sc Bg.} $x_{v} \oplus \tilde{t}$}
        \end{minipage}
        \caption*{(A) Decomposition of BadNets}
        \label{fig:decomp_badnet}
    \end{minipage}
    
    \begin{minipage}{0.5\textwidth}
    \centering
        \begin{minipage}[t]{.22\textwidth}
            \centering
            \includegraphics[width=1.0\textwidth]{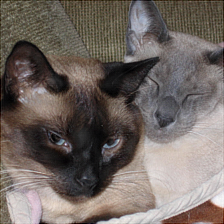}
            \caption*{(a) $x \oplus t$}
        \end{minipage}
        \begin{minipage}[t]{.22\textwidth}
            \centering
            \includegraphics[width=1.0\textwidth]{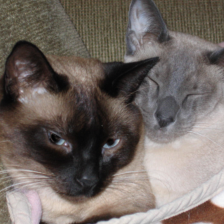}
            \caption*{(b) $x$}
        \end{minipage}
        \begin{minipage}[t]{.22\textwidth}
            \centering
            \includegraphics[width=1.0\textwidth]{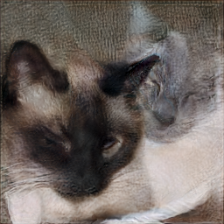}
            \caption*{(c) Fb. $\tilde{x}$}
        \end{minipage}
        \begin{minipage}[t]{.22\textwidth}
            \centering
            \includegraphics[width=1.0\textwidth]{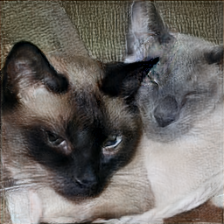}
            \caption*{(d) {\sc Bg.} $\tilde{x}$}
        \end{minipage}
    \end{minipage}
    \begin{minipage}{0.50\textwidth}
    \centering
        \begin{minipage}[t]{.22\textwidth}
            \centering
            \includegraphics[width=1.0\textwidth]{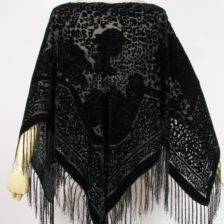}
            \caption*{(e) $x_{v}$}
        \end{minipage}
        \begin{minipage}[t]{.22\textwidth}
            \centering
            \includegraphics[width=1.0\textwidth]{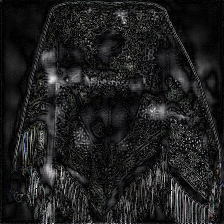}
            \caption*{(f) $x_{v} \oplus t - x_v$}
        \end{minipage}
        \begin{minipage}[t]{.22\textwidth}
            \centering
            \includegraphics[width=1.0\textwidth]{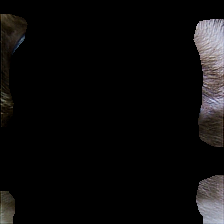}
            \captionsetup{font=footnotesize}
            \caption*{(g) Fb. $x_{v} \oplus \tilde{t} - x_{v}$}
        \end{minipage}
        \begin{minipage}[t]{.22\textwidth}
            \centering
            \includegraphics[width=1.0\textwidth]{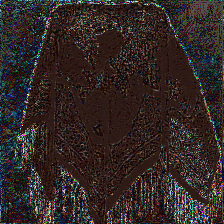}
            \captionsetup{font=footnotesize}
            \caption*{(h) {\sc Bg.}\! $x_{v}\! \oplus \tilde{t}\! - x_{v}$}
        \end{minipage}
        \caption*{(B) Decomposition of Invisible}
        \label{fig:decomp_invisible}
    \end{minipage}
    %
    \begin{minipage}{0.5\textwidth}
    \centering
        \begin{minipage}[t]{.22\textwidth}
            \centering
            \includegraphics[width=1.0\textwidth]{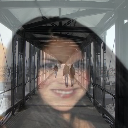}
            \caption*{(a) $x \oplus t$}
        \end{minipage}
        \begin{minipage}[t]{.22\textwidth}
            \centering
            \includegraphics[width=1.0\textwidth]{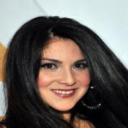}
            \caption*{(b) $x$}
        \end{minipage}
        \begin{minipage}[t]{.22\textwidth}
            \centering
            \includegraphics[width=1.0\textwidth]{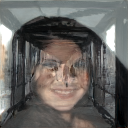}
            \caption*{(c) Fb. $\tilde{x}$}
        \end{minipage}
        \begin{minipage}[t]{.22\textwidth}
            \centering
            \includegraphics[width=1.0\textwidth]{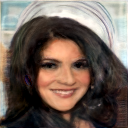}
            \caption*{(d) {\sc Bg.} $\tilde{x}$}
        \end{minipage}
    \end{minipage}
    \begin{minipage}{0.5\textwidth}
    \centering
        \begin{minipage}[t]{.22\textwidth}
            \centering
            \includegraphics[width=1.0\textwidth]{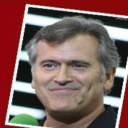}
            \caption*{(e) $x_{v}$}
        \end{minipage}
        \begin{minipage}[t]{.22\textwidth}
            \centering
            \includegraphics[width=1.0\textwidth]{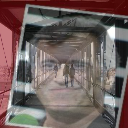}
            \caption*{(f) $x_{v} \oplus t$}
        \end{minipage}
        \begin{minipage}[t]{.22\textwidth}
            \centering
            \includegraphics[width=1.0\textwidth]{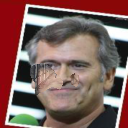}
            \caption*{(g) Fb. $x_{v} \oplus \tilde{t}$}
        \end{minipage}
        \begin{minipage}[t]{.22\textwidth}
            \centering
            \includegraphics[width=1.0\textwidth]{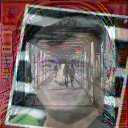}
            \caption*{(h) {\sc Bg.} $x_{v} \oplus \tilde{t}$}
        \end{minipage}
        \caption*{(C) Decomposition of Reflection}
        \label{fig:decomp_reflection}
    \end{minipage}
    \caption{
    Attack Decomposition. The decomposition of each attack is visualized in a block of 2 $\times$ 4 images. Image (a) shows the given trojaned image $x \oplus t$, (b) its ground-truth clean version $x$, (c) the decomposed clean version $\tilde{x}$ of Februus (Fb.), and (d) the decomposed clean version $\tilde{x}$ of \Tech{} ({\sc Bg.}) respectively. Image (e) shows a validation clean image $x_{v}$, (f) its ground-truth trojaned version $x_{v} \oplus t$, (g) the trojaned version stamped with the decomposed trigger $x_{v} \oplus \tilde{t}$ by Februus (Fb.), (h) and the trojaned version stamped with the decomposed trigger $x_{v} \oplus \tilde{t}$ by \Tech{} ({\sc Bg.}) respectively. For Invisible attack, we visualize the difference between the clean version and trigger-stamped one ($x_v \oplus t - x_v$) as the trigger effect is invisible.}
    \label{fig:decomposition}
\end{figure}

Observe that in most cases, \Tech{} outperforms Februus in visual quality, classification accuracy, and ASR.
As Februus is designed to handle patch attacks, we can see for BadNets and TrojNN on ImageNet and CelebA, in some cases, Februus outperforms ours with slightly better visual quality for decomposed triggers. While Februus directly removes the trigger area, we optimize a trigger mask, which may potentially induce some noise as the mask is continuous ranging from 0 to 1, causing some trigger pixels not fully extracted.
Figure~\ref{fig:decomposition} (A) shows a case for BadNets. The first row shows the decomposition results for a trojaned sample. Subfigure (a) is the trojaned sample, (b) the ground-truth clean version of the trojaned sample, (c) the decomposed clean image by Februus, and (d) the decomposed clean image by \Tech{}. The second row shows the quality of the decomposed trigger. Subfigure (e) is a clean image, (f) the clean image with the ground-truth trigger, (g) the clean image with Februus's decomposed trigger, and (h) the clean image with our decomposed trigger. Observe that the trigger extracted by Februus contains part of the ground-truth trigger. Observe that \Tech{} captures almost all the trigger features. However, using the metrics in Figure~\ref{fig:decompose_imagenet} on this case shows that Februus outperforms \Tech{}. This is because the trigger extracted by \Tech{} contains some (imperceptible) noise, degrading the measured values.
For Invisible and WaNet, Februus has better visual quality on decomposed clean images than ours. This is because we apply adaptive normalization to trojaned images before reconstruction, which may induce a slightly different distribution compared to the original one. It is reasonable as we have no knowledge of the original distribution. Besides, note that the difference between trojaned samples and their clean counterparts is small (invisible perturbation). Februus removes the entire area and directly reconstructs the input using GAN, which leads to unfaithful decomposed inputs and triggers. This is evidenced by that the ACC of decomposed clean images and the ASR of the decomposed triggers are both low for Februus. \Tech{}, on the other hand, faithfully reconstructs the clean images and approximates the trigger injection functions, achieving the ACC of 100\% and the ASR of larger than 90\%.
Figure~\ref{fig:decomposition} (B) shows a case for Invisible attack. Our recovered image is only slightly different from the source image and the decomposed trigger is similar to the original injected one.
For other attacks, such as Reflection, SIG, and Gotham, \Tech{} significantly outperforms Februus in both visual quality and classification results in Figure~\ref{fig:decompose_celebA}.
Figure~\ref{fig:decomposition} (C) shows a case of the reflection attack. \Tech{} is able to achieve high visual quality in reconstructing the clean image and extracting the trigger, whereas Februus fails.

\subsection{Evaluation on Attack Clustering and  Summarization} \label{sec:summarize}

A critical assumption of \Tech{} is that we are able to model existing backdoors by two mathematical forms (see Section~\ref{sec:decomposition}):
patching and transforming. Here we validate this assumption.
We conduct an experiment on CIFAR-10 with VGG-11 for seven backdoor attacks: BadNets, Dynamic, Reflection, SIG, Invisible, WaNet, and Gotham. We train 30 trojaned models for each attack and then randomly sample 5 from each attack to form a pool of attack instances. We then use \Tech{} to cluster these instances. \Tech{}   generates 11 clusters, 3 for the reflection attack corresponding to the three trigger images used, 3 for SIG (due to the 3 different trigger images), and one for each of the remaining attack types.
Table~\ref{tab:trigger_type} shows the summarization results. 
The first row shows the attacks.  The second row describes the clusters including their forms (patching or transforming) and coefficient distributions (e.g., binomial and constant).
The third row shows the ASR of the decomposd triggers.
Observe that BadNets and Dynamic belong to the patching category and their masks have a binomial distribution, indicating that they replace pixel values. Reflection and SIG belong to the patching category with constant mask values, meaning that they merge images. Gotham belongs to the transforming type and the coefficients are simple (mostly 0). Invisible and WaNet on the other hand have complex grid coefficients.
Observe that all the decomposed triggers have high ASRs, supporting that our backdoor modeling can cover all these backdoors.
We also validate our assumption on TrojAI rounds 2 and 3. Details can be found in the supplementary document Section B~\cite{beagle}.

\subsection{Impact of Attack Sample Bias} \label{sec:sample}

As \Tech{} leverages a small set of inputs (clean and trojaned) and trojaned models, we study the impact of sampling biases in such data. For sampled inputs, we include a few naturally misclassified inputs without any injected backdoors. This simulates the real world scenario where classification models do not usually achieve 100\% accuracy. For sampled models, we intentionally introduce biases to the number of trojaned models with different attack types.
Our experiments show that \Tech{} is robust to sample biases. Details can found in the supplementary document Section C~\cite{beagle}.

\begin{table}[t]
    \centering
    \scriptsize
    \tabcolsep=3.0pt
    \caption{Summarization of different attacks}
    \begin{tabular}{cccccccc}
        \toprule
         \textbf{Attack} & BadNets & Dynamic & Reflection(3) & SIG(3) & Invisible & WaNet & Gotham \\
         \toprule
         \textbf{Cluster} & P(binm) & P(binm) & P(cnst) & P(cnst) & T(cmplx) & T(cmplx) & T(smpl) \\
         \midrule
         \textbf{ASR} & 1.00 & 1.00 & 0.98 & 0.94 & 0.97 & 0.91 & 1.00 \\
         \bottomrule
    \end{tabular}
    \label{tab:trigger_type}
\end{table}

\subsection{Adaptive Attack} \label{sec:adaptive}

We study three attack scenarios where the adversary has the knowledge of \Tech{}.
Our results show that \Tech{} does not have performance degradation in most cases. For those that it does degrade, the adaptive attack is not effective. Details can be found in the supplementary document Section D~\cite{beagle}.

\subsection{Ablation Study} \label{sec:ablation}

This section studies different design choices of \Tech{} in attack sample decomposition and scanner synthesis. The results show that \Tech{} has a robust design. Details can be found in the supplementary document Section E~\cite{beagle}.

\section{Limitation} \label{sec:limitation}

\begin{table}[t]
    \centering
    \footnotesize
    \tabcolsep=8pt
    \caption{Transferability Evaluation of \Tech{}}
    \label{tab:transfer}
    \begin{tabular}{cccccc}
        \toprule
        \textbf{Scanner} & \textbf{TP} & \textbf{FN} & \textbf{TN} & \textbf{FP} & \textbf{ACC} \\
        \midrule
        \textbf{Patch Scanner} & 0 & 20 & 20 & 0 & 0.50 \\
        \textbf{Filter Scanner} & 2 & 18 & 19 & 1 & 0.53 \\
        \textbf{WaNet Scanner} & 18 & 2 & 1 & 19 & \textbf{0.93} \\
        \bottomrule
    \end{tabular}
\end{table}

One limitation of \Tech{} is lack of transferability.
The goal of BEAGLE is to synthesize scanners based on seen attack instances and detect attacks of the same type, with which existing scanners have difficulties. 
BEAGLE is likely ineffective if a new attack is categorically quite different from those it has seen.

We conduct an experiment to evaluate the transferability of \Tech{} in Table~\ref{tab:transfer}, which shows a limitation of \Tech{} mentioned in Section~\ref{sec:limitation}.
We leverage two scanners synthesized based on patch and filter attack samples in Section~\ref{sec:scan}, and evaluate on 20 trojaned models by WaNet mixed with 20 clean models. The patch scanner achieves 50\% accuracy and the filter scanner achieves 53\%, much lower than BEAGLE's WaNet scanner that achieves 93\% accuracy.
Scanning attacks of unseen categories is a hard challenge and we will leave this as the future work.

\section{Related Work}
\smallskip
\noindent
\textbf{Backdoor Attack.}
There are a large number of existing backdoor attacks.
Some attach small patches/watermarks~\cite{badnet, trojnn, dynamic, inputaware, saha2020hidden, tao2022backdoor} as the trigger. Some blend the input image with another image~\cite{blend, sig, reflection, composite}. Others leverage an input transformation function to directly inject the trigger into the input image~\cite{invisible, wanet, dfst}.

\smallskip
\noindent
\textbf{Backdoor Defense.} Backdoor detection aims to determine whether a model is trojaned~\cite{kolouri2020universal,xu2019detecting,qiao2019defending,guo2020towards,huang2019neuroninspect}.
Scanners such as TABOR~\cite{guo2020towards} are based on NC and inherit similar limitations.
Some leverage NC's method to invert a trigger~\cite{karm}, and others use more complex input transformation function~\cite{ABS,liu2021ex}.
We have discussed and enhanced both techniques.
Another type of defenses focuses on detecting poisoned data instead of models~\cite{ma2019nic,tang2019demon,strip,chen2018detecting,li2020rethinking,liu2017neural,chou2020sentinet,tran2018spectral,fu2020detecting,chan2019poison,du2019robust,veldanda2020nnoculation}. 
AC~\cite{chen2018detecting} and STRIP~\cite{strip}.
There are also backdoor elimination~\cite{liu2018fine,borgnia2020strong,zeng2020deepsweep,li2021neural,deck,zhang2022flip} and certified robustness against backdoors~\cite{mccoyd2020minority,xiang2021patchguard,xiang2021patchcleanser,jia2020certified}.

\smallskip
\noindent
\textbf{Traceback of data-poisoning attack.}
Researchers have proposed a traceback technique on data-poisoning attack~\cite{traceback}. It mainly focuses on separating trojaned data from clean data given the whole training dataset. Its goal is different from ours. 

\section{Conclusions}

We propose a novel DL backdoor forensics technique. It can decompose attack samples to clean inputs and triggers. It can automatically synthesize scanners from the forensics results such that other instantiations of the same type of backdoor can be identified (without the trojaned inputs). Our results show that the technique substantially outperforms the state-of-the-art.

\section*{Acknowledgements}
We thank the anonymous reviewers for their constructive comments. This research was supported, in part by IARPA TrojAI W911NF-19-S-0012, NSF 1901242 and 1910300, ONR N000141712045, N000141410468 and N000141712947. Any opinions, findings, and conclusions in this paper are those of the authors only and do not necessarily reflect the views of our sponsors.

\bibliographystyle{IEEEtranS}
\bibliography{reference}

\section{Appendix}

\subsection{Details of Backdoor Attacks and Datasets/Models} \label{sec:append_setup}

BadNets~\cite{badnet} uses a small patch at the corner as an universal trigger.
TrojNN~\cite{trojnn} uses a watermark.
Dynamic~\cite{dynamic} places different patches based on different input images to inject input-specific backdoor.
Reflection~\cite{reflection} leverages an universal reflection image as the trigger.
Blend~\cite{blend} performs a static random perturbation on the input image to inject a backdoor.
SIG~\cite{sig} produces some strip-like effects on input images.
Invisible~\cite{invisible} leverages an encoder-decoder structure to perform a complex input transformation.
WaNet~\cite{wanet} uses complex wrapping functions.
Gotham~\cite{ABS} uses the Gotham instagram filter to transform input images.
DFST~\cite{dfst} leverages a GAN-based generator to inject some visual effects to input images.

\smallskip \noindent
\textbf{ImageNet}~\cite{ILSVRC15} is a popular, large scale object classification dataset with 1,281,167 images of 1,000 classes. The task is to predict the correct class label for an image.
We resize the images to $224\times224$ for evaluation.
Two different networks are utilized for ImageNet: VGG-16~\cite{vgg16} and ResNet-50~\cite{he2016deep}.

\smallskip \noindent
\textbf{CelebA}~\cite{celeba} is a face attributes dataset that contains 10,177 identities with 202,599 face images, each image with 40 attribute annotations.
We resize the images to $128\times128$ for evaluation.
Three networks are used for this dataset: VGG-13~\cite{vgg16}, ResNet-18~\cite{he2016deep} and ResNet-34~\cite{he2016deep}.

\smallskip \noindent
\textbf{CIFAR-10}~\cite{cifar10} is an object recognition dataset with 10 classes. It consists of 60,000 images and is divided into a training set (48,000 images), a validation set (2,000 images), and a test set (10,000 images).
CIFAR-10 images are all $32\times32$ and we don't need to resize them.
We leverage four networks for this dataset: VGG-11~\cite{vgg16}, VGG-13~\cite{vgg16}, ResNet-18~\cite{he2016deep} and ResNet-34~\cite{he2016deep}.

\smallskip \noindent
\textbf{GTSRB}~\cite{gtsrb} is a German traffic sign recognition dataset with 51,840 images of road signs in 43 classes. The set contains images of more than 1700 traffic sign instances. We split the dataset into a training set, validation set, and a test set.
All the images are resized to $32\times32$ before evaluation.
We leverage the same four networks for GTSRB as CIFAR-10.

\smallskip \noindent
\textbf{TrojAI}~\cite{trojai} rounds 2 and 3 consist of 1104 and 1008 pre-trained image classification models, respectively. The models were trained on synthetic images, of size $224\times224$, containing artificial traffic signs and realistic street view background from KITTI dataset~\cite{kitti}, Cityscapes dataset~\cite{cityscapes} and Swedish Roads dataset~\cite{swedish}.
Random transformation, e.g., shifting, rotating, lighting, blurring, weather effects, are applied for data augmentation.
This raises the difficulty in the decomposition process of \Tech{} since the collected trojaned images or validation images can be diverse.
Altogether 22 network are leveraged for training the models including several complex structures with a large number of parameters, e.g., DenseNet121~\cite{densenet}, InceptionV3~\cite{inception}, and MobileNetV2~\cite{mobilenet}.
Half of the models have been poisoned with some backdoor which causes model misclassification. These backdoors include polygon patches and filters.
Some are universal (causing any images with trigger to be misclassified to the target label) and the others are label-specific (only causing images of a victim class to be misclassified).
Compared to round 2, round 3 models leverage adversarial training to suppress natural trojans.

\subsection{Details of Backdoor Scanners}
\label{sec:scanner_appendix}
\noindent \textbf{Tabor}~\cite{Tabor} formulates trojan detection as a non-convex optimization problem, guided by explainable AI and other heuristics to increase detection accuracy. Similar to NC, it only supports universal patch backdoors. 

\noindent \textbf{K-Arm}~\cite{karm} leverages K-Arm bandit originally proposed in Reinforcement Learning to iteratively and stochastically select the most promising labels for trigger inversion.
Its stochastic selection ensures that even if the true target label is not selected for the current round, it still has a good chance to be selected later. It supports both universal and label-specific patch backdoors.

\noindent \textbf{SRI Trinity}~\cite{sri}.
This is a technique from TrojAI that 
can scan both patch and filter backdoors. It has two components, trigger inversion and backdoor classification that classifies a given model to clean or trojaned based on inversion results. We only use its trigger inversion component as the other component is orthogonal.

\subsection{Function Selection During Attack Decomposition} \label{sec:select_func}

\begin{table}[t]
    \centering
    \footnotesize
    \tabcolsep=10pt
    \caption{Function selection during attack decomposition.}
    \label{tab:select_func}
    \begin{tabular}{ccccc}
        \toprule
        \multirow{2}{*}{\textbf{Attack}} & \multicolumn{2}{c}{\textbf{Patching}} & \multicolumn{2}{c}{\textbf{Transforming}} \\
        \cmidrule(lr){2-3} \cmidrule(lr){4-5}
        ~ & Binomial & Uniform & Simple & Complex \\
        \midrule
        BadNets & \textbf{0.99} & 0.87 & 0.42 & 0.82 \\
        Reflection & 0.76 & \textbf{0.96} & 0.71 & 0.74 \\
        Instagram & 0.52 & 0.85 & \textbf{0.98} & 0.68 \\
        WaNet & 0.84 & 0.78 & 0.74 & \textbf{0.97} \\
        \bottomrule
    \end{tabular}
\end{table}

Given a model for forensics, since \Tech{} doesn't know which function to use during decomposition, it first decomposes the attack samples using both functions (patching and transforming) for a few steps and then choose the one with better performance, as discussed in Section~\ref{sec:decomposition}.
We apply two functions and two distributions for patching and transforming parameters, altogether four types to perform attack decomposition on trojaned instances of BadNets, Reflection, Instagram filter, and WaNet.
The results are shown in Table~\ref{tab:select_func}, where the first column denotes the attack and the following columns denote the ASR of decomposed trigger on clean validation images by different functions.
Observe that for each attack, one of the functions stands out with an obviously higher ASR compared with others. For instance, binomial patching achieves 99\% ASR, outperforming uniform patching and transforming by at least 12\%.
This is reasonable and expected because \Tech{}'s functions are summarized based on existing backdoor injection functions, and different functions have little overlapping.
For example, a complex transformation function is designed to approximate the localized pixel warping of WaNet, which is hard to realize leveraging patch functions or simple transformation. Therefore, only the complex transformation function achieves a high ASR of decomposed trigger.

\subsection{Generalization to State-of-the-art Attacks} \label{sec:new_attack}

\begin{table}
    \centering
    \scriptsize
    \tabcolsep=3.5pt
    \caption{Attack decomposition of state-of-the-art attacks}
    \label{tab:new_attack_decomp}
    \begin{tabular}{cggccggcc}
        \toprule
        \multirow{2}{*}{\textbf{Attack}} & \multicolumn{4}{c}{\textbf{Decomposed Clean Images}} & \multicolumn{4}{c}{\textbf{Decomposed Trigger}} \\ \cmidrule(lr){2-5} \cmidrule(lr){6-9}
        ~ & \cellcolor{white}{L1 $\downarrow$} & \cellcolor{white}{PSNR $\uparrow$} & \cellcolor{white}{SSIM $\uparrow$} & \cellcolor{white}{ACC $\uparrow$} & \cellcolor{white}{L1 $\downarrow$} & \cellcolor{white}{PSNR $\uparrow$} & \cellcolor{white}{SSIM $\uparrow$} & \cellcolor{white}{ASR $\uparrow$} \\
        \midrule
        \textbf{Composite} & 0.211 & 17.64 & 0.63 & 1.0 & 0.025 & 26.28 & 0.97 & 0.94 \\
        \textbf{LIRA} & 0.098 & 27.56 & 0.95 & 1.0 & 0.149 & 23.40 & 0.81 & 0.91 \\
        \bottomrule
    \end{tabular}
\end{table}

\begin{table}[t]
    \centering
    \footnotesize
    \tabcolsep=4.5pt
    \caption{Effectiveness of synthesized scanner on state-of-the-art attack}
    \label{tab:new_attack_scan}
    \begin{tabular}{cggcccggccc}
        \toprule
        \multirow{2}{*}{\textbf{Attack}} & \multicolumn{5}{c}{Original} & \multicolumn{5}{c}{\Tech{}} \\ \cmidrule(lr){2-6} \cmidrule(lr){7-11}
        ~ & \cellcolor{white}{TP} & \cellcolor{white}{FN} & \cellcolor{white}{TN} & \cellcolor{white}{FP} & \cellcolor{white}{ACC} & \cellcolor{white}{TP} & \cellcolor{white}{FN} & \cellcolor{white}{TN} & \cellcolor{white}{FP} & \cellcolor{white}{ACC} \\
        \midrule
        Composite & 5 & 15 & 20 & 0 & 0.63 & 17 & 3 & 18 & 2 & \textbf{0.88} \\
        LIRA & 4 & 16 & 20 & 0 & 0.60 & 20 & 0 & 19 & 1 & \textbf{0.98} \\
        \bottomrule
    \end{tabular}
\end{table}

There are some state-of-the-art backdoor attacks that introduce new trigger types, e.g., semantic and hidden backdoors.
Semantic backdoors~\cite{composite} leverage semantic features as secret triggers.
For example, the composite backdoor attack~\cite{composite} uses the co-presence of  natural features as the trigger. For example, the presence of an airplane in a truck image causes the truck to be misclassified as a bird.
Hidden backdoors~\cite{lira, doan2021backdoor, li2019invisible, li2020invisible} inject invisible patterns into input images as backdoor trigger by constrained optimization or using a network.
For example, LIRA~\cite{lira} trains a trigger injection network and the trojaned classifier simultaneously. In addition, the trigger perturbation is constrained to a small range to ensure invisibility.
We conduct experiments following the setup in Section~\ref{sec:scan} and evaluate \Tech{}'s effectiveness against the composite attack and LIRA.
We train 20 trojaned models for each attack and use 20 additional clean models (half VGG-11 and half ResNet18) on CIFAR-10 for the experiments.
We assume \Tech{} has access to 3 additional trojaned models for attack decomposition and utilize ABS as the base scanner during synthesis.
Table~\ref{tab:new_attack_decomp} shows the results of attack decomposition.
Observe that for both attacks, the decomposed clean images resemble the source images and the decomposed trigger is similar to the injected one, with low $\normlone$ error and high PSNR and SSIM scores.
Besides, the decomposed clean images are correctly classified as the ground-truth label, with 100\% accuracy, and the decomposed trigger achieves high ASR on clean validation images (higher  than 90\%).
These results are consistent with the existing attacks shown in Table~\ref{tab:decomposition}.
Moreover, Table~\ref{tab:new_attack_scan} shows the scanning results of the vanilla base scanner and \Tech{}'s synthesized scanner.
Observe that \Tech{}'s synthesized scanner outperforms the original version by 25\% on the composite attack and by 38\% on LIRA.
This delineates that \Tech{} is effective against state-of-the-art backdoor attacks.
Specifically, the composite backdoor falls into the patching category, which is defined as $x \oplus t = x \cdot (1 - m) + t \cdot m$, where $x$ is the source image, $t$ is the trigger pattern, and $m$ is the region that the attacker stamps the trigger.
Although it leverages semantic information as the trigger, these triggers are directly stamped on victim images.
For example, assume the attacker stamps a piece of green clothing on a source image to attack the frog class, \Tech{} can recognize the region $m$ where the attacker stamps the garment and $t$ the garment itself.
In addition, LIRA and some other hidden backdoors~\cite{doan2021backdoor, li2019invisible, li2020invisible} are similar to Invisible and WaNet discussed in Section~\ref{sec:decomposition} which inject hidden/invisible perturbation into images and \Tech{} is effective in approximating their trigger injection algorithm using a transformation function. Hence \Tech{} decomposes triggers from trojaned samples and detects backdoors well.

\clearpage
\newpage

\section{Supplementary Document}

\subsection{More Results on Validating Decomposed Clean Inputs and Triggers}
\label{sec:append_decomp}

\begin{table*}[t]
    \centering
    \scriptsize
    \tabcolsep=3pt
    \caption{Validation on decomposed clean images and triggers
    }
    \label{tab:decomposition}
    \begin{tabular}{cccgggcccgggccgggcccgggcc}
        \toprule
        \multirow{4}{*}{\textbf{DS}} & \multirow{4}{*}{\textbf{Nk}} & \multirow{4}{*}{\textbf{Attack}} & \multicolumn{11}{c}{\textbf{Decomposed Clean Images}} & \multicolumn{11}{c}{\textbf{Decomposed Trigger}} \\ \cmidrule(lr){4-14} \cmidrule(lr){15-25}
        ~ & ~ & ~ & \multicolumn{3}{c}{L1 $\downarrow$} & \multicolumn{3}{c}{PSNR $\uparrow$} & \multicolumn{3}{c}{SSIM $\uparrow$} & \multicolumn{2}{c}{ACC $\uparrow$} & \multicolumn{3}{c}{L1 $\downarrow$} & \multicolumn{3}{c}{PSNR $\uparrow$} & \multicolumn{3}{c}{SSIM $\uparrow$} & \multicolumn{2}{c}{ASR $\uparrow$} \\ \cmidrule(lr){4-14} \cmidrule(lr){15-25}
        ~ & ~ & ~ & $x \oplus t$ & Fb. & {\sc Bg.} & $x \oplus t$ & Fb. & {\sc Bg.} & $x \oplus t$ & Fb. & {\sc Bg.} & Fb. & {\sc Bg.} & $x$ & Fb. & {\sc Bg.} & $x$ & Fb. & {\sc Bg.} & $x$ & Fb. & {\sc Bg.} & Fb. & {\sc Bg.} \\
        \midrule
        \multirow{5}{*}{\rotatebox{90}{ImageNet}} & \multirow{5}{*}{\rotatebox{90}{VGG-16}} & BadNets & 0.037 & 0.128 & \textbf{0.088} & 22.65 & 22.86 & \textbf{26.93} & 0.95 & 0.64 & \textbf{0.75} & 1.0 & \textbf{1.0} & 0.038 & \textbf{0.023} & 0.042 & 22.51 & 24.30 & \textbf{24.37} & 0.95 & \textbf{0.97} & 0.91 & 0.63 & \textbf{1.00} \\
        ~ & ~ & TrojNN & 0.065 & 0.157 & \textbf{0.083} & 19.41 & 21.60 & \textbf{28.69} & 0.82 & 0.54 & \textbf{0.75} & 0.5 & \textbf{1.0} & 0.067 & 0.077 & \textbf{0.044} & 19.17 & 19.15 & \textbf{23.60} & 0.84 & 0.84 & \textbf{0.85} & 0.97 & \textbf{1.00} \\
        ~ & ~ & Invisible & 0.075 & \textbf{0.151} & 0.166 & 29.46 & \textbf{23.26} & 23.04 & 0.89 & 0.56 & \textbf{0.57} & 0.1 & \textbf{1.0} & 0.093 & 0.222 & \textbf{0.140} & 27.43 & 17.48 & \textbf{24.08} & 0.90 & \textbf{0.77} & 0.76 & 0.13 & \textbf{0.95} \\
        ~ & ~ & WaNet & 0.024 & \textbf{0.112} & 0.134 & 38.70 & \textbf{25.36} & 25.01 & 0.98 & 0.67 & \textbf{0.67} & 0.7 & \textbf{1.0} & 0.034 & 0.223 & \textbf{0.217} & 33.18 & 16.83 & \textbf{20.39} & 0.97 & \textbf{0.72} & 0.70 & 0.10 & \textbf{0.91} \\
        ~ & ~ & Gotham & 0.409 & 0.443 & \textbf{0.214} & 15.25 & 14.49 & \textbf{20.70} & 0.77 & 0.50 & \textbf{0.59} & 0.0 & \textbf{1.0} & 0.428 & 0.643 & \textbf{0.200} & 15.05 & 11.62 & \textbf{20.35} & 0.76 & 0.40 & \textbf{0.75} & 0.21 & \textbf{0.99} \\
        \midrule
        \multirow{5}{*}{\rotatebox{90}{CelebA}} & \multirow{5}{*}{\rotatebox{90}{ResNet-18}} & BadNets & 0.047 & 0.090 & \textbf{0.036} & 20.60 & 20.55 & \textbf{32.59} & 0.94 & 0.87 & \textbf{0.94} & 1.0 & \textbf{1.0} & 0.045 & \textbf{0.027} & 0.079 & 20.94 & \textbf{24.07} & 19.88 & 0.94 & \textbf{0.96} & 0.94 & 0.95 & \textbf{1.00} \\
        ~ & ~ & TrojNN & 0.062 & 0.113 & \textbf{0.041} & 20.28 & 21.47 & \textbf{31.06} & 0.81 & 0.75 & \textbf{0.94} & 0.0 & \textbf{1.0} & 0.061 & \textbf{0.060} & 0.078 & 20.37 & 20.77 & \textbf{20.77} & 0.81 & 0.83 & \textbf{0.86} & 0.78 & \textbf{1.00} \\
        ~ & ~ & Reflection & 0.469 & 0.472 & \textbf{0.243} & 13.98 & 13.92 & \textbf{19.83} & 0.43 & 0.40 & \textbf{0.62} & 0.0 & \textbf{1.0} & 0.443 & 0.450 & \textbf{0.177} & 14.71 & 14.51 & \textbf{21.17} & 0.49 & 0.46 & \textbf{0.72} & 0.23 & \textbf{0.98} \\
        ~ & ~ & SIG & 0.380 & 0.384 & \textbf{0.209} & 16.43 & 16.24 & \textbf{21.14} & 0.67 & 0.59 & \textbf{0.73} & 0.0 & \textbf{1.0} & 0.390 & 0.374 & \textbf{0.167} & 16.09 & 16.41 & \textbf{23.25} & 0.64 & 0.56 & \textbf{0.85} & 0.20 & \textbf{0.93} \\
        ~ & ~ & Blend & 0.293 & 0.306 & \textbf{0.254} & 18.77 & 18.48 & \textbf{19.09} & 0.56 & 0.52 & \textbf{0.62} & 0.4 & \textbf{1.0} & 0.248 & 0.268 & \textbf{0.175} & 20.07 & 19.45 & \textbf{20.58} & 0.63 & 0.59 & \textbf{0.73} & 0.44 & \textbf{1.00} \\
        \midrule
        \multirow{11}{*}{\rotatebox{90}{CIFAR-10}} & \multirow{11}{*}{\rotatebox{90}{VGG-11}} & BadNets & 0.040 & 0.082 & \textbf{0.03} & 22.04 & 21.69 & \textbf{31.14} & 0.95 & 0.93 & \textbf{0.97} & 0.8 & \textbf{1.0} & 0.038 & 0.028 & \textbf{0.016} & 22.75 & 25.94 & \textbf{28.87} & 0.95 & \textbf{0.98} & 0.96 & 1.00 & \textbf{1.00} \\
        ~ & ~ & TrojNN & 0.070 & 0.156 & \textbf{0.025} & 18.96 & 19.16 & \textbf{33.41} & 0.75 & 0.76 & \textbf{0.98} & 0.1 & \textbf{1.0} & 0.067 & 0.221 & \textbf{0.008} & 19.37 & 16.53 & \textbf{30.58} & 0.75 & 0.70 & \textbf{0.98} & 0.84 & \textbf{1.00} \\
        ~ & ~ & Dynamic & 0.021 & 0.061 & \textbf{0.054} & 25.83 & \textbf{26.28} & 25.83 & 0.95 & 0.94 & \textbf{0.94} & 1.0 & \textbf{1.0} & 0.022 & 0.150 & \textbf{0.081} & 25.61 & 18.02 & \textbf{20.18} & 0.95 & 0.77 & \textbf{0.81} & 0.99 & \textbf{1.00} \\
        ~ & ~ & Reflection & 0.436 & 0.441 & \textbf{0.212} & 15.07 & 14.96 & \textbf{21.08} & 0.53 & 0.49 & \textbf{0.81} & 0.0 & \textbf{1.0} & 0.446 & 0.430 & \textbf{0.164} & 14.93 & 15.35 & \textbf{22.92} & 0.56 & 0.52 & \textbf{0.89} & 0.17 & \textbf{0.98} \\
        ~ & ~ & SIG & 0.274 & 0.297 & \textbf{0.191} & 19.82 & 18.58 & \textbf{21.89} & 0.59 & 0.51 & \textbf{0.84} & 0.3 & \textbf{1.0} & 0.272 & 0.270 & \textbf{0.077} & 18.57 & 16.48 & \textbf{27.73} & 0.67 & 0.52 & \textbf{0.96} & 0.44 & \textbf{0.94} \\
        ~ & ~ & Blend & 0.183 & 0.220 & \textbf{0.196} & 22.60 & 19.63 & \textbf{21.92} & 0.81 & 0.66 & \textbf{0.84} & 0.2 & \textbf{1.0} & 0.187 & 0.291 & \textbf{0.118} & 22.43 & 17.85 & \textbf{26.08} & 0.82 & 0.53 & \textbf{0.94} & 0.21 & \textbf{1.00} \\
        ~ & ~ & Invisible & 0.061 & \textbf{0.080} & 0.091 & 31.07 & \textbf{28.38} & 28.29 & 0.97 & 0.94 & \textbf{0.96} & 0.1 & \textbf{1.0} & 0.061 & 0.123 & \textbf{0.099} & 30.93 & 21.62 & \textbf{27.12} & 0.96 & 0.82 & \textbf{0.88} & 0.10 & \textbf{0.92} \\
        ~ & ~ & WaNet & 0.057 & \textbf{0.102} & 0.116 & 29.96 & 23.90 & \textbf{26.20} & 0.97 & 0.92 & \textbf{0.95} & 0.0 & \textbf{1.0} & 0.059 & 0.337 & \textbf{0.101} & 29.56 & 15.41 & \textbf{27.68} & 0.96 & 0.50 & \textbf{0.97} & 0.12 & \textbf{0.90} \\
        ~ & ~ & Gotham & 0.457 & 0.503 & \textbf{0.183} & 14.84 & 13.59 & \textbf{22.49} & 0.80 & 0.64 & \textbf{0.91} & 0.8 & \textbf{1.0} & 0.488 & 0.654 & \textbf{0.185} & 14.45 & 11.49 & \textbf{20.96} & 0.78 & 0.34 & \textbf{0.90} & 0.29 & \textbf{0.97} \\
        ~ & ~ & DFST & 0.593 & 0.623 & \textbf{0.454} & 13.48 & 13.01 & \textbf{15.83} & 0.65 & 0.59 & \textbf{0.74} & 0.2 & \textbf{1.0} & 0.632 & 0.780 & \textbf{0.383} & 12.82 & 10.51 & \textbf{16.41} & 0.61 & 0.18 & \textbf{0.68} & 0.16 & \textbf{0.92} \\
        \midrule
        \multirow{11}{*}{\rotatebox{90}{GTSRB}} & \multirow{11}{*}{\rotatebox{90}{ResNet-18}} & BadNets & 0.043 & 0.087 & \textbf{0.044} & 21.88 & \textbf{27.67} & 26.39 & 0.94 & 0.92 & \textbf{0.95} & 1.0 & \textbf{1.0} & 0.041 & 0.237 & \textbf{0.013} & 22.14 & 17.02 & \textbf{29.81} & 0.93 & 0.62 & \textbf{0.98} & 1.00 & \textbf{1.00} \\
        ~ & ~ & TrojNN & 0.072 & 0.127 & \textbf{0.020} & 18.81 & 19.40 & \textbf{37.94} & 0.71 & 0.74 & \textbf{0.99} & 0.6 & \textbf{1.0} & 0.070 & 0.076 & \textbf{0.001} & 18.81 & 20.45 & \textbf{49.01} & 0.69 & 0.82 & \textbf{1.00} & 1.00 & \textbf{1.00} \\
        ~ & ~ & Dynamic & 0.023 & 0.078 & \textbf{0.034} & 25.19 & 27.73 & \textbf{30.21} & 0.93 & 0.89 & \textbf{0.95} & 1.0 & \textbf{1.0} & 0.026 & 0.640 & \textbf{0.065} & 24.63 & 11.64 & \textbf{21.16} & 0.94 & 0.16 & \textbf{0.84} & 1.00 & \textbf{1.00} \\
        ~ & ~ & Reflection & 0.565 & 0.567 & \textbf{0.315} & 12.70 & 12.64 & \textbf{17.60} & 0.32 & 0.27 & \textbf{0.59} & 0.0 & \textbf{1.0} & 0.547 & 0.458 & \textbf{0.277} & 12.88 & 13.64 & \textbf{18.39} & 0.22 & 0.37 & \textbf{0.77} & 0.14 & \textbf{0.97} \\
        ~ & ~ & SIG & 0.275 & 0.306 & \textbf{0.188} & 19.18 & 17.91 & \textbf{22.57} & 0.47 & 0.37 & \textbf{0.72} & 0.7 & \textbf{1.0} & 0.275 & 0.287 & \textbf{0.081} & 19.09 & 18.65 & \textbf{28.36} & 0.37 & 0.59 & \textbf{0.96} & 0.15 & \textbf{0.89} \\
        ~ & ~ & Blend & 0.208 & 0.220 & \textbf{0.213} & 21.47 & 20.76 & \textbf{20.79} & 0.71 & 0.71 & \textbf{0.72} & 0.2 & \textbf{1.0} & 0.206 & 0.308 & \textbf{0.126} & 21.67 & 17.79 & \textbf{25.39} & 0.68 & 0.54 & \textbf{0.90} & 0.16 & \textbf{1.00} \\
        ~ & ~ & Invisible & 0.058 & \textbf{0.123} & 0.152 & 31.70 & 23.78 & \textbf{24.71} & 0.94 & 0.80 & \textbf{0.92} & 0.0 & \textbf{1.0} & 0.055 & \textbf{0.123} & 0.149 & 32.18 & 17.02 & \textbf{24.25} & 0.94 & 0.50 & \textbf{0.92} & 0.43 & \textbf{0.96} \\
        ~ & ~ & WaNet & 0.048 & \textbf{0.088} & 0.094 & 33.11 & 26.10 & \textbf{28.38} & 0.97 & 0.88 & \textbf{0.95} & 0.0 & \textbf{1.0} & 0.042 & 0.152 & \textbf{0.072} & 33.42 & 20.14 & \textbf{30.52} & 0.97 & 0.68 & \textbf{0.96} & 0.42 & \textbf{0.97} \\
        ~ & ~ & Gotham & 0.427 & 0.450 & \textbf{0.222} & 14.68 & 14.18 & \textbf{21.08} & 0.74 & 0.68 & \textbf{0.87} & 0.8 & \textbf{1.0} & 0.446 & 0.645 & \textbf{0.153} & 14.48 & 11.83 & \textbf{23.45} & 0.77 & 0.32 & \textbf{0.81} & 0.22 & \textbf{1.00} \\
        ~ & ~ & DFST & 0.597 & 0.592 & \textbf{0.582} & 13.69 & \textbf{13.80} & 13.88 & 0.62 & 0.59 & \textbf{0.62} & 0.0 & \textbf{1.0} & 0.709 & 0.826 & \textbf{0.788} & 11.58 & 9.99 & \textbf{10.47} & 0.48 & 0.22 & \textbf{0.51} & 0.42 & \textbf{0.97} \\
        \midrule
        \multicolumn{3}{c}{\textbf{Average}} & 0.202 & 0.245 & \textbf{0.155} & 21.29 & 19.97 & \textbf{25.12} & 0.78 & 0.69 & \textbf{0.82} & 0.38 & \textbf{1.0} & 0.207 & 0.316 & \textbf{0.147} & 21.02 & 16.90 & \textbf{23.66} & 0.77 & 0.58 & \textbf{0.83} & 0.45 & \textbf{0.96} \\
        \bottomrule
    \end{tabular}\vspace{-10pt}
\end{table*}

We provide additional results in Table~\ref{tab:decomposition} to validate decomposed clean inputs and triggers in Section~\ref{sec:decomposition}.
The first column denotes the dataset, the second column the network structure, and the third column the backdoor attack type.
There are two large column blocks, presenting the decomposition results of clean inputs and triggers, respectively.
In each block, we show the $\normlone$ distance, PSNR, SSIM, and ACC/ASR. For $\normlone$, PSNR, and SSIM, the first column denotes the difference between trojaned images and their clean counterparts. The second column shows the difference between original clean images and decomposed images by Februus (Fb) and the third column the difference by \Tech{} (Bg).
Each entry presents the average value for the given 10 trojaned images and 100 additional clean test images.
For ACC/ASR, the first column is for Februus and the second for $\Tech{}$. Each entry in the ACC columns denotes the average clean accuracy for the decomposed clean images from the given 10 trojaned samples. As the 100 validation clean images have been used in optimization, we use the images from the test set and stamp the decomposed trigger to calculate the ASR.
The last row in Table~\ref{tab:decomposition} shows the average results. Better results between Februus and \Tech{} are highlighted with the bold font.

\begin{figure}[h]
    \centering
    \begin{minipage}[t]{.3\textwidth}
        \includegraphics[width=0.9\textwidth]{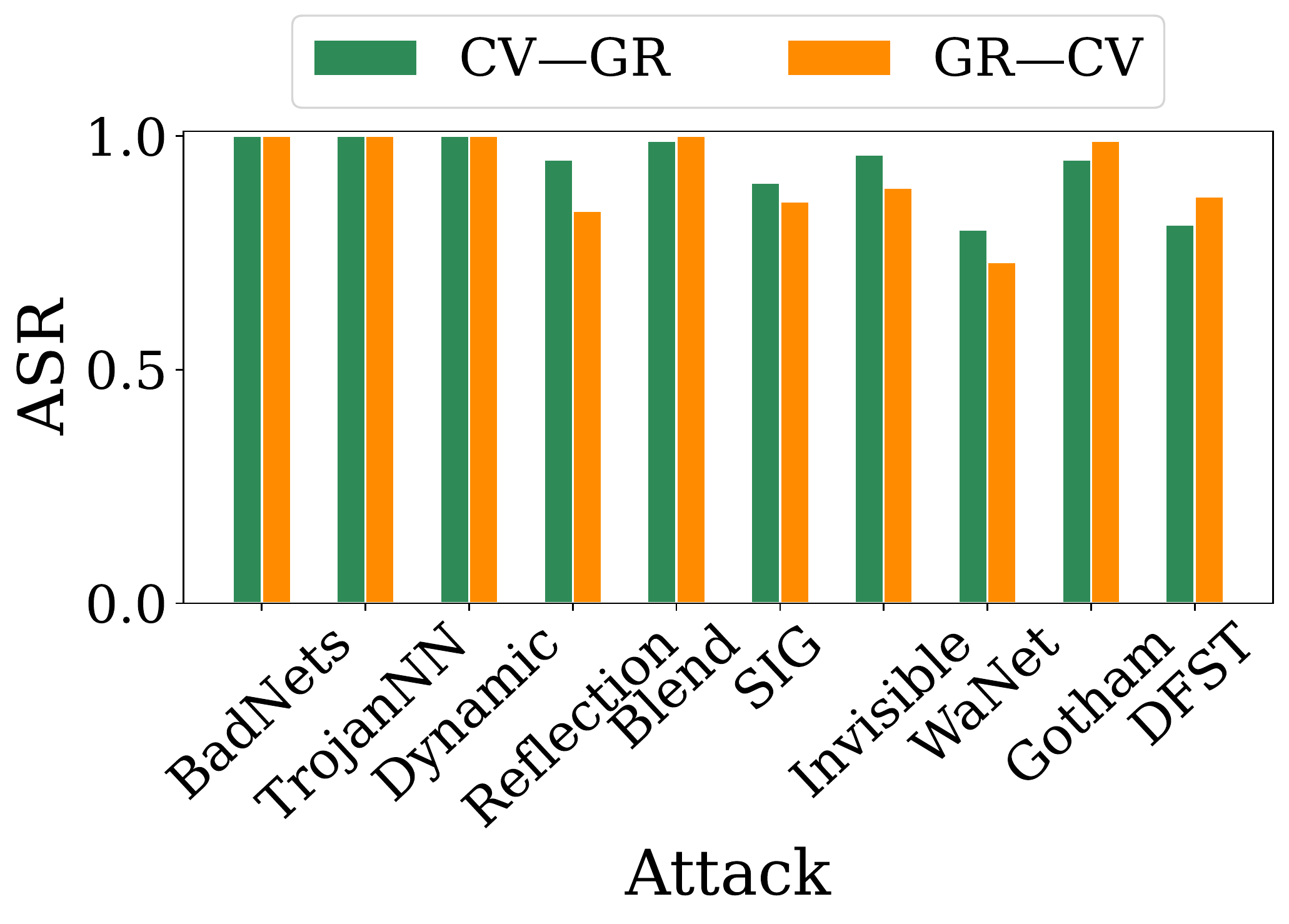}
        \captionsetup{width=0.8\textwidth}
        \caption{Validation on decomposed triggers across datasets/models}
        \label{fig:diff_dataset}
    \end{minipage}\kern-0.2em
    \begin{minipage}[t]{.20\textwidth}
        \includegraphics[width=.9\textwidth]{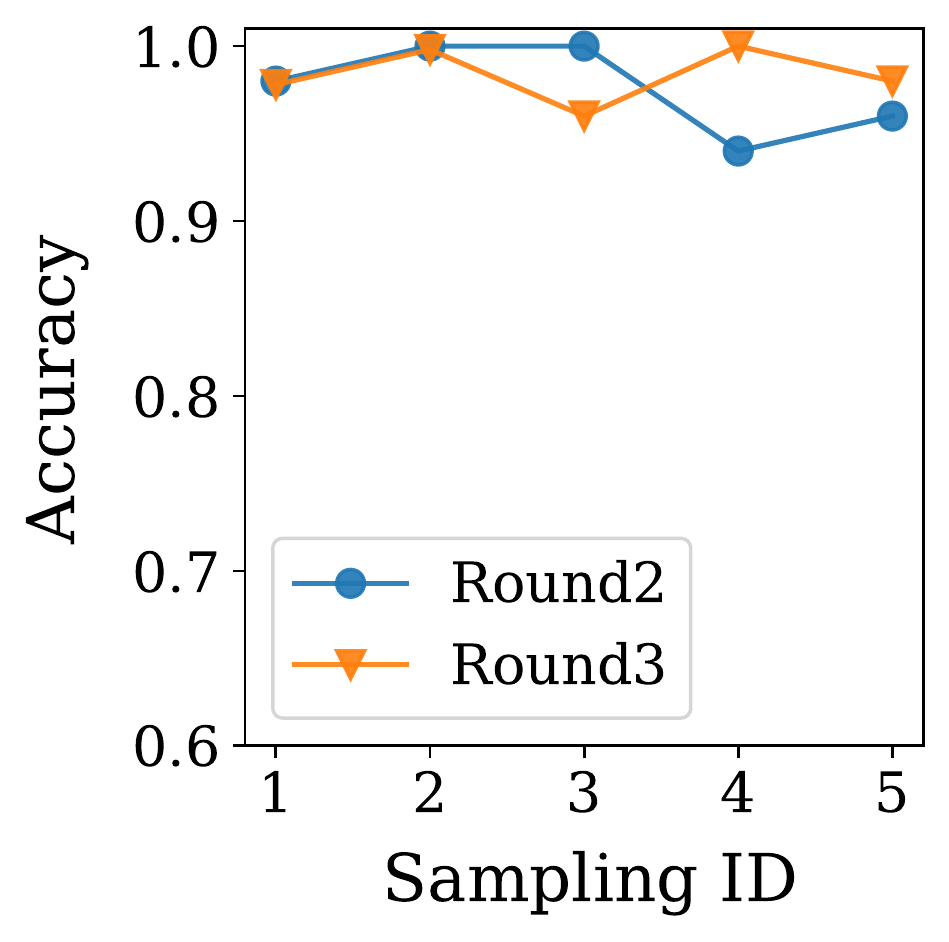}
        \captionsetup{width=0.85\textwidth}
        \caption{Summarization of different attacks in TrojAI}
        \label{fig:auto_trigger_type_trojai}
    \end{minipage}
\end{figure}

\smallskip
\noindent
\textbf{Validating Decomposed Triggers Across Datasets/Models.}
In the previous experiments, the triggers are decomposed and validated on the same dataset and model. \Tech{} can effectively extract the trigger with high visual quality and classification accuracy.
In real-world scenarios, the trojaned model and its dataset that are available for forensics might be different from the subject model/dataset. Here, we study the performance of \Tech{} in decomposing triggers across different datasets and model architectures.
Particularly, we use the same injected trigger to poison a VGG-11 model on CIFAR-10 and a ResNet-18 model on GTSRB. We apply \Tech{} to extract the trigger from the trojaned VGG-11 model and then test it on the ResNet-18 (CV-GR) and vice versa.
Figure~\ref{fig:diff_dataset} shows the results. The x-axis denotes the different backdoor attacks, and the y-axis the ASR of the decomposed trigger on the other model. Observe that almost all the ASRs are high for 10 different attacks, meaning the decomposed triggers are highly effective even when applied on a different dataset/model. This demonstrates the decomposition of \Tech{} is general and does not overfit on specific datasets/models.

\subsection{More Results of Evaluation on Attack Clustering and Summarization}
\label{sec:appendix_summary}

In Section~\ref{sec:summarize}, we evaluate \Tech{}'s attack clustering and summarization performance on CIFAR-10 for seven backdoor attacks.
Here, we evaluate on TrojAI rounds 2 and 3. Specifically, we randomly sample 50 trojaned models (out of 1000 total) that are injected with polygon or (one of the 5 types of)  filter triggers and apply \Tech{} to identify the attack types. We conduct the experiment for five times (trials) and report the results in Figure~\ref{fig:auto_trigger_type_trojai}. The x-axis denotes the trial ID and the y-axis the accuracy of correctly identifying the attack type. Observe that the recognition accuracies are nearly 100\% for both rounds across the five random trials. We make use of four well-known clustering methods: (1) K-means~\cite{kmeans} equipped with Silhouette Score~\cite{silhouette}, abbreviated as K-means-S; (2) K-means~\cite{kmeans} equipped with Elbow method~\cite{elbow}, abbreviated as K-means-E; (3) Gaussian Mixture Model (GMM)~\cite{gmm} equipped with Silhouette Score~\cite{silhouette}, abbreviated as GMM-S; and (4) DBSCAN~\cite{dbscan}.
Since K-means and GMM require to pre-set the number of clusters, we test different numbers of clusters, and leverage the Silhouette Score and Elbow methods to determine the optimal number (with the highest score). DBSCAN can automatically determine the optimal number of clusters. Their clustering results have negligible differences (all having 4-6 clusters). The downstream synthesized scanners have similar performance too (see Table~\ref{tab:diff_cluster}).

\subsection{Impact of Attack Sample Bias} \label{sec:sample_appendix}

As \Tech{} leverages a small set of inputs (clean and trojaned) and trojaned models, we study the impact of sampling biases in such data. For sampled inputs, we include a few naturally misclassified inputs without any injected backdoors. This simulates the real world scenario where classification models do not usually achieve 100\% accuracy. For sampled models, we intentionally introduce biases to the number of trojaned models with different attack types.

\begin{figure}[t]
    \centering
    \includegraphics[height=38mm]{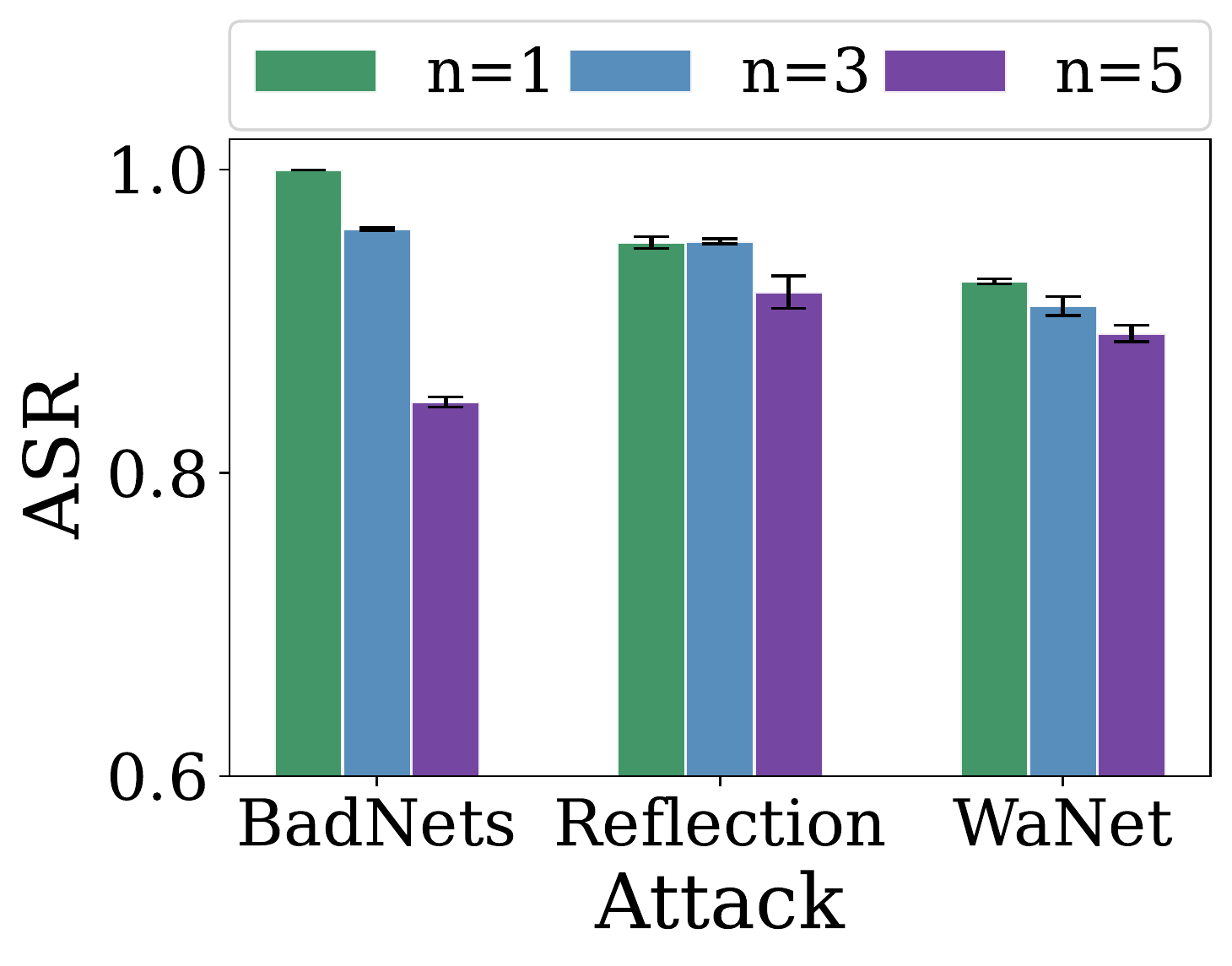}
    \caption{Impact of including naturally miclassified images}
    \label{fig:num_mis}
\end{figure}

\smallskip \noindent \textbf{Attack Samples including Naturally Misclassified Images.}
We mix different numbers of misclassified images with trojaned samples and then measure the ASRs of the decomposed triggers (extracted from these inputs) on the test set. A high ASR means the decomposed trigger is effectively decomposed and hence useful to synthesize scanners. We conduct the experiments on CIFAR-10 with VGG-11 and three attacks: BadNets, Reflection, and WaNet. We randomly select 1, 3 and 5 misclassified images and mix them with 9, 7 and 5 trojaned samples, respectively.
We run the experiment of each setting for five times and show the results in Figure~\ref{fig:num_mis}. The x-axis denotes different attacks and the y-axis the ASR. Each bar shows the ASR of the decomposed trigger and the whisker denotes the variance. Observe that the ASR decreases with the increase of the number of misclassified images, which is expected. But with half of the inputs being naturally misclassified ones (5 out of 10), the ASR of the decomposed trigger is still around 90\%, delineating the robustness of \Tech{} in attack decomposition.
The reason is that  the natural samples unlikely have  consistent misclassification-inducing features and hence they hardly affect the decomposition results.

\begin{figure}[t]
    \centering
    \begin{minipage}[b]{0.23\textwidth}
        \includegraphics[width=1.0\textwidth]{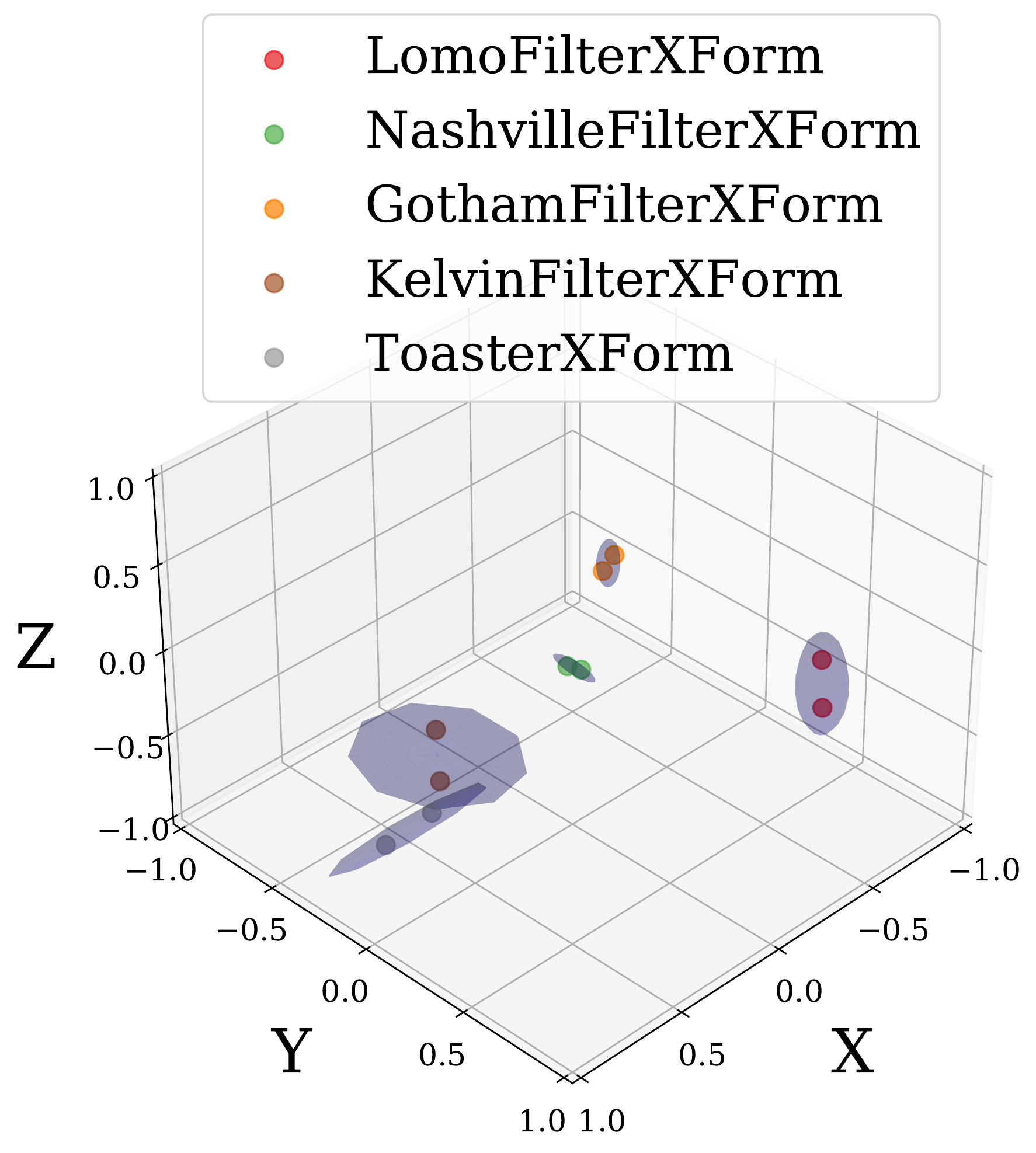}
        \caption*{(a) Uniform}
    \end{minipage}
    \begin{minipage}[b]{0.23\textwidth}
        \includegraphics[width=1.0\textwidth]{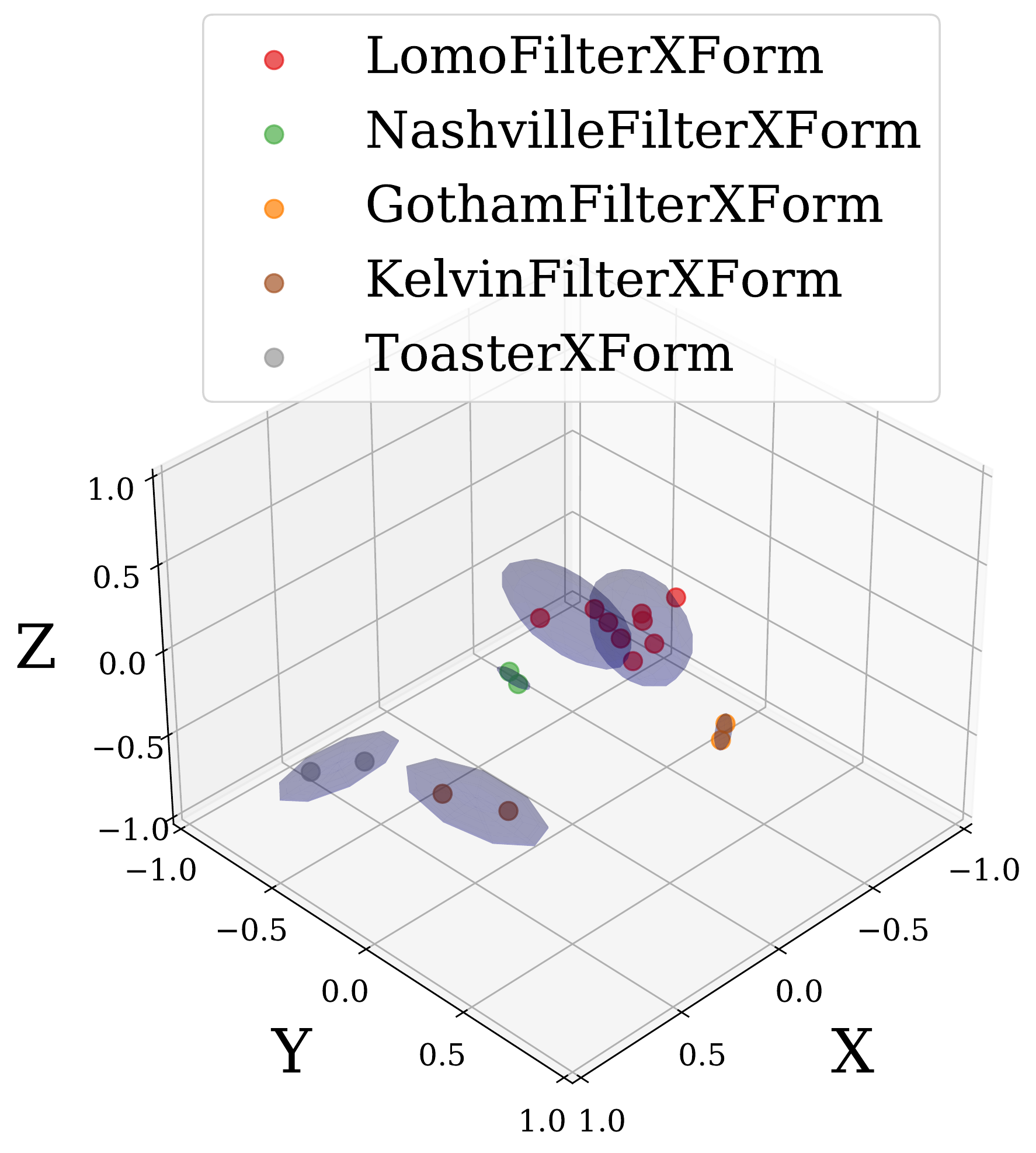}
        \caption*{(b) Highly Biased}
    \end{minipage}
    \begin{minipage}[b]{0.23\textwidth}
        \includegraphics[width=1.0\textwidth]{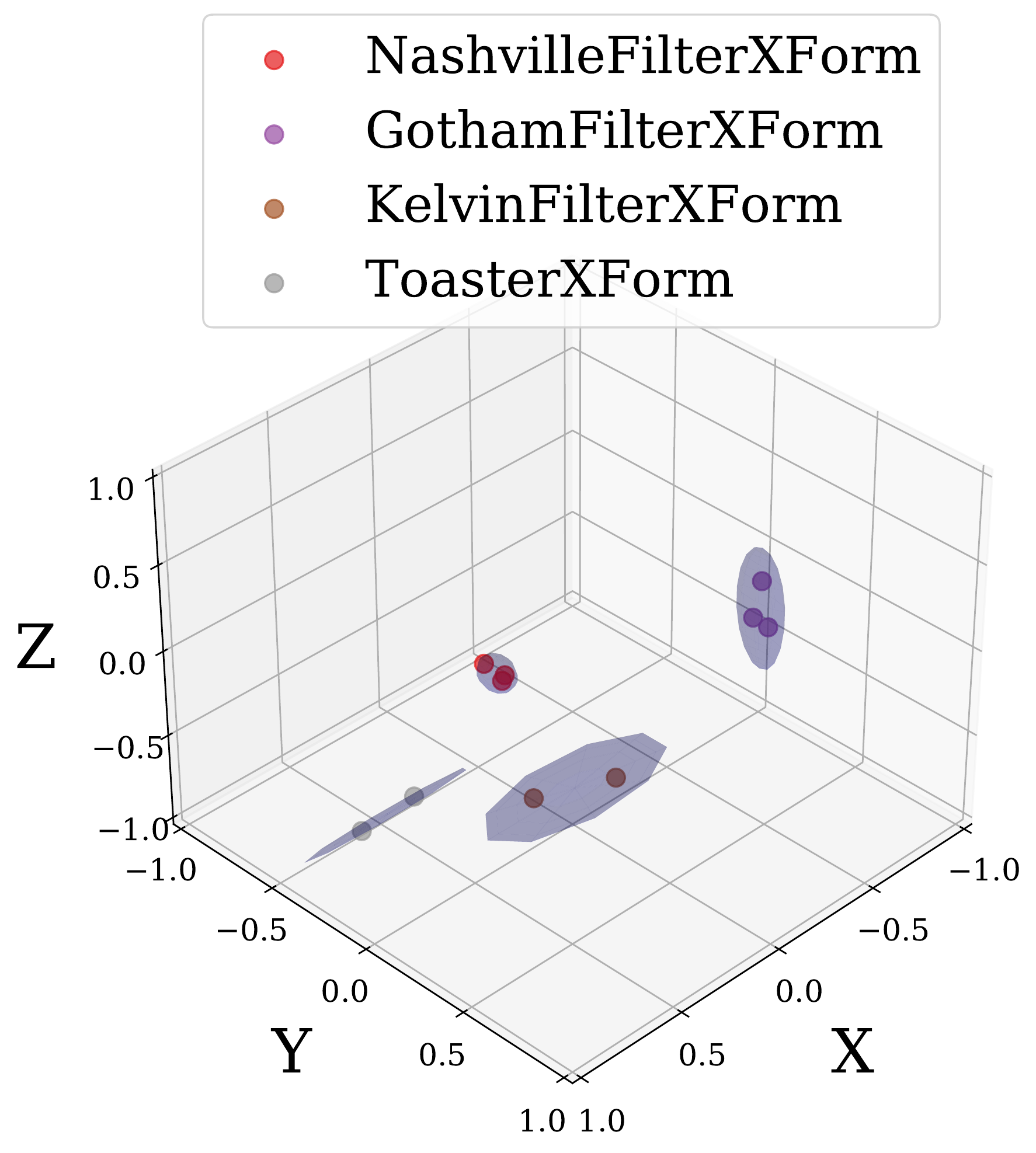}
        \caption*{(c) Missing}
    \end{minipage}
    \caption{Partitioning results under different sampling settings}
    \label{fig:sample_bias_model}
\end{figure}

\smallskip \noindent \textbf{Biases in Attack Types.}
We leverage five filter attacks in TrojAI round 3 and study three different sampling settings:
(1) \textit{uniform} sampling that samples the same number of trojaned models from each attack type;
(2) \textit{highly-biased} sampling that samples a large number of models for some attack type;
(3) \textit{missing} sampling that does not include some attack type. We then apply \Tech{} to partition the sampled trojaned models.
Figure~\ref{fig:sample_bias_model} shows the partitioning results using K-means-S. Each point denotes the reduced attack features for each trojaned model. Observe that in Figure~\ref{fig:sample_bias_model} (b) with the majority of models trojaned by the Lomo filter, \Tech{} can still  accurately partition them. Similar observation can be made in Figure~\ref{fig:sample_bias_model} (c).
We also evaluate the detection accuracy of the synthesized scanners and report the results in Table~\ref{tab:sample_bias_model}. Observe that the scanning accuracies are more than 90\% for the uniform and highly-biased cases. The accuracy for the missing attack setting is slightly lower (87\%) as \Tech{} could not synthesize the scanner for the unseen attack.

\begin{table}[t]
    \centering
    \footnotesize
    \caption{Scanning accuracy under different sampling settings}
    \begin{tabular}{cccc}
        \toprule
         Sampling Setting & Uniform & Biased & Missing \\
         \toprule
         Number of clusters & 5 & 5 & 4 \\
         \midrule
         Accuracy & 0.93 & 0.92 & 0.87 \\
         \bottomrule
    \end{tabular}
    \label{tab:sample_bias_model}
\end{table}

\subsection{Adaptive Attack} \label{sec:adaptive_appendix}
\begin{table}[t]
    \centering
    \scriptsize
    \renewcommand{\arraystretch}{1.2}
    \tabcolsep=1.9pt
    \caption{Results on adaptive attack}
    \label{tab:adaptive}
    \begin{tabular}{cccggccggcc}
        \toprule
        \multirow{2}{*}{\textbf{Backdoor}} & \multirow{2}{*}{\textbf{ACC}} & \multirow{2}{*}{\textbf{ASR}} & \multicolumn{4}{c}{\textbf{Decomposed Clean Images}} & \multicolumn{4}{c}{\textbf{Decomposed Trigger}} \\ \cmidrule(lr){4-7} \cmidrule(lr){8-11}
        ~ & ~ & ~ & \cellcolor{white}{L1$\downarrow$} & \cellcolor{white}{PSNR$\uparrow$} & SSIM$\uparrow$ & ACC$\uparrow$ & \cellcolor{white}{L1$\downarrow$} & \cellcolor{white}{PSNR$\uparrow$} & SSIM$\uparrow$ & ASR$\uparrow$ \\
        \midrule
        \multirow{2}{*}{BadNets} & 0.919 & 1.000 & 0.030 & 31.14 & 0.97 & 1.0 & 0.016 & 28.87 & 0.96 & 1.00 \\
        ~ & 0.911 & 1.000 & 0.029 & 31.51 & 0.98 & 1.0 & 0.018 & 28.19 & 0.96 & 0.99 \\
        \midrule
        \multirow{2}{*}{TrojNN} & 0.917 & 1.000 & 0.025 & 33.41 & 0.98 & 1.0 & 0.008 & 30.58 & 0.98 & 1.00 \\
        ~ & 0.908 & 0.997 & 0.024 & 34.91 & 0.99 & 1.0 & 0.004 & 34.86 & 0.99 & 1.00 \\
        \midrule
        \multirow{2}{*}{Dynamic} & 0.919 & 1.000 & 0.054 & 25.83 & 0.94 & 1.0 & 0.081 & 20.18 & 0.81 & 1.00 \\
        ~ & 0.911 & 1.000 & 0.056 & 26.20 & 0.94 & 1.0 & 0.088 & 19.88 & 0.79 & 1.00 \\
        \midrule
        \multirow{2}{*}{Reflection} & 0.918 & 0.991 & 0.212 & 21.08 & 0.81 & 1.0 & 0.164 & 22.92 & 0.89 & 0.98 \\
        ~ & 0.917 & \textbf{0.850} & 0.213 & 21.03 & 0.81 & 1.0 & 0.169 & 22.53 & 0.88 & \textbf{0.81} \\
        \midrule
        \multirow{2}{*}{SIG} & 0.914 & 0.952 & 0.191 & 21.89 & 0.84 & 1.0 & 0.077 & 27.73 & 0.96 & 0.94 \\
        ~ & 0.908 & \textbf{0.566} & 0.198 & 21.53 & 0.84 & 1.0 & 0.090 & 25.71 & 0.93 & \textbf{0.21} \\
        \midrule
        \multirow{2}{*}{Blend} & 0.920 & 1.000 & 0.196 & 21.92 & 0.84 & 1.0 & 0.118 & 26.08 & 0.94 & 1.00 \\
        ~ & 0.909 & 1.000 & 0.196 & 21.84 & 0.84 & 1.0 & 0.118 & 26.02 & 0.93 & 0.99 \\
        \midrule
        \multirow{2}{*}{Invisible} & 0.918 & 1.000 & 0.091 & 28.29 & 0.96 & 1.0 & 0.099 & 27.12 & 0.88 & 0.92 \\
        ~ & 0.907 & 0.998 & 0.089 & 28.39 & 0.96 & 1.0 & 0.089 & 28.14 & 0.91 & 0.89 \\
        \midrule
        \multirow{2}{*}{WaNet} & 0.908 & 0.989 & 0.116 & 26.20 & 0.95 & 1.0 & 0.101 & 27.68 & 0.97 & 0.90 \\
        ~ & 0.882 & \textbf{0.639} & 0.117 & 26.14 & 0.95 & 1.0 & 0.097 & 27.91 & 0.96 & \textbf{0.61} \\
        \midrule
        \multirow{2}{*}{Gotham} & 0.913 & 1.000 & 0.183 & 22.49 & 0.91 & 1.0 & 0.185 & 20.96 & 0.90 & 0.97 \\
        ~ & 0.911 & 0.991 & 0.182 & 22.53 & 0.91 & 1.0 & 0.188 & 20.70 & 0.89 & 0.89 \\
        \midrule
        \multirow{2}{*}{DFST} & 0.889 & 0.996 & 0.454 & 15.83 & 0.74 & 1.0 & 0.383 & 16.41 & 0.68 & 0.92 \\
        ~ & 0.878 & 0.996 & 0.425 & 16.22 & 0.75 & 1.0 & 0.383 & 16.53 & 0.67 & 0.77 \\
        \bottomrule
    \end{tabular}
\end{table}

\begin{table}[t]
    \centering
    \footnotesize
    \tabcolsep=8pt
    \caption{Impact of mixing targeted adversarial examples.}
    \label{tab:target_adv}
    \begin{tabular}{cccc}
        \toprule
        \textbf{Number} & \textbf{BadNets} & \textbf{Reflection} & \textbf{WaNet} \\
        \midrule
        \textbf{1} & 0.999 & 0.959 & 0.921 \\
        \textbf{3} & 0.978 & 0.939 & 0.865 \\
        \textbf{5} & 0.859 & 0.925 & 0.809 \\
        \bottomrule
    \end{tabular}
\end{table}

We study three attack scenarios where the adversary has the knowledge of \Tech{}.

\smallskip
\noindent
\textbf{Robust Trigger Injection.}
Particularly, the adversary first trains a trojaned model that has high clean accuracy and ASR. He then follows \Tech{} to extract the injected trigger and stamps the decomposed trigger on clean samples with their original labels. Poisoned in such a way, the trojaned model is not sensitive to the decomposed trigger and has a more robust injected backdoor.
We apply \Tech{} on those trojaned models and present the results in Table~\ref{tab:adaptive}. Column 1 denotes the attacks. Columns 2-3 show the clean accuracy (ACC) and attack success rate (ASR). The remaining columns show the results on visual quality and classification accuracy as introduced in Section~\ref{sec:visual}. For each attack in the table, we report the results on the original trojaned models and on the models by the adaptive attack in the first and the second rows, respectively.
Observe that for most attacks, the adaptive attack induces slight degradation on clean accuracy and ASR. 
For those cases, \Tech{} can still decompose high quality clean images and triggers. 
In contrast for Reflection, SIG, and WaNet, the decomposed triggers' ASRs have nontrivial degradation,
indicating that \Tech{} becomes less effective. However, observe that the ASRs of the injected triggers also degrade a lot, suggesting the adaptive attack is ineffective either.
The reason is that the decomposed triggers are so similar to the injected ones that they cancel each others out.

\smallskip
\noindent
\textbf{Mixing Targeted Adversarial Examples.}
The adversary in this scenario mixes targeted adversarial examples (i.e., adversarial examples misclassified to the target class)
with the collected trojaned instances to affect the decomposition process of \Tech{}.
There are two cases.
First given a clean model, the adversary provides 10 targeted adversarial examples. In this case, we evaluate whether \Tech{} can still correctly classify clean models.
We generate a set of targeted adversarial samples using PGD with a reasonable bound $L_{\infty}=16/256$ and feed them to \Tech{}. \Tech{} can only decompose a trigger that achieves 69.7\% ASR on the clean validation samples.
We use the synthesized scanner to scan 10 clean models, and the scanning accuracy is 100\%, which means no model is considered trojaned.
The reason is that targeted adversarial attack, not like backdoor attack, does not have a universal secret trigger.
Therefore, \Tech{} cannot summarize a trigger with a high ASR, and hence the synthesized scanner will not classify clean models as trojaned.
In the second case, the adversary mixes targeted adversarial examples with real trojaned samples to affect \Tech{}'s decomposition process, which is similarly to that of mixing naturally misclassified samples~\ref{sec:sample_appendix}.
We conduct experiments on CIFAR-10 with VGG-11 and three attacks, BadNets, Reflection, and WaNet. We randomly mix 1, 3, and 5 targeted adversarial examples with 9, 7, and 5 real trojaned samples respectively.
Table~\ref{tab:target_adv} shows the results where the numbers in the table denote the ASR of decomposed trigger.
Observe that the ASR slightly decreases with the increase of the
number of targeted adversarial examples as expected.
However, even when half of the given samples are adversarial examples, \Tech{} can still synthesize effective triggers.

\smallskip
\noindent
\textbf{Injecting Multiple Trojans.}
The adversary may inject multiple backdoors into a subject model to affect the decomposition process of \Tech{}.
Although this scenario is beyond our threat model~\ref{sec:intro}, where we assume all the trojaned inputs in an attack instance used in forensics are exploiting the same backdoor, we conduct the experiment to evaluate \Tech{} in this extreme scenario.
We suppose the subject model is injected with three backdoors, BadNets, Instagram filter, and WaNet.
There are two cases for this multiple trojan scenario.
In the first case, the three backdoors aim to attack three different classes. Then it is not much different from the model having only one backdoor, because it is easy to distinguish the trojaned samples of the three backdoors by checking their output labels.
In the second case, the three backdoors aim to attack the same class. The problem is hence reduced to whether \Tech{} can decompose a trigger from a set of mixed trojaned samples.
We have trained 10 trojaned models with 3 backdoors and 10 clean VGG-11 models on CIFAR-10.
We assume \Tech{} has access to 3 additional trojaned models for decomposition and scanner synthesis using ABS.
For each trojaned model, we use 10 trojaned samples with 4 BadNets, 3 Instagram filter and 3 WaNet. We leverage the patching function to decompose a trigger from these trojaned images. Note that the decomposed patch trigger achieves an average ASR of 91.7\%, outperforming the decomposed transforming trigger with an ASR of 87.3\%.
The result shows that \Tech{}'s synthesized scanner detects the backdoor well with 100\% accuracy and no FP or FN.
Although injecting multiple backdoors affects the decomposition process, it causes the trigger features to be more distinguishable and thus easy to detect by the scanner.

\begin{table}[t]
    \centering
    \scriptsize
    \renewcommand{\arraystretch}{1.2}
    \tabcolsep=2.1pt
    \caption{Ablation study of different number of given samples}
    \label{tab:ablation_number}
    \begin{tabular}{cccggccggcc}
        \toprule
        \multirow{2}{*}{\textbf{Backdoor}} & \multirow{2}{*}{\textbf{N\_P}} & \multirow{2}{*}{\textbf{N\_C}} & \multicolumn{4}{c}{\textbf{Decomposed Clean Images}} & \multicolumn{4}{c}{\textbf{Decomposed Trigger}} \\ \cmidrule(lr){4-7} \cmidrule(lr){8-11}
        ~ & ~ & ~ & \cellcolor{white}{L1$\uparrow$} & \cellcolor{white}{PSNR$\uparrow$} & SSIM$\uparrow$ & ACC$\uparrow$ & \cellcolor{white}{L1$\uparrow$} & \cellcolor{white}{PSNR$\uparrow$} & SSIM$\uparrow$ & ASR$\uparrow$ \\
        \midrule
        \multirow{6}{*}{BadNets} & 2 & 100 & 0.052 & 24.11 & 0.95 & 1.0 & 0.032 & 24.89 & 0.90 & 0.99 \\
        ~ & 5  & 100 & 0.040 & 27.91 & 0.97 & 1.0 & 0.020 & 27.64 & 0.92 & 1.00 \\
        ~ & 10 & 100 & 0.032 & 31.74 & 0.97 & 1.0 & 0.018 & 28.17 & 0.94 & 1.00 \\
        ~ & 10 & 50  & 0.050 & 26.28 & 0.95 & 1.0 & 0.043 & 23.96 & 0.86 & 1.00 \\
        ~ & 10 & 20  & 0.061 & 27.71 & 0.95 & 1.0 & 0.116 & 20.15 & 0.75 & 1.00 \\
        ~ & 10 & 10  & 0.087 & 25.15 & 0.92 & 1.0 & 0.248 & 16.55 & 0.60 & 1.00 \\
        \midrule
        \multirow{6}{*}{Dynamic} & 2 & 100 & 0.057 & 26.29 & 0.94 & 1.0 & 0.113 & 18.54 & 0.77 & 1.00 \\
        ~ & 5  & 100 & 0.060 & 25.06 & 0.94 & 1.0 & 0.104 & 18.39 & 0.79 & 1.00 \\
        ~ & 10 & 100 & 0.058 & 26.70 & 0.93 & 1.0 & 0.088 & 20.03 & 0.78 & 1.00 \\
        ~ & 10 & 50  & 0.066 & 24.97 & 0.92 & 1.0 & 0.103 & 19.34 & 0.74 & 1.00 \\
        ~ & 10 & 20  & 0.090 & 24.08 & 0.90 & 1.0 & 0.190 & 17.49 & 0.67 & 1.00 \\
        ~ & 10 & 10  & 0.107 & 23.06 & 0.88 & 1.0 & 0.297 & 15.63 & 0.55 & 1.00 \\
        \midrule
        \multirow{6}{*}{Reflection} & 2 & 100 & 0.459 & 14.55 & 0.58 & 1.0 & 0.177 & 22.37 & 0.84 & 0.97 \\
        ~ & 5  & 100 & 0.342 & 17.48 & 0.77 & 1.0 & 0.199 & 22.00 & 0.87 & 0.96 \\
        ~ & 10 & 100 & 0.217 & 20.87 & 0.80 & 1.0 & 0.170 & 22.35 & 0.85 & 0.98 \\
        ~ & 10 & 50  & 0.225 & 20.60 & 0.80 & 1.0 & 0.217 & 20.63 & 0.78 & 0.94 \\
        ~ & 10 & 20  & 0.243 & 19.90 & 0.76 & 1.0 & 0.268 & 18.64 & 0.69 & 0.90 \\
        ~ & 10 & 10  & 0.255 & 19.44 & 0.74 & 1.0 & 0.287 & 18.61 & 0.69 & 0.85 \\
        \midrule
        \multirow{6}{*}{WaNet} & 2 & 100 & 0.109 & 26.25 & 0.96 & 1.0 & 0.090 & 28.10 & 0.95 & 0.94 \\
        ~ & 5  & 100 & 0.119 & 25.83 & 0.94 & 1.0 & 0.113 & 26.69 & 0.94 & 0.91 \\
        ~ & 10 & 100 & 0.118 & 25.98 & 0.95 & 1.0 & 0.108 & 27.17 & 0.96 & 0.90 \\
        ~ & 10 & 50  & 0.120 & 25.84 & 0.94 & 1.0 & 0.106 & 27.27 & 0.96 & 0.78 \\
        ~ & 10 & 20  & 0.112 & 25.99 & 0.94 & 1.0 & 0.089 & 28.36 & 0.96 & 0.50 \\
        ~ & 10 & 10  & 0.131 & 24.94 & 0.93 & 1.0 & 0.111 & 26.75 & 0.96 & 0.28 \\
        \midrule
        \multirow{6}{*}{Gotham} & 2 & 100 & 0.206 & 21.62 & 0.93 & 1.0 & 0.221 & 19.69 & 0.88 & 0.98 \\
        ~ & 5  & 100 & 0.187 & 22.27 & 0.91 & 1.0 & 0.199 & 20.10  & 0.89 & 0.99 \\
        ~ & 10 & 100 & 0.184 & 22.38 & 0.91 & 1.0 & 0.186 & 20.90 & 0.89 & 0.98 \\
        ~ & 10 & 50  & 0.189 & 22.15 & 0.90 & 1.0 & 0.196 & 20.37 & 0.90 & 0.96 \\
        ~ & 10 & 20  & 0.214 & 21.18 & 0.88 & 1.0 & 0.222 & 19.50 & 0.88 & 0.96 \\
        ~ & 10 & 10  & 0.222 & 20.68 & 0.87 & 1.0 & 0.207 & 20.14 & 0.88 & 0.95 \\
        \bottomrule
    \end{tabular}
\end{table}

\begin{table}[t]
    \centering
    \footnotesize
    \tabcolsep=5pt
    \caption{Scanning accuracy using different clustering methods}
    \begin{tabular}{ccccc}
        \toprule
         Method & K-means-S & K-means-E & GMM-S & DBSCAN \\
         \toprule
         Number of Clusters & 5 & 6 & 4 & 4 \\
         \midrule
         Accuracy & 0.94 & 0.93 & 0.94 & 0.94 \\
         \bottomrule
    \end{tabular}
    \label{tab:diff_cluster}
\end{table}

\subsection{Ablation Study} \label{sec:ablation_study}

\begin{table*}[t]
    \centering
    \scriptsize
    \renewcommand{\arraystretch}{1.2}
    \tabcolsep=3pt
    \caption{Ablation study on normalization and trigger inversion}
    \label{tab:ablation}
    \begin{tabular}{ccggccggccggccggcc}
        \toprule
        \multirow{4}{*}{\textbf{Technique}} & \multirow{4}{*}{\textbf{Backdoor}} & \multicolumn{8}{c}{\textbf{Decomposed Clean Images}} & \multicolumn{8}{c}{\textbf{Decomposed Trigger}} \\ \cmidrule(lr){3-10} \cmidrule(lr){11-18}
        ~ & ~ & \multicolumn{2}{c}{L1 $\downarrow$} & \multicolumn{2}{c}{PSNR $\uparrow$} & \multicolumn{2}{c}{SSIM $\uparrow$} & \multicolumn{2}{c}{ACC $\uparrow$} & \multicolumn{2}{c}{L1 $\downarrow$} & \multicolumn{2}{c}{PSNR $\uparrow$} & \multicolumn{2}{c}{SSIM $\uparrow$} & \multicolumn{2}{c}{ASR $\uparrow$} \\ \cmidrule(lr){3-10} \cmidrule(lr){11-18}
        ~ & ~ & Baseline & {\sc Bg.} & Baseline & {\sc Bg.} & Baseline & {\sc Bg.} & Baseline & {\sc Bg.} & Baseline & {\sc Bg.} & Baseline & {\sc Bg.} & Baseline & {\sc Bg.} & Baseline & {\sc Bg.} \\
        \midrule
        \multirow{7}{*}{Normalization} & Reflection & 0.662 & \textbf{0.212} & 11.97 & \textbf{21.08} & 0.62 & \textbf{0.81} & 1.0 & \textbf{1.0} & 0.425 & \textbf{0.164} & 15.31 & \textbf{22.92} & 0.70 & \textbf{0.89} & 0.92 & \textbf{0.98} \\
        ~ & SIG & \textbf{0.142} & 0.191 & \textbf{24.25} & 21.89 & \textbf{0.86} & 0.84 & 1.0 & \textbf{1.0} & \textbf{0.063} & 0.077 & \textbf{29.62} & 27.73 & 0.96 & \textbf{0.96} & 0.86 & \textbf{0.94} \\
        ~ & Blend & 0.304 & \textbf{0.196} & 18.47 & \textbf{21.92} & 0.81 & \textbf{0.84} & 1.0 & \textbf{1.0} & 0.214 & \textbf{0.118} & 21.61 & \textbf{26.08} & 0.88 & \textbf{0.94} & 0.99 & \textbf{1.00} \\
        & Invisible & \textbf{0.069} & 0.091 & \textbf{30.32} & 28.29 & 0.96 & \textbf{0.96} & 1.0 & \textbf{1.0} & \textbf{0.071} & 0.099 & \textbf{29.39} & 27.12 & \textbf{0.93} & 0.88 & 0.74 & \textbf{0.92} \\
        ~ & WaNet & \textbf{0.067} & 0.116 & \textbf{29.25} & 26.20 & \textbf{0.96} & 0.95 & 1.0 & \textbf{1.0} & \textbf{0.054} & 0.101 & \textbf{31.49} & 27.68 & 0.97 & \textbf{0.97} & 0.89 & \textbf{0.90} \\
        ~ & Gotham & 0.414 & \textbf{0.183} & 15.68 & \textbf{22.49} & 0.81 & \textbf{0.91} & 1.0 & \textbf{1.0} & 0.432 & \textbf{0.185} & 15.05 & \textbf{20.96} & 0.75 & \textbf{0.90} & 0.82 & \textbf{0.97} \\
        ~ & DFST & 0.593 & \textbf{0.454} & 13.52 & \textbf{15.83} & 0.65 & \textbf{0.74} & 1.0 & \textbf{1.0} & 0.734 & \textbf{0.383} & 11.94 & \textbf{16.41} & 0.61 & \textbf{0.68} & 0.92 & \textbf{0.92} \\
        \midrule
        
        \midrule
        \multirow{3}{*}{NC} & BadNets & 0.040 & \textbf{0.030} & 28.33 & \textbf{31.14} & 0.97 & \textbf{0.97} & 1.0 & \textbf{1.0} & 0.082 & \textbf{0.016} & 17.84 & \textbf{28.87} & 0.78 & \textbf{0.96} & 1.00 & \textbf{1.00} \\
        ~ & TrojNN & 0.031 & \textbf{0.025} & 29.84 & \textbf{33.41} & 0.97 & \textbf{0.98} & 1.0 & \textbf{1.0} & 0.109 & \textbf{0.008} & 15.93 & \textbf{30.58} & 0.58 & \textbf{0.98} & 1.00 & \textbf{1.00} \\
        ~ & Dynamic & 0.055 & \textbf{0.054} & \textbf{27.29} & 25.83 & 0.94 & \textbf{0.94} & 1.0 & \textbf{1.0} & 0.105 & \textbf{0.081} & 18.15 & \textbf{20.18} & 0.73 & \textbf{0.81} & 1.00 & \textbf{1.00} \\
        
        \midrule
        \multirow{4}{*}{ABS-filter} & Invisible & 0.104 & \textbf{0.091} & 27.00 & \textbf{28.29} & 0.95 & \textbf{0.96} & 1.0 & \textbf{1.0} & 0.366 & \textbf{0.099} & 15.41 & \textbf{27.12} & 0.55 & \textbf{0.88} & 0.41 & \textbf{0.92} \\
        ~ & WaNet & 0.122 & \textbf{0.116} & 25.74 & \textbf{26.20} & 0.95 & \textbf{0.95} & 1.0 & \textbf{1.0} & 0.207 & \textbf{0.101} & 21.83 & \textbf{27.68} & 0.84 & \textbf{0.97} & 0.21 & \textbf{0.90} \\
        ~ & Gotham & \textbf{0.177} & 0.183 & \textbf{22.69} & 22.49 & 0.91 & \textbf{0.91} & 1.0 & \textbf{1.0} & 0.195 & \textbf{0.185} & 20.86 & \textbf{20.96} & 0.89 & \textbf{0.90} & 0.97 & \textbf{0.97} \\
        ~ & DFST & \textbf{0.412} & 0.454 & \textbf{16.47} & 15.83 & \textbf{0.75} & 0.74 & 1.0 & \textbf{1.0} & 0.549 & \textbf{0.383} & 13.72 & \textbf{16.41} & 0.59 & \textbf{0.68} & 0.68 & \textbf{0.92} \\

        \bottomrule
    \end{tabular}
\end{table*}


\smallskip \noindent \textbf{Effect of Normalization in Attack Decomposition.}
Normalization helps ensure the values of decomposed clean images are within the distribution of clean validation images as defined in Eq.~\ref{equ:normalization}.
We conduct the experiments on CIFAR-10 with VGG-11. Similar to Table~\ref{tab:decomposition}, we report the visual quality using $\normlone$ distance, PSNR, SSIM, and classification accuracy using ACC, ASR for decomposed clean images and triggers. The top half of Table~\ref{tab:ablation} shows the results. For each evaluation metric, we show the results without the normalization on the left and the results with normalization on the right. Observe that for most cases, using normalization can effectively improve the performance of the decomposition.

\smallskip \noindent \textbf{Comparison with Trigger Inversion Methods.}
In this study, we compare the quality of the decomposed triggers by \Tech{} with those directly inverted from trojaned models using existing trigger inversion techniques.
The comparison is not to claim our results are better as \Tech{} leverages attack samples, but rather to provide a reference.
Specifically, we compare with NC on patching attacks and ABS-filter on transforming attacks.
The bottom half of Table~\ref{tab:ablation} presents the results. Observe that \Tech{} outperforms NC on visual quality for patching backdoors and ABS-filter for transforming backdoors, especially for complex triggers such as Invisible, WaNet, and DFST. The decomposed triggers by \Tech{} can achieve >90\% ASR, while those by ABS-filter can only achieve 20-70\% ASR. This demonstrates that \Tech{} can effectively approximate the trigger injection function comparing to existing works.

\begin{figure}[t]
    \centering
    \includegraphics[width=0.48\textwidth]{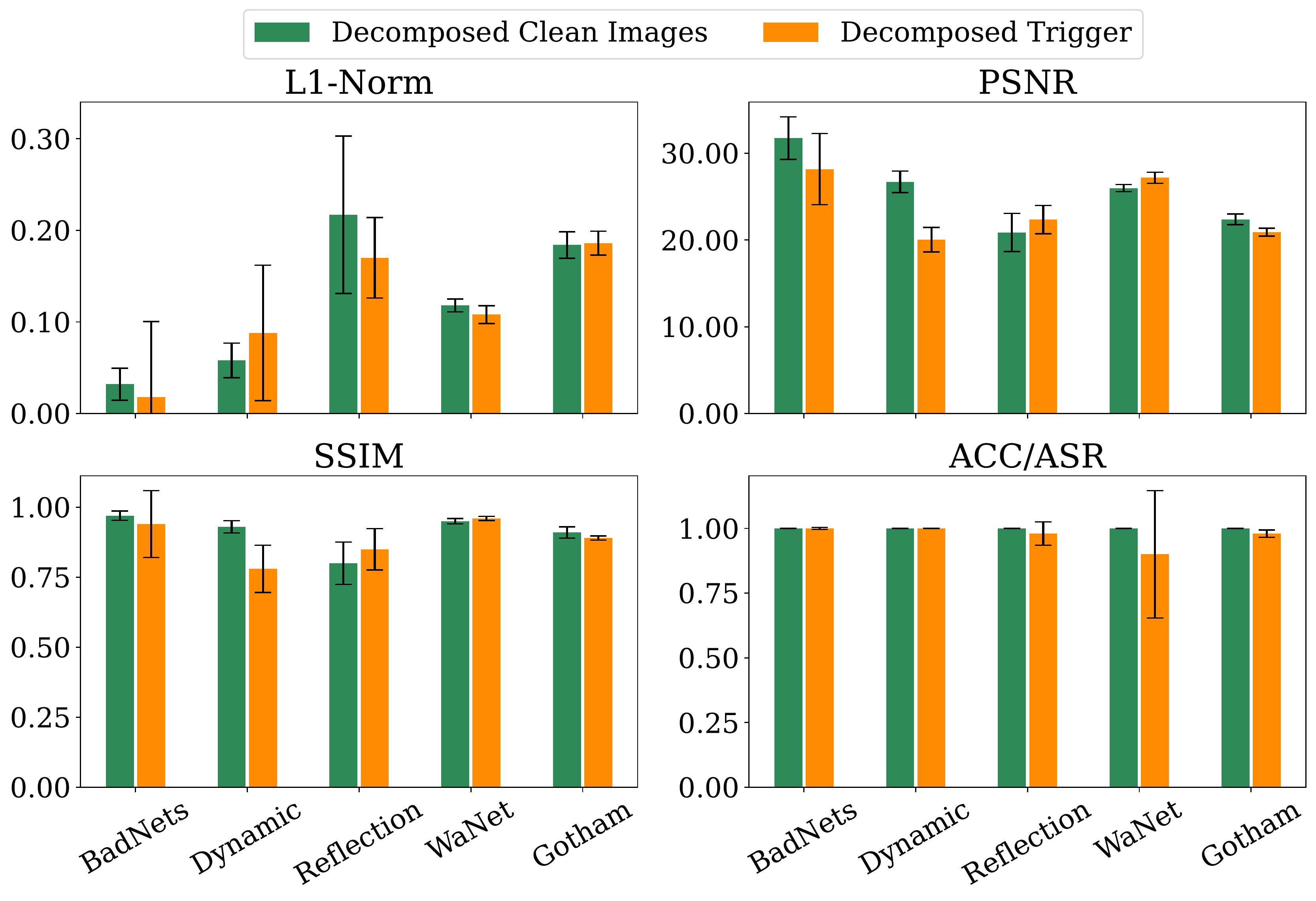}
    \caption{Effect of different numbers of available inputs}
    \label{fig:ablation_number}
\end{figure}

\smallskip
\noindent
\textbf{Effect of Different Numbers of Available Inputs.}
We study the effect of using different numbers of trojaned and clean samples. We evaluate on using 2, 5, 10 trojaned images, and 10, 20, 50, 100 clean images. The other settings are kept the same and five attacks are used in the experiments.
Figure~\ref{fig:ablation_number} shows the results (details can be found in Table~\ref{tab:ablation_number}). The x-axis in each subfigure denotes different attacks and the y-axis the value of the corresponding evaluation metric. The green bars are for the decomposed clean images and the orange bars for the decomposed triggers.
The bars present the results for the original setting (10 trojaned images + 100 clean images), and the whiskers denote the results of using other settings. Observe that the variance is normally small in most cases, except for Reflection and WaNet.
As \Tech{} leverages normalization for attack decomposition, the number of images may affect the approximation of trojaned and clean distributions and hence the final decomposition. The experiments show the number of available inputs has minor impact on \Tech{}.

\begin{table}[t]
    \centering
    \footnotesize
    \tabcolsep=8pt
    \caption{Ablation study of different choices of distribution percentile}
    \label{tab:percentile}
    \begin{tabular}{cccccc}
        \toprule
        \textbf{Percentile} & \textbf{TP} & \textbf{FN} & \textbf{TN} & \textbf{FP} & \textbf{ACC} \\
        \midrule
        \textbf{2\%-98\%} & 95 & 5 & 90 & 10 & 0.93 \\
        \textbf{15\%-85\%} & 91 & 9 & 96 & 4 & 0.94 \\
        \textbf{30\%-70\%} & 91 & 9 & 96 & 4 & 0.94 \\
        \bottomrule
    \end{tabular}
\end{table}

\begin{table}
    \centering
    \scriptsize
    \tabcolsep=3.5pt
    \caption{Ablation study of using different pre-trained GANs}
    \label{tab:gan}
    \begin{tabular}{cggccggcc}
        \toprule
        \multirow{2}{*}{\textbf{Dataset}} & \multicolumn{4}{c}{\textbf{Decomposed Clean Images}} & \multicolumn{4}{c}{\textbf{Decomposed Trigger}} \\ \cmidrule(lr){2-5} \cmidrule(lr){6-9}
        ~ & \cellcolor{white}{L1 $\downarrow$} & \cellcolor{white}{PSNR $\uparrow$} & \cellcolor{white}{SSIM $\uparrow$} & \cellcolor{white}{ACC $\uparrow$} & \cellcolor{white}{L1 $\downarrow$} & \cellcolor{white}{PSNR $\uparrow$} & \cellcolor{white}{SSIM $\uparrow$} & \cellcolor{white}{ASR $\uparrow$} \\
        \midrule
        \textbf{CIFAR-10} & 0.212 & 21.08 & 0.81 & 1.0 & 0.164 & 22.92 & 0.89 & 0.98 \\
        \textbf{GTSRB} & 0.321 & 17.84 & 0.58 & 1.0 & 0.294 & 18.13 & 0.76 & 0.96 \\
        \textbf{CelebA} & 0.373 & 15.66 & 0.42 & 1.0 & 0.263 & 16.64 & 0.63 & 0.97 \\
        \bottomrule
    \end{tabular}
\end{table}

\begin{table}
    \centering
    \scriptsize
    \tabcolsep=3.5pt
    \caption{Ablation study of different choices of $\alpha$}
    \label{tab:alpha}
    \begin{tabular}{cggccggcc}
        \toprule
        \multirow{2}{*}{\textbf{$\alpha$}} & \multicolumn{4}{c}{\textbf{Decomposed Clean Images}} & \multicolumn{4}{c}{\textbf{Decomposed Trigger}} \\ \cmidrule(lr){2-5} \cmidrule(lr){6-9}
        ~ & \cellcolor{white}{L1 $\downarrow$} & \cellcolor{white}{PSNR $\uparrow$} & \cellcolor{white}{SSIM $\uparrow$} & \cellcolor{white}{ACC $\uparrow$} & \cellcolor{white}{L1 $\downarrow$} & \cellcolor{white}{PSNR $\uparrow$} & \cellcolor{white}{SSIM $\uparrow$} & \cellcolor{white}{ASR $\uparrow$} \\
        \midrule
        \textbf{$0.1$} & 0.702 & 12.78 & 0.39 & 0.7 & 0.650 & 11.99 & 0.32 & 0.99 \\
        \textbf{$10^{2}$} & 0.217 & 20.84 & 0.80 & 1.0 & 0.186 & 21.83 & 0.85 & 0.98 \\
        \textbf{$10^{5}$} & 0.221 & 20.75 & 0.80 & 0.8 & 0.175 & 22.95 & 0.88 & 0.82 \\
        \bottomrule
    \end{tabular}
\end{table}

\smallskip
\noindent
\textbf{Effect of Different Choices of Percentile of Distribution.}
We study the effect of different choices of percentile of distribution in the regularization term in Eq.~\ref{equ:distrib} of the synthesized scanner.
Take TrojAI round 3 as an example. We randomly select 100 clean models and 100 models attacked by polygon triggers, where the random seed is 1024.
We follow the experiment setup presented in Section~\ref{sec:scan} and use ABS as \Tech{}'s downstream scanner.
Besides the default choice of percentile range 15\%-85\%, we conduct two more experiments with percentile range 2\%-98\% and 30\%-70\%.
Table~\ref{tab:percentile} shows the results.
Observe that in all cases, \Tech{} can achieve high accuracy over 93\%, and the choice of percentile is a trade-off between low false negative rate and low false positive rate.
When the percentile is set at a large extent, e.g., 2\%-98\%, \Tech{} has a slightly high false positive rate 10\%, and a low false negative rate 5\%.
On the other hand, when the percentile is constrained in a small extent, e.g., 30\%-70\%, the false positive rate is reduced to 4\% while the false negative rate increases to 9\%.
This is reasonable as a large percentile range poses less regularization penalty on trigger inversion and thus more corner case triggers can be inverted. However, a small percentile range penalizes more on trigger inversion and hence fewer natural trojans (false positive cases) are induced.
The trade-off here is that whether the user prefers a low false positive rate or a low false negative rate.
Typically, we set the default percentile as 15\%-85\% for a balanced performance.

\smallskip
\noindent
\textbf{Effect of Using Different Pre-trained GANs in Attack Decomposition.}
In this study, we aim to show that \Tech{}'s decomposition process may not require that the GAN is well-trained on the model input domain.
We conduct an experiment in which we leverage a pre-trained GAN on CIFAR-10 to decompose 10 Reflection attack instances whose source images come from CIFAR-10, GTSRB and CelebA.
Note that GTSRB and CelebA have no overlapping classes with CIFAR-10.
Table~\ref{tab:gan} shows the results.
The first column denotes the dataset and the following columns measure the decomposition performance, same as Table~\ref{tab:decomposition}.
The first row shows the result when the decomposition process leverages the GAN pre-trained on CIFAR-10, the model input domain. Observe that the decomposition performance is very good with a small $\normlone$ error, high PSNR, SSIM scores and high accuracy/ASR for both decomposed clean images and triggers.
The second row illustrates the decomposition performance of GTSRB instances using the same GAN. We can see that there is a medium degradation on the decomposed clean images, with -0.23 SSIM score difference, and a small degradation on the decomposed trigger, with -0.13 SSIM score difference.
In a more extreme case that we use the GAN trained on CIFAR-10 with an input size 32$\times$32 to generate high-resolution images from CelebA with an input size 128$\times$128.
The decomposition performance is further reduced, with -0.39 SSIM score difference on the decomposed clean images and -0.26 on the decomposed trigger, compared to the results on CIFAR-10 instances.
However, in both cases, the decomposed clean images can achieve 100\% accuracy and the decomposed triggers can achieve > 95\% ASR.
We argue that the degradation on visual quality of decomposed clean images is acceptable as finally the decomposed trigger is used for scanner synthesis.
Besides, since the decomposed triggers achieve a high ASR, they are sufficient for further attack summarization.

\smallskip
\noindent
\textbf{Effect of Different Choices of Parameter $\alpha$ in Attack Decomposition.}
Parameter $\alpha$ in Eq.~\ref{e:loss} controls the trade-off between the reconstruction quality and the trigger effectiveness (Accuracy/ASR) during attack decomposition.
We take the decomposition of Reflection backdoor on CIFAR-10 as an example to study the effect of setting different values of $\alpha$.
Table~\ref{tab:alpha} shows the results.
The first column denotes the choice of $\alpha$ and the following columns show the decomposition performance.
Observe that when $\alpha$ is small, e.g., 0.1, the reconstruction quality is low, achieving near 0.3 SSIM score while the ASR of decomposed trigger is high, reaching nearly 100\%.
On the other hand, when $\alpha$ is large, e.g., $10^5$, the reconstruction quality tends to be high, increasing to an SSIM score of 0.8. However, the trigger effectiveness is reduced, i.e., 0.16 ASR degradation.
Overall, we find that $\alpha=1e^{2}$ provides the best trade-offs between the visual reconstruction quality and accuracy/ASR. Hence, we set $\alpha=1e^2$ for Eq.~\ref{e:loss}.

\end{document}